\documentclass[11pt,a4paper]{article}
\usepackage[T1]{fontenc}
\usepackage[utf8]{inputenc}%{accenti}
\usepackage[italian,greek,english]{babel}
\usepackage{amsmath}
\usepackage{mathrsfs}
\usepackage{graphicx}
\usepackage{amssymb}%{simboli come h tagliato}
\usepackage{mathtools}
\usepackage{bm}%{evidenzia-nero-in ambiente matematico}
\usepackage[output-decimal-marker={,}]{siunitx}
\DeclareGraphicsExtensions{.png,.jpg}
\DeclarePairedDelimiter{\abs}{\lvert}{\rvert}
\DeclareMathOperator{\Tr}{Tr}
\usepackage{braket}
\usepackage[square,numbers]{natbib}
\usepackage{natbib}
\usepackage{booktabs}%{aspetto-tabelle}
\usepackage[a4paper,top=2cm,bottom=2cm,left=2cm,right=2cm]{geometry}
\usepackage [svgnames]{xcolor}
\usepackage{bbm}
\usepackage{empheq}
\usepackage{footmisc}
\usepackage{caption}
\usepackage[margin=10pt, font=small, labelfont=bf, format=hang, font=up ]{caption}
\usepackage[footnote,printonlyused]{acronym}
\usepackage{scrextend}
\usepackage{pifont}%{simboli}
\usepackage[sep=5pt]{simpler-wick}
\usepackage[titletoc]{appendix}
\usepackage{xcolor}
\usepackage{cancel}
\usepackage{url}
\usepackage{authblk}%{per il preambolo pre-abstract}
\deffootnote{1.2em}{1em}{\textsuperscript{\thefootnotemark}\,\enskip}
\addto\captionsenglish{}
\numberwithin{equation}{section}

\newcommand{\DP}{\partial}
\newcommand{\al}{\alpha'}
\newcommand{\Hp}{_{\mathcal H^+}}

\newcommand{\R}{{\cal R}}
\newcommand{\T}{\textbf{T}}

\begin{document}

 \begin{center}
 {\Large \bf{Closed superstring moduli tree-level two-point scattering amplitudes \\
in Type IIB orientifold on $T^6/(Z_2\times Z_2)$}}
 \end{center}
 
\vspace{0.1cm}
 \begin{center}
{\large Alice Aldi\footnote{E-mail: alice.aldi@roma2.infn.it} and
  Maurizio Firrotta\footnote{E-mail: maurizio.firrotta@roma2.infn.it}}
  
\vspace{0.5cm}
{\it Dipartimento di Fisica, Universit\`a di Roma Tor Vergata, \\
I.N.F.N. Sezione di Roma Tor Vergata, \\
Via della Ricerca Scientifica, 1 - 00133 Roma, ITALY}
\end{center}

\vspace{3cm}
\begin{abstract}
We reconsider the two-point string scattering amplitudes of massless Neveu-Schwarz--Neveu-Schwarz states of Type IIB orientifold superstring theory on the disk and projective plane in ten dimensions and analyse its $\al$ expansion. We also discuss the unoriented Type IIB theory on $T^6/\mathbb{Z}_2\times\mathbb{Z}_2$ where two-point string scattering amplitudes of the complex K\"ahler moduli and complex structures of the untwisted sector are computed on the disk and projective plane. New results are obtained together with known ones. Finally, we compare string scattering amplitudes results at $\al^2$-order with the (curvature)$^2$ terms in the low energy effective action of D-branes and $\Omega$-planes in both cases. 
\end{abstract}

\newpage
\tableofcontents
%%%%%%%%%%%%
%%%%%%%%%%%%
%%%%%%%%%%%%%
\section{Introduction}\label{sec1}
Perturbative string theories are characterised by a genus expansion controlled by the string coupling, $g_s= e^{\langle \Phi \rangle}$. This expansion in worldsheet topologies where vertex operators are inserted allows to calculate n-point string scattering amplitudes at a given perturbative level. When strings propagate in a non-trivial background, the worldsheet action describes an interacting 2$d$ field theory that usually is not exactly solvable, but can be explored perturbatively as an expansion in powers of $\al$, i.e.\ as an  expansion in derivatives. For each fixed worldsheet topology, string scattering amplitudes encompass an infinite series in powers of $\al$, corresponding to terms of increasing order in momenta that lead to an infinite tower of higher and higher derivative terms in the Low Energy Effective Action (LEEA). This double perturbative expansion can be exploited to investigate possible corrections to the LEEA. On the other hand, corrections to the LEEA can be induced not only by the inclusion of the ``physical'' D-branes and $\Omega$-planes, but also in the non-perturbative regime by worldsheet instantons and D-branes instantons. 

D-branes are extended objects~\cite{Polchinski:1995mt} to which R-R $p$-forms can couple\,\footnote{The NS-NS 2-form, which is common to all the string theories, is in general coupled to the fundamental string $F_1$.} and also where open strings can end, i.e.\ the loci where gauge groups in the sense of quantum field theory appear. $\Omega$-planes, instead, are objects typical of unoriented string theory, in the sense that, when the orientifold projection $\Omega$ (worldsheet parity operator) is performed, part of the original spectrum is truncated since only states that are invariant under the exchange of left movers with right movers survive. Type-I string theory is the result of the $\Omega$-gauging of Type IIB. However, when compactification schemes and T-duality are taken into account, more general orientifold projections can be performed (as discussed below), producing Type II orientifold (or open descendant) models~\cite{Ortf1,
Ortf3,Bianchi:1990yu,Pradisi:1988xd,Gimon:1996rq,Bianchi:1990tb,Bianchi:1996sz,Ortf6,Ortfws1,Ortfws2} (see for a review \cite{Angelantonj:2002ct, OrtfRev2, Sagnotti:1996eb, Blumenhagen:2006ci}). 

As anticipated worldsheet instantons are non-perturbative in $\al$ expansion but perturbative in $g_s$ since they can appear at tree level in the $g_s$ expansion. In particular they appear in four dimensional string vacuum configurations when closed string worldsheets wrap non-trivial 2-cycles of the internal compactification manifold~\cite{Dine:1986zy,Dine:1987bq,Bianchi:2009ij,Bianchi:2012ud,Bianchi:2007fx}. On the other hand, objects that are non-perturbative in $g_s$, such as D-branes instantons (or ED-branes) and NS5-branes instantons, can wrap different non-trivial cycles which characterised the geometry of the internal compactified space~\cite{Johnson:1997ib,Evans:1997hk,Polchinski:1996fm,Bianchi:1997gt}. 

Compactification is a necessary step that one has to take in order to make contact with four dimensional physics starting from theories with extra dimensions. The role of the compactification process and the structure of the internal geometry enters not only in the study of non-perturbative contributions to the LEEA as we will discuss. In the LEEA only massless fields are taken as starting point in the construction of the field configuration that solves the e.o.m.\ of the effective theory. The standard assumption about the overall space-time geometry is that it can be factorised into a four dimensional Minkowskian geometry, $\mathcal M_4$, times a six dimensional compact manifold, $X_6$, where the specific properties of the latter affect in a non-trivial way the four dimensional physics. The simplest internal manifold that one can consider is a $T^6$-torus, which is a generalisation of the Kaluza-Klein compactification on a circle~\cite{Kaluza:1921tu,Klein:1926tv}. It falls in the class of \textit{toroidal} compactification. The main problem with this kind of compactification is the too large amount of supersymmetry in four-dimensions. One gets, in fact, the maximal $\mathcal N = 8 $ (in Type II) and $\mathcal N= 4$ (in Heterotic and Type-I), because all the initial supercharges are conserved. Going ahead with this purely geometrical point of view, where only the internal metric is non-trivial yet constant and all the other fields vanish, promising compactified models with less supersymmetry in four-dimensions are provided by Calabi-Yau manifolds~\cite{Candelas:1985en}. In fact, a $CY_3$ internal manifold admits a Ricci-flat K\"ahler metric~\cite{Eugenio Calabi,Yau:1977ms} thus satisfying the Einstein equations in vacuum. $CY$ compactifications preserve one quarter of the supersymmetries of the initial ten dimensional string models. As a result, the field content can be organised in $\mathcal N=2$ supermultiplets (Type II string) or $\mathcal N=1$ supermultiplets (Heterotic string). 

Compactifications produce a rather large number of light neutral scalar fields (experimentally unobserved up to now) called \textit{moduli} fields. Their role is to parametrise the size and shape of the compactification manifold $X_6$ or the position of the D-branes. Moreover their vev's directly affect several parameters of the four dimensional EA like gauge couplings and masses. The main difference between Heterotic  and Type II compactifications is the origin of moduli, namely whether they come from NS-NS states only (NS states Heterotic) or there is a mixing with the RR part (Type II). In the context of $CY$ compactifications the origin of the moduli fields gives constraints on the moduli space geometry which characterises the kinetic terms for the moduli (at tree level).
The moduli space for Type II theories on $CY_3$ is a direct product space \cite{Ferrara:1989ik, BLT} 
\begin{equation}
\mathscr M = \mathcal Q \times \mathcal S_K 
\label{eq:modulispace}
\end{equation}
where $\mathcal S_K$ a special K\"ahler manifold and $\mathcal Q$ a quaternionic manifold. The special K\"ahler manifold $\mathcal S_K$ describes only geometric moduli fields (moduli coming from NS-NS states) which fall into vector multiplets of $\mathcal N = 2$ supersymmetry. While, the quaternionic manifold $\mathcal Q$ contains both geometric and non-geometric moduli (coming from RR states) and the universal dilaton, all collected in hypermultiples. The number of these $\mathcal N=2$ supermultiplets dictates the complex or real dimension of the corresponding manifold, namely $h^{2,1}$ ($h^{1,1})$ complex dimension for $\mathcal S_K$ manifold in Type IIB(A), 4$(h^{1,1} + 1)$ (4$(h^{2,1} + 1)$) real dimension for $\mathcal Q$ Type IIB(A), where $h^{1,1}$ and $h^{2,1}$ are the non-trivial Hodge numbers of the Calabi-Yau threefolds, related to the K\"ahler moduli and the complex structure moduli respectively. In both Type II theories $\mathcal S_K$ vector multiplet moduli space doesn't receive any kinds of $g_s$ corrections, as it is exact at string tree level. In Type IIB this moduli space is also exact, at tree level, in the $\al$ expansion\,\footnote{In Type IIA some $\al$ string effects can shape the $\mathcal S_K$ moduli space as dictated by \textit{classical} geometry.}. On the other hand the hypermultiplet moduli space $\mathcal Q$, receives perturbative and non-perturbative string corrections. Perturbative $g_s$ corrections, like worldsheet instantons wrapping two-cycles in the $CY$ manifold, enter in Type IIA. While non-perturbative $D$-brane and $NS$-brane instantons wrapping supersymmetric cycles\,\footnote{All the even-cycles are supersymmetric while, among odd-cycles, only 3-cycles are.} inside the $CY$, enter both Type II theories. More precisely ED(-1), ED1, ED3, ED5 and ENS5 brane instantons enter in Type IIB, while ENS5 and ED2 brane instantons in Type IIA. Something special happens for fixed values of non-geometric moduli, when they become non dynamical fields the quaternionic manifold $\mathcal Q$ can be written as the direct product
\begin{equation}
\mathcal Q \sim \frac{SU(1,1)}{U(1)}\times \mathcal {\hat S}_K
\end{equation}
with $\mathcal {\hat S}_K$ a special K\"ahler manifold describing only the $h^{1,1}(h^{2,1})$ hypermultiplets that contain the geometric moduli and $SU(1,1)/U(1)$ the moduli space of the universal complex axion-dilaton hypermultiplet of Type IIB(A)~\cite{Ferrara:1989ik,Ferrara:1989py}. In the case where geometric moduli are fixed, the resulting moduli space for the axion-dilaton and the non-geometrical moduli would be $SU(1,h^{1,1}+2)/(U(1)\times SU(h^{1,1} +2))$ with $h^{1,1}$ ($h^{2,1}$) the number of these hypermultiplets in Type IIB  (Type IIA)~\cite{Ferrara:1989ik,Ferrara:1989py}. 

From the string point of view this kind of compactifications can be used as a reliable four dimensional approximation of the ten dimensional supergravity theory, when only the tree-level term in the $\al$-expansion is retained, i.e.\ in the large volume regime where the $R$ length scale which characterises the size of $CY$ manifold is large. This is because, as we said, in the presence of a non trivial curved background, the $2d$-field theory becomes an interacting theory. In this situation one loses the powerful tools represented by the vertex operator formulation of the free $2d$-CFT, two-point functions definition that are fundamental for the construction of a four dimensional effective field theories by a string $S$-matrix approach. Compactifications suitable for an $S$-matrix approach are, for instance, the Toroidal \textit{orbifolds}~\footnote{For completeness, we recall that there are also several \textit{abstract} CFT constructions that provide solutions to these problems of 2d-theory in curved background as for instance the Gepner models, not discussed in this paper \cite{Angelantonj:1996mw}.}. Essentially they are the result of the quotient of a toroidal background, $T^d$, under the action of some discrete group of isometries (orbifold group). Examples are 
\begin{equation}
X_6 = \frac{T^6}{\mathbb Z_N}\,\quad,\,\quad X_6 = \frac{T^6}{\mathbb Z_N\times \mathbb Z_M}
\label{eq:Torbifold}
\end{equation}
These quotients act identifying some points and leaving fixed others on the six-torus $T^6$ lattice. At the fixed points, orbifolds are singular.
%fail to be manifolds, because they are a singular points on $X^6$. 
But outside the singularities, the resulting geometry is locally flat, thus one can use CFT methods. Sometimes orbifolds are referred to as $CY$ limits because the singularities can be removed using a \textit{blowing up} process. Naturally the field content of orbifold compactifications, is a truncation of the six-torus $T^6$ compactification to invariant states under the orbifold action. They fall into $\mathcal N = 2$ supermultiples (Type II) or $\mathcal N=1$ supermultiplets (Heterotic) in four-dimensions, like in the case of compactification on $CY_3$. The field content of Type-I superstring both on toroidal orbifolds and on $CY_3$ can also be organised in $\mathcal N=1$ supermultiplets in four-dimensions. We warn that in the literature sometimes the Type-I appears as a special case of more general class of Type II \textit{orientifold} compactifications, where in general the final amount of supersymmetries in four-dimensions is $\mathcal N=1$ due to the non-trivial orientifold projection, and the resulting spectrum is a truncation to states that are invariant under the orientifold action. 

In Type II orientifold compactifications extended objects like $D$-branes and $\Omega$-planes need to be included for consistency. Moduli are collected into chiral multiplets, and orientifold projection at tree level doesn't destroy the structure of closed string moduli space~\eqref{eq:modulispace} unless open strings moduli (Wilson lines) coming from $D$-branes that wrap internal directions, enter the spectrum~\footnote{We don't consider this general case, but when $D$-branes moduli are taken as dynamical fields the geometry of moduli space needs to be described by closed as well as open string moduli, since in principle there is no factorisation, and the structure of the moduli space is more complicated.}. 

Even if there is yet no phenomenological, evidence of supersymmetry, the main attention in the literature is focused on theories with $\mathcal N = 1$ supersymmetry in four-dimensions because the presence of chiral matter is needed in order to bridge string theories and the Standard Model. The information on the structure of any $\mathcal N=1$ supergravity Lagrangian in four-dimensions are encoded in the K\"ahler potential $\mathcal K(\phi,\bar \phi)$, the superpotential $\mathcal W (\phi)$ and the (matrix) gauge kinetic function $f(\phi)$. In the context of string compactification these functions depend non-trivially on the complex scalar fields $\phi$ which arise from compactification process itself. The resulting four dimensional LEEA for the bosonic fields can be written in the form
\begin{equation}
\mathcal{L}^{SG}_{_{\mathcal N=1}} = \frac{R}{2k_4^2} - \mathcal{K}_{I\bar J}(\phi,\bar\phi)\nabla_{\mu}\phi^I\nabla^{\mu}\phi^{\bar J} - V(\phi,\bar\phi) - \frac{1}{8}\mathbb{R}e(f_{ab}(\phi))F^a_{\mu\nu}F^{b\,\mu\nu} - \frac{1}{8}\mathbb{I}m(f_{ab}(\phi))F^{a}_{\mu\nu}\tilde{F}^{b\,\mu\nu}+{\dots}
\label{eq:effective.act}
\end{equation}
where $a,b$ label gauge group representations and $\mathcal{K}_{I\bar J}$ is the K\"ahler metric 
\begin{equation}
\mathcal{K}_{I\bar J}(\phi,\bar\phi) = \frac{\DP^2 \mathcal{K}(\phi,\bar\phi)}{\DP\phi^I\DP\phi^{\bar J}}
\label{eq:Kmanifold}
\end{equation}
with $I,\bar J$ running both on geometrical and non-geometrical moduli. The scalar potential $V(\phi,\bar\phi)$ with it's $F$- and $D$-term is given by
\begin{equation}
V(\phi,\bar\phi) = e^{k_4^2\mathcal K}\left(\mathcal{K}^{I\bar J}\nabla_{I}\mathcal{W}\nabla_{\bar J}\bar{\mathcal{W}} - 3 k_4^2\abs{\mathcal W}^2\right) + \frac{1}{2}\left(\mathbb{R}e(f^{-1})\right)_{ab}D^aD^b
\label{eq:scalar potential}
\end{equation}
In the above equation $\nabla_I \equiv \DP_I + k_4^2(\DP_I\mathcal K)$ is the K\"ahler covariant derivative.\,\footnote{The fermionic terms can be obtained by supersymmetry.} Upon computing tree-level string scattering amplitudes, the explicit form of the terms in the Lagrangian can be derived and compared with the terms coming from the dimensional reduction of the higher dimensional string model in consideration. Up to the point where $\mathcal N=1$ supersymmetry is unbroken, it is also known which kinds of corrections the three functions $\mathcal K$, $\mathcal W$ and $f_{ab}$ can receive. The K\"ahler potential $\mathcal K$ can in principle have both perturbative and non-perturbative corrections in $\al$ and $g_s$. The superpotential $\mathcal W$, by general non-renormalization theorems, can receives only non-perturbative $g_s$ corrections while, in the gauge kinetic function, $f_{ab}$, can appear corrections both perturbative (up to one-loop) and non-perturbative in $g_s$. This kind of corrections are needed in order to arrive at some string solutions which can describe a positive vacuum energy configuration (de Sitter vacua) and \textit{stabilise} the moduli to positive mass-squared values. If (some of) the moduli remain massless they can mediate long range forces and, from the cosmological point of view, \textit{overclose} the universe. This last problem emerges from the fact that at the tree-level of the $\al$ and $g_s$ expansions, the scalar potential $V(\phi,\bar \phi)$~\eqref{eq:scalar potential} is of \textit{non-scale} type. It is, in fact, identically zero both for the K\"ahler moduli (geometrical and non-geometrical) and open moduli (when present), leaving all them unfixed~\footnote{The axion-dilaton and complex structure moduli can in general be stabilised turning on \textit{fluxes}~\cite{Blumenhagen:2006ci}.}. The problem of these unfixed moduli in Type IIB occurs in general because the superpotential $\mathcal W$ at tree-level can only depend on the complex axion-dilaton and the complex structure moduli (geometric moduli). Thus not all the desired F-terms can be generated. The~KKLT \cite{Kachru:2003aw} and the LVS~\cite{Balasubramanian:2005zx} scenarios are the two principal ones that took into account the K\"ahler moduli stabilisation by adding non-perturbative contributions (to both) and perturbative ones (to LVS). The challenges of building cosmological string models which can link the \textit{inflation} phase to the \textit{cosmological standard model}(CSM) beyond the solutions of the classical theory, are based on these ideas and considerations.\,\footnote{An exhaustive overview on inflation models by string theories approaches can be found in~\cite{Baumann:2014nda}.} 

Our work is based on the computation of tree-level string scattering amplitudes useful to extract information on the structure of the LEEA terms at specific orders in the $g_s$ expansion, i.e.\ coming from disk and projective plane worldsheet.
\\\\
The paper is organized as follows. In Section~\ref{sec1} an introduction with motivations and general aspects related to the topic are given. In Section~\ref{sec2} closed string scattering amplitudes in $D$ $\leq 10$ dimensions on the disk $D_2$ and real projective plane $RP_2$ are computed. We give a detailed review of earlier disk amplitude computations and results obtained in \cite{Garousi:1996ad,Hashimoto:1996kf,Hashimoto:1996bf, Bachas:1999um}. We give also a detailed review of earlier real projective plane amplitude computations and results obtained in \cite{Garousi:2006zh}. An introduction on the role of real projective plane is provided for completeness to recall the geometry of extended objects like $\Omega_p$-planes in unoriented theories. In Section~\ref{sec3} we give an overview on Type IIB orientifold model and focus on Type IIB orientifold on $\T^6/{\mathbb Z_2}\times{\mathbb Z_2}$. In this framework we review the construction of the vertex operators for the closed untwisted K\"ahler moduli ($T$) and complex structure moduli ($U$) in presence of $D$-brane and the new construction in presence of $\Omega$-plane. Tool blocks for the two-point functions on $D_2$ and $RP_2$ are included. In Section~\ref{sec4} two-point string scattering amplitudes with untwisted K\"ahler moduli ($T$) and complex structure moduli ($U$) of Type IIB orientifold on $\T^6/{\mathbb Z_2}\times{\mathbb Z_2}$ on the disk $D_2$ are reviewed and new computations on the projective plane $RP_2$ are performed and discussed. We have verified that there are no corrections to the tree-level K\"ahler potential $\mathcal K$ of the analyzed model and no order $\al$-corrections to the Einstein-Hilbert term $R$ when also the scattering amplitudes on $RP_2$ are considered.\footnote{In general there are arguments like in \cite{Berg:2005ja,Berg:2014ama} where do not excluded the presence of these corrections at tree-level.} In Section~\ref{sec5} we perform a comparison at $\al^2$-order between string scattering amplitudes results and (curvature)$^2$ terms in the LEEA of $D$-branes and $\Omega$-planes, starting from a generic Type IIB orientifold model in high dimensions and going to the specific case of Type IIB orientifold on $\T^6/{\mathbb Z_2}\times{\mathbb Z_2}$. Finally, in Section~\ref{sec6}, discussions and perspectives for future lines of investigation are given.

\section{Scattering Amplitudes from $D_P$-brane and ${\Omega_P}$-plane}
\label{sec2}

The purpose of this Section is to review ten (or less) dimensional two-point closed string scattering amplitudes, involving massless states, on the disk and real projective plane and their leading $\alpha'$ contributions (in particular $\alpha'^2$), in the context of unoriented string theory where objects like $D_{P}$-branes and ${\Omega}_{P}$-planes are necessary for consistency reasons. 
There are many works that approach these topics \cite{Garousi:1996ad,Hashimoto:1996kf,Hashimoto:1996bf, Bachas:1999um},but for our purposes it is instructive to reproduce these results and point out precisely the $\alpha'^2$ contributions that will be present in low energy effective action as already mentioned in the works \cite{Bachas:1999um, Fotopoulos:2001pt, Fotopoulos:2002wy} for the disk and \cite{Garousi:2006zh} for the projective plane. 
In the following one can find the general setup of scattering amplitudes including the specific choice of fixing gauge for the invariance under the specific conformal killing group together with results and implications. All the computational technicalities are relegated to the appendices.

\subsection{${\cal A}_{D_2}\left(NS{-}NS,NS{-}NS\right)$}
The starting point is the computation of scattering amplitude of two massless NS-NS states from $D_P$-branes, which at tree-level involves the disk $D_2$ as worldsheet. To this aim  one is forced to use vertex operators different pictures in order to cancel the vacuum extra charge of the superghosts. In addition, the presence of the $D_p$-brane is taken into account by introducing the reflection matrix $\cal R$, a diagonal matrix with entries $+1$ in the Neumann (Poincarè preserving) directions and $-1$ in the Dirichlet (Poincarè non-preserving) directions.  This information is encoded in the vertex operators as follows:
\begin{equation}
\begin{split}
{\cal W}_{{_{NS{-}NS(q,\bar q)}}}(E,k,z,\bar{z})&=E_{M N} :{\cal V}^{M}_{(q)}(k,z)\,{\cal V}^{N}_{(\bar{q})}(k,\bar{z}):\,= E_{M N}\, {\cal R}^{N}_{Q} :{\cal V}^{M}_{(q)}(k,z): :{\cal V}^{Q}_{(\bar{q})}(k \R,\bar{z}):\\
\end{split}
\label{vertex1}
\end{equation}
In particular, the form of the vertices in the two standard choices $(0,0)$ and $(-1,-1)$ of the superghost picture is
\begin{equation}
\begin{split}
&{\cal W}_{{_{NS{-}NS(-1,-1)}}}(E,k,z,\bar{z})= E_{MN}\,\R^{N}_{P} :e^{-\phi}\psi^{M}e^{ikX}(z):\,:e^{-\phi}\psi^{P}e^{i k\R X}(\bar{z}): \\
%%%
&{\cal W}_{{_{NS{-}NS(0,0)}}}(E,k,z,\bar{z})= E_{MN}\,\R^{N}_{P} {:}\left(i\DP X^{M}{+} \frac{\al}{2} (k\psi)\,\psi^{M}\right)e^{ikX}(z){::}\left( i\bar\DP X^{P} {+} \frac{\al}{2} (k\R\psi \big)\psi^{P}\right) e^{ik\R X}(\bar{z}){:}
\end{split}
\label{vertex2}
\end{equation}
The scattering amplitude is then given by
\begin{equation}
\begin{aligned}
&{\cal A}_{D_2}^{_{\left(NS{-}NS,NS{-}NS\right)}}= g_c^2C_{_{D_2}}\int_{_{\mathcal H^+}}\frac{d^2z_1d^2z_2}{V_{CKG}} \langle {\cal W}_{{_{NS{-}NS(-1,-1)}}}(E_1,k_1,z_1,\tilde z_1){\cal W}_{{_{NS{-}NS(0,0)}}}(E_2,k_2,z_2,\tilde z_2)\rangle_{ \mathcal {H^+}}\\
%%%%%%%%%%%%%
&=\frac{2g_c^2C_{_{D_2}}}{\al}\int_{_{\Hp}}\frac{d^2z_1d^2z_2}{V_{CKG}}\,E^1_{M_1N_1}E^2_{M_2N_2}\R^{N_1}_{P_1}\R^{N_2}_{P_2} \, \langle:e^{-\phi}\psi^{M_1} e^{ik_1X}(z_1)::e^{-\phi}\psi^{P_1}e^{ik_1\R X}(\bar{z}_1): \\
%%%%%%%%%%%%%
&\hspace{2.5cm}:\Big(i\DP X^{M_2} + \frac{\al}{2}\big(k_2\psi\big)\psi^{M2}\Big)e^{ik_2X}(z_2)::\Big(i\bar\DP X^{P_2} + \frac{\al}{2}\big(k_2\R\psi\big)\psi^{P_2}\Big)e^{ik_2\R X}(\bar{z}_2):\rangle\\
%%%%%%%%
&=\frac{2g_c^2C_{_{D_2}}}{\al}\int_{_{\Hp}}\frac{d^2z_1d^2z_2}{V_{CKG}} \,\langle :e^{-\phi}(z_1): :e^{-\phi}(\bar{z}_1):\rangle\, \left( {\cal M}^{(1)}+{\cal M}^{(2)}+{\cal M}^{(3)}+{\cal M}^{(4)}\right)
\end{aligned}
\label{amplD2_1}
\end{equation}
where $V_{CKG}$ is the conformal killing volume and $\mathcal M$'s indicate the different sub-amplitudes that one has to calculate, as for instance    
\begin{equation}
{\cal M}^{(1)}=E^1_{M_1N_1}E^2_{M_2N_2}\R^{N_1}_{P_1}\R^{N_2}_{P_2}\, \langle:\psi^{M_1}e^{ik_1X}(z_1){::}\psi^{P_1}e^{ik_1\R X}(\bar z_1){::}i\DP X^{M_2}e^{ik_2X}(z_2){::}i\bar\DP X^{P_2}e^{ik_2\R X}(\bar z_2):\rangle\,\,.
\label{M1D2}
\end{equation}
In order to compute all the contractions the relevant two-point functions are
\begin{equation}
\boxed{
\begin{aligned}
&\langle e^{-\phi}(z)e^{-\phi}(\bar z)\rangle_{D_2} = \frac{1}{(z- \bar z)}\,,\quad \langle X^M(z)X^N(\bar{z})\rangle_{D_2}=-{\alpha' \over 2} \eta^{M N} \log|z-\bar{z}|\, \\
&\langle X^M(z) X^N(\bar{w})\rangle_{D_2}=-{\alpha' \over 2} \eta^{M N} \log ( z-\bar{w})\,,\quad \langle X^M(z) X^N(w)\rangle_{D_2}=-{\alpha' \over 2} \eta^{M N} \log(z-w)\\
&\langle \psi^M(z)\psi^N(\bar{z})\rangle_{D_2} = \frac{\eta^{M N}}{(z- \bar{z})}\,,\quad\langle\psi^M(z) \psi^N(\bar{w})\rangle_{D_2} = \frac{\eta^{M N}}{(z- \bar{w})}\,,\quad\langle \psi^M(z) \psi^N(w) \rangle_{D_2} = \frac{\eta^{M N}}{(z - w)} 
\end{aligned}}\label{eq:correlator10D_disk}
\end{equation}
The explicit result of the $\mathcal M^{(i)}$'s calculation, being quite cumbersome, is reported in Appendix A.2 of ref.\,\cite{AM:appendice} . 
%%%%%%%%%%%%%%%%%%%%%%%%%%%%%%
%%%%%%%%%%%%%%%%%%%%%%%%%%%%%%
First of all, one has to remove the redundancy related to the Conformal Killing Group of $D_2$.  Identifying the disk with $\mathcal H^+ = S_2/\mathfrak{I}_{D_{2}}$, the CKG of the residual symmetry is $PSL(2,\mathbb{R}) = SL(2,\mathbb{R})/\mathbb{Z}_2$, i.e. the subgroup of $SL(2,\mathbb{C})$ that preserves the involution $\mathfrak{I}_{D_2}(z)=\bar z$. The finite transformations are \cite{BLT,Tanii,Kitaev:2017hnr}
\begin{equation}
\begin{aligned}
&z\mapsto z' = \frac{az + b}{cz +d}\\
&M=\begin{pmatrix}a&b\\c&d \end{pmatrix}\in SL(2,\mathbb{R})\qquad\text{i.e.}\,\,\,\begin{cases}a,b,c,d\in\mathbb{R}\\ad - cb =1\rightarrow\,\text{det}M = 1\end{cases}\,\,\,.
\end{aligned}\label{eq:PSL2t}
\end{equation}
The corresponding Lie algebra is $\mathfrak{sl}_2(\mathbb R)$, whose generators are $2\times2$ traceless real matrices \cite{Gilmore:2008zz}. The  infinitesimal transformation can be obtained from the finite transformation expanding it around $a=d=1; c=b=0$ as follows
\begin{equation}
\begin{aligned}
\delta(z') &= \delta\Big(\frac{az + b}{cz + d}\Big)\Big|_{a=d=1; c=b=0}\\&=\frac{z}{cz + d}\delta a + \frac{1}{cz + d}\delta b - \frac{az + b}{(cz + d)^2}\delta d - \frac{az^2 + bz}{(cz +d)^2}\delta c\,\,\Big|_{a=d=1; c=b=0}\\
&= z\delta a + \delta b -z\delta d - z^2\delta c
\overset{\delta a + \delta d  = 0}{\rightarrow} 
\delta b + 2\delta a\, z - \delta c\,z^2\\
& = p + qz + mz^2\qquad p,q,m \in \mathbb{R}\,\,\,.
\end{aligned}
\label{Psl(2,R)}
\end{equation}
where $p =\delta b\,,q=2\delta a$ and $m =-\delta c$. Setting $z= x + iy$ we have 

\begin{equation}
\begin{aligned}
&\delta (x' + iy') = p + q(x + iy) + m(x + iy)^2\rightarrow\begin{cases}\delta x' = p + qx + m(x^2 - y^2)\\\delta y' = qy + 2mxy\end{cases}\,\,\,.\label{eq:PSL22t}
\end{aligned}
\end{equation}
The $PSL(2,\mathbb R)$ symmetry allows to fix the positions of three chiral vertex operators. In the case with two closed string one can fix two chiral(antichiral) and one antichiral(chiral). In the case of two closed strings, using the doubling trick, the amplitude can be written in the form
\begin{equation}
\begin{aligned}
&\int \frac{dz_1d\bar z_1d_2d\bar z_2}{V_{CKG}}\big<:V^1_L(z_1)::V^1_L(\bar z_1)::V^2_L(z_2)::V^2_L(\bar z_2):\big>\,;\quad  V_{CKG} = \int dp\,dq\,dm
\end{aligned}
\label{gen_2ptD2}
\end{equation}
where the integral over the parameters is divergent in the absence of gauge fixing. To do so we write the surface element $dz\,d\bar z$ in terms of its real components \cite{BLT}
\begin{equation}
dz_1d\bar z_1dz_2d\bar z_2 = 4dx_1dy_1dx_2dy_2
\label{element_vol}
\end{equation}
and use \eqref{eq:PSL2t} to fix, for instance,  $x_1,x_2$ and $y_1$:  
\begin{equation}
\begin{aligned}
&4dx_1dy_1dx_2dy_2 = 4\,\abs{J}\,dp\,dq\,dm\,dy_2\,;\quad J = \det\begin{vmatrix} 1&0&1\\x_1&y_1&x_2\\(x_1^2 {-} y_1^2)&2x_1y_1&(x_2^2 {-} y_2^2) \end{vmatrix}=y_1(y^2_1 {-} y_2^2) + y_1(x_1 {-} x_2)^2 \label{eq:Jacob1}
\end{aligned}
\end{equation}
The same result can be obtained by inserting three unintegrated vertices, i.e putting $cV$ in \eqref{gen_2ptD2} instead of $\int V$, where $c$ is the reparametrization ghost. This means that the $\langle\abs{ccc}\rangle$ ghost correlator has to reproduce exactly the Jacobian of the $PSL(2,\mathbb R)$ transformation. After fixing the symmetry, directly using the $PSL(2,\mathbb R)$ transformations or the correct $\langle\abs{ccc}\rangle$ correlator must give the same result. In \cite{BLT} one can check that in the most common examples such as the case of three unintegrated closed vertices on the sphere or three unintegrated open vertices on the disk, the Jacobian of the transformation can be written as a $\langle\abs{ccc}\rangle\langle\abs{\bar c\bar c\bar c}\rangle$ and $\langle\abs{ccc}\rangle$ ghost correlator, respectively. The cases of purely closed or open/closed amplitudes on the disk are less straightforward to discuss. The reason is that $PSL(2,\mathbb R)$ acts on the complex variable $z$ in non trivial way \eqref{eq:PSL2t} and with the freedom of fixing only three real paramiters. In the purely closed amplitudes, the Jacobian \eqref{eq:Jacob1} can not be reproduced by a single $\langle\abs{cc\bar c}\rangle$ insertion, but one needs to insert the  linear combination 
\begin{equation}
\begin{aligned}
\abs{\langle c(z_1)c(\bar z_1)c(z_2)\rangle + \langle c(z_1)c(\bar z_1)c(\bar z_2)\rangle}&= \abs{(z_1 - \bar z_1)(\bar z_1 - z_2)(z_1 - z_2) + (z_1 - \bar z_1)(\bar z_1 - \bar z_2)(z_1 - \bar z_2)}\\
%%%%%%
%&= |(x_1 {+} i y_1 {-} x_1 {+} iy_1)(x_1{-}iy_1 {-}x_2 {-}iy_2)(x_1 {+} iy_1 {-} x_2 {-}iy_2)\\&+(x_1 {+} iy_1 {-}x_1 {+} iy_1)(x_1 {-}iy_1 {-}x_2 {+}iy_2)(x_1 {+}iy_1 {-} x_2 {+} iy_2)|\\
&=\abs{ 4y_1(y_1^2 - y_2^2) +4y_1(x_1 - x_2)^2}
\end{aligned}
\label{eq:ghost1}
\end{equation}
that takes into account both the position of the vertices in $\mathcal H^+$ as well as those of the images in $\mathcal H^-$. One gets in this way
\begin{equation}
dz_1d\bar z_1dz_2d\bar z_2 = 4\,\abs{J}\,dp\,dq\,dm\,dy_2 \equiv \abs{\langle c(z_1)c(\bar z_1)c(z_2)\rangle + \langle c(z_1)c(\bar z_1)c(\bar z_2)\rangle}\,dp\,dq\,dm\,dy_2
\label{eq:cambio}
\end{equation}
One can see that specializing conveniently the points $z_1, z_2$ to be %
\begin{equation}
\begin{aligned}
&z_1\mapsto z'_1 = i; \quad z_2\mapsto z'_2 = iy; \quad \Leftrightarrow \quad x_1 = 0; \quad y_1= 1; \quad x_2 = 0, \quad y_2= y\\
&\bar z'_1 = -i; \quad \bar z'_2 = -iy \quad  \Leftrightarrow \quad x_1 = 0, \quad y_1= 1; \quad x_2 = 0,\quad y_2= y
\end{aligned}
\label{eq:fixpoints1}
\end{equation}
and inserting them in \eqref{eq:cambio}, with the help of \eqref{eq:Jacob1} and \eqref{eq:ghost1}, the final integration measure takes the form
\begin{equation}
\begin{aligned}
dz_1d\bar z_1dz_2d\bar z_2 &= 4(1 {-}y^2)\,dp\, dq\, dm\, dy\,.%&\equiv \abs{\big(i {-} (-i)\big)\big((-i) {-} iy\big)(i {-} iy) {+} \big(i {-} (-i)\big)\big((-i) {-} (-iy)\big)\big(i {-} (-iy)\big)}\,dp\,dq\,dm\,dy\\&\equiv4(1 -y^2)\,dp\, dq\, dm\, dy
\end{aligned}
\label{eq:cghostD2}
\end{equation}
%%%%%%%%%%%%%%%%%%%%%%%%%%%%%%%
%%%%%%%%%%%%%%%%%%%%%%%%%%%%%%%
%%%%%%%%%%%%%%%%%%%%%%%%%%%%%%%

Putting everything together one gets for the amplitude \eqref{amplD2_1}
\begin{equation}
\begin{aligned}
{\cal A}_{D_2}^{\left(NS{-}NS,NS{-}NS\right)}= \frac{4g_c^2C_{_{D_2}}}{\al}\int_0^1dy\,&\left(\frac{4y}{(1 {+} y)^2}\right)^{-\al s}\left(\frac{1 {-} y}{1 {+} y}\right)^{-\al \frac{t}{2}}4(1{-}y^2)\\&\left\{\frac{a_1}{(1{ -} y^2)^2} {+} \frac{a_2}{4y(1 {+} y)^2} {+} \frac{a_3(1 {+} \al s)}{16y^2}\right\}
\label{eq:10D-prel-res}
\end{aligned}
\end{equation}
where $s$ and $t$ are the Mandelstam variables,\footnote{Our definition of Mandelstam variable $t$ is different from the one in \cite{Hashimoto:1996bf}.} see Appendix \ref{KIN}, and  
\begin{equation}
\begin{split}
a_1 =& -\frac{\al^2}{4}\bigg\{%\frac{\Tr({E}_2^{_{_T}}{E}_1)k_2{\R}k_2}{2} 
\frac{\Tr({E}_1^{_{_T}}{E}_2)k_1{\R}k_1}{2}
 %+ \Tr({\R}{E}_2)k_2{E}_1k_2 
+ \Tr({\R}{E}_1)k_1{E}_2k_1%+\\&\hspace{1.2cm} - k_1{E}_2{\R}{E}_1k_2
 - k_2{E}_1{\R}{E}_2k_1 
 %+ k_1{\R}{E}_2^{_{_T}}{E}_1k_2 
 + k_2{\R}{E}_1^{_{_T}}{E}_2k_1 
 %+\frac{k_1{E}_2^{_{_T}}{E}_1k_2}{2}
 + \frac{k_2{E}_1^{_{_T}}{E}_2k_1}{2} %+ k_1{\R}{E}_2{E}_1^{_{_T}}k_2 
 \\&\hspace{1.2cm}+ k_2{\R}{E}_1{E}_2^{_{_T}}k_1 %+ \frac{k_1{E}_2{E}_1^{_{_T}}k_2}{2}
  + \frac{k_2{E}_1{E}_2^{_{_T}}k_1}{2} + (1\leftrightarrow 2)\bigg\}\\
a_2 = & - \frac{\al^2}{4}\bigg\{%\frac{\Tr({E}_2^{_{_T}}{E}_1)k_2{\R}k_2}{2} + 
\frac{\Tr({E}_1^{_{_T}}{E}_2)k_1{\R}k_1}{2} %- \frac{\Tr({E}_2{\R}{E}_1{\R})k_1{\R}k_1}{2}
 - \frac{\Tr({E}_1{\R}{E}_2{\R})k_2{\R}k_2}{2}%\\&\hspace{1.2cm} 
 %- \Tr({\R}{E}_2)k_2{E}_1{\R}k_1
%%%%%%%
- \Tr({\R}{E}_1)k_1{E}_2{\R}k_2 %+ \Tr({\R}{E}_2)k_1{\R}{E}_1{\R}k_2 
+ \Tr({\R}{E}_1)k_2{\R}{E}_2{\R}k_1\\&\hspace{1.2cm} %-k_2{\R}{E}_2{\R}{E}_1{\R}k_1
 - k_1{\R}{E}_1{\R}{E}_2{\R}k_2
%%%%%%
%+\frac{k_2{\R}{E}_2^{_{_T}}{E}_1{\R}k_1}{2} 
+ \frac{k_1{\R}{E}_1^{_{_T}}{E}_2{\R}k_2}{2} 
%+ \frac{k_2{\R}{E}_2{E}_1^{_{_T}}{\R}k_1}{2} +\\&\hspace{1.2cm}
+ \frac{k_1{\R}{E}_1{E}_2^{_{_T}}{\R}k_2}{2} + (1\leftrightarrow2)\bigg\}\\
a_3 =&\,\frac{\al}{2}\Tr({\R}{E}_1)\Tr({\R}{E}_2)\,.
\end{split}
\label{kinemat_ai}
\end{equation}
The complete derivation of the $a_i$ coefficients can be found in Appendix A.3 of ref.\cite{AM:appendice}. Following \cite{Garousi:1996ad,Hashimoto:1996kf,Hashimoto:1996bf} it is convenient to perform in \eqref{eq:10D-prel-res} the change of variable 
\begin{equation}
\begin{aligned}
&y = \frac{1 - \sqrt{x}}{1 + \sqrt{x}}
\end{aligned}\label{sub}
\end{equation}

%one can write the integrals as 

%\begin{equation}
%\begin{aligned}
%{\cal A}_{D2}^{\left(NS{-}NS,NS{-}NS\right)}&=\frac{4g_c^2C_{_{D2}}}{\al}\int_0^1dx\,(1 - x)^{-\al s}  \,x^{-\al \frac{t}{4}}\,\left\{\frac{a_1}{ x} + \frac{a_2}{(1 - x)} + \frac{a_3(1 {+} \al s)}{(1 - x)^{2}}\right\}
%\end{aligned}
%\end{equation}
which makes the result a manifest combination of Euler Beta function (B), that is
\begin{equation}
\small{\begin{aligned}
&{\cal A}_{D_2}^{\left(NS{-}NS,NS{-}NS\right)}=\\
&=\frac{4g_c^2C_{_{D_2}}}{\al}\Big\{a_1B[(-\al s {+}1); (-\al t/4)]+ a_2B[(-\al s); (-\al t/4{+}1)]+a_3(1 {+} \al s)B[(-\al s {-} 1);(-\al t/4{+} 1)]\Big\}\,.
\end{aligned}}
\end{equation}

Using the $\Gamma$ function properties, one can write the previous equation in the compact form
%\begin{equation}
%\begin{split}
%\Gamma(1 + b)= b\Gamma(b)
%&\frac{2g_c^2C_{_{D2}}}{\al}\bigg\{a_1(-\al s) {-} a_2(\al {t/4}){-} a_3(\al t/4)(\al s {+} \al t/4)\bigg\}\frac{\Gamma(-\al s)\Gamma(-\al t/4)}{\Gamma({-}\al s {-} \al t/4 {+}1)}
%\end{split}\label{GammaPr}
%\end{equation}
\begin{equation} \boxed{
{\cal A}_{D_2}^{\left(NS{-}NS,NS{-}NS\right)}=\frac{4g_c^2C_{_{D_2}}}{\al}\left(-\al s\,a_1 - \al \frac{t}{4}\,\hat{a}_2\right)\frac{\Gamma(-\al s)\Gamma(-\al t/4)}{\Gamma(-\al s - \al t/4 +1)}
}
\label{eq:D2res}
\end{equation}
which is explicitly symmetric under the $(1\leftrightarrow 2)$ exchange \cite{Garousi:2006zh,Garousi:1996ad,Hashimoto:1996kf,Hashimoto:1996bf} with
%%%%%
\begin{equation}
\hat{a}_2 =\,a_2 +\bigg(\al s + \al \frac{t}{4}\bigg)\,a_3\,\,.
\label{a2hat}
\end{equation}
In order to identify the leading $\alpha'$ corrections one expands the combination of gamma functions in the limit $\alpha'\,\rightarrow \,0$ obtaining
\begin{equation}
\frac{\Gamma(-\al s)\Gamma(-\al t/4)}{\Gamma(-\al s - \al t/4 +1)}={4 \over \alpha' s \,\alpha' t} - \zeta(2) - \left(\alpha' s + \alpha' {t\over 4}\right)\zeta(3) + O(\alpha'^2)\ .
\label{eq:gammaexp}
\end{equation}

Combining this with the terms above, one can find that the amplitude \eqref{eq:D2res} exhibits open string poles in the s-channel and closed string poles in the t-channel as expected. There are also terms of order $(\al)^0$ proportional to $Tr({\R}{E}_1)Tr({\R}{E}_2)$. The  $(\alpha')^2$ terms coming from \eqref{eq:D2res} read

\begin{equation}
g_c^2C_{_{D_2}}\zeta(2)\left\{ 4s\, a_1+t\,a_2+\left( \alpha' s + {\alpha' t \over 4} \right) t \,a_3   \right\}\,.
\label{eq:alpha2}
\end{equation}
%(that are not a single valued projectable terms).  
%%%%%%%%%%%%%%%%%%%QUI

%%%%%%%%%%%%%%%%%%%%%%%%%%%%%%%%%%%
%%%%%%%%%%%%%%%%%%%%%%%%%%%%%%%%%%%
%%%%%%%%%%%%%%%%%%%%%%%%%%%%%%%%%%%
%%%%%%%%%%%%%%%%%%%%%%%%%%%%%%%%%%%
%%%%%%%%%%%%%%%%%%%%%%%%%%%%%%%%%%%
%%%%%%%%%%%%%%%%%%%%%%%%%%%%%%%%%%%
%%%%%%%%%%%%%%%%%%%%%%%%%%%%%%%%%%%
%%%%%%%%%%%%%%%%%%%%%%%%%%%%%%%%%%%
%%%%%%%%%%%%%%%%%%%%%%%%%%%%%%%%%%%

\subsection{${\cal A}_{RP_2}\left(NS{-}NS,NS{-}NS\right)$}
Scattering amplitudes of two massless NS-NS states from $\Omega_p$-planes involve at tree level a worldsheet with the topology of the real projective plane. They can be dealt with in analogy to the disk amplitudes.  ${\Omega_P}$-planes are the fixed loci of the space-time involution whose combined action with the worldsheet parity operator $\Omega$ (possibly dressed with suitable action on fermions) realizes the unoriented projection along with $(9{-}p)$ T-dualities. The real projective plane is a quotient of the Riemann sphere via the anti-conformal involution $\mathfrak {I}_{RP_2}(z)=-1/\bar z$. Possible choices for the fundamental region are thus the upper-half-plane or the unit disk \cite{Polchinski:1998rr,green2}. Vertex operators on the real projective plane must be defined consistently with the involution. 
The combination that takes into account all of these features is
\begin{equation}
{\cal W}_{NS{-}NS(q,\bar{q})}(E,k,z,\bar z)\, \underset{RP_2}{\rightarrow} \,  {\cal W}^{\otimes}_{NS{-}NS(q,\bar{q})}(E,k,z,\bar z)
\label{vertexRp2}
\end{equation}
where the vertex operator on $RP_2$ is given by
\begin{equation}
{\cal W}^{\otimes}_{{_{NS{-}NS(q,\bar q)}}}(E,k,z,\bar z)={1\over 2}E_{MN}\left\{:{\cal V}_{(q)}^{M}(k,z) {\cal V}_{(\bar{q})}^{N}(k,\bar{z}): + \R^{M}_{P}\R^{N}_{S}{:}{\cal V}_{(\bar{q})}^{P}(k\R,\bar{z})     {\cal V}_{(q)}^{S}(k\R ,z) : \right\}\,
\label{vertexRp2_expli}
\end{equation}
and in particular the two $\mathcal{R}$ matrices in the second term are due to the $T$-dualities. Using the doubling trick it is possible to arrive at the final form of the vertex operator
\begin{equation}
\begin{aligned}
&{\cal W}^{\otimes}_{{_{NS{-}NS(q,\bar q)}}}(E,k,z,\bar z) =\frac{1}{2}E_{MN}\bigg\{\R^{N}_{P}{:}{\cal V}_{(q)}^{M}(k,z){::}{\cal V}_{(\bar q)}^{P}(k\R,\bar z){:} + \R^{M}_{Q}\R^{N}_{S}\R^{Q}_{P}{:}{\cal V}_{(\bar q)}^{P}(k,\bar z){::}{\cal V}_{(q)}^{S}(k\R, z):\bigg\}\\
\end{aligned}
\label{vertexRp2_pictures}
\end{equation}
In the standard pictures, one gets from \eqref{vertexRp2_pictures}
\begin{equation}
\begin{aligned}
&{\cal W}^{\otimes}_{{_{NS{-}NS(-1,-1)}}}(E,k,z,\bar z) =\frac{E_{MN}\R^{N}_{P}}{2}\bigg\{{:}e^{-\phi}\psi^{M}e^{ikX}(z){::}e^{-\phi}\psi^{P}e^{ik\R X}(\bar z){:} + (z \leftrightarrow\bar z)
%e^{-\phi}\psi^{M}e^{ikX}(\bar z){::}e^{-\phi}\psi^{P}e^{ik\R X}(z)
\bigg\} \\
&{\cal W}^{\otimes}_{{_{NS{-}NS(0,0)}}}(E,k,z,\bar z) = \frac{E_{MN}\R^{N}_{P}}{2}\bigg\{{:}\Big(i\DP X^{M} + \frac{\al}{2}\big(k\psi\big)\psi^{M}\Big)e^{ikX}(z):\\&\hspace{6cm}:\Big(i\bar\DP X^{P} + \frac{\al}{2}\big(k\R\psi\big)\psi^{P}\Big)e^{ik\R X}(\bar z)\,{:}+%\Big(i\bar\DP X^{M} + \frac{\al}{2}\big(k\psi\big)\psi^{M}\Big)e^{ikX}(\bar z){::}\Big(i\DP X^{P} + \frac{\al}{2}\big(k \R\psi\big)\psi^{P}\Big)e^{ik\R X}(z)
(z\leftrightarrow\bar z)\bigg\}\,.
\end{aligned}
\end{equation}
At this point one can start with the computation of the two-point scattering amplitude from ${\Omega_P}$-plane of massless NS-NS states.
%, i.e. the tree-level worldsheet involved is the real projective plane RP2.  
Owing to the involution of the real projective plane, there exist several sub-amplitudes $\Lambda_i$ that one needs to consider  
\begin{equation}
\begin{aligned}
{\cal A}_{RP_2}^{\left(NS{-}NS,NS{-}NS\right)}=&\,g^2_{c}C_{_{RP_2}}\int_{{\abs{z}\leq 1}}\frac{d^2z_1d^2z_2}{V_{CKG}}\langle{\cal W}^{\otimes}_{_{NS{-}NS({-}1,{-}1)}}(E_1,k_1,z_1,\bar z_1){\cal W}^{\otimes}_{_{NS-NS(0,0)}}(E_2,k_2,z_2,\bar z_2)\rangle_{_{RP_2}}
\\=& \Lambda_1 + \Lambda_2 + \Lambda_3 + \Lambda_4
\end{aligned}\label{eq:RP1100}
\end{equation}
where for a representative sub-amplitude, like $\Lambda_i$, one has
\begin{equation}
\begin{aligned}
\Lambda_1 =&\,g^2_{c}C_{_{RP_2}}\frac{1}{4}\frac{2}{\al}\int_{{\abs{z}\leq 1}}\frac{d^2z_1d^2z_2}{V_{CKG}}E^1_{M_1N_1}E^2_{M_2N_2}\R^{N_1}_{P_1}\R^{N_2}_{P_2}\langle{:}e^{-\phi}\psi^{M_1}e^{ik_1X}(z_1){::}e^{-\phi}\psi^{P_1}e^{ik_1\R X}(\bar z_1):\\&\hspace{3cm}{:}\Big(i\DP X^{M_2} + \frac{\al}{2}\big(k_2\psi\big)\psi^{M2}\Big)e^{ik_2X}(z_2){::}\Big(i\bar\DP X^{P_2} + \frac{\al}{2}\big(k_2\R\psi\big)\psi^{P_2}\Big)e^{ik_2\R X}(\bar z_2){:}\rangle\\
%%%%%%%%%
=&\,g^2_{c}C_{_{RP_2}}\frac{1}{4}\frac{2}{\al}\int_{{\abs{z}\leq 1}}\frac{d^2z_1d^2z_2}{V_{CKG}}\langle{:}e^{-\phi}(z_1){::}e^{-\phi}(\bar z_1){:}\rangle\bigg({\cal M}^{(1)}_{_{\Lambda_1}} + {\cal M}^{(2)}_{_{\Lambda_1}} + {\cal M}^{(3)}_{_{\Lambda_1}} + {\cal M}^{(4)}_{_{\Lambda_1}}\bigg)\\
%%%%%%%%
\end{aligned}\label{eq:lambda1-1100}
\end{equation}
%%%%
%%%%%%%%%%

and once again can be separated like the disk amplitude in ${\cal M}^{(i)}_{\Lambda}$'s terms. The complete set of sub-amplitudes can be found in  Appendix A.4 of ref.\cite{AM:appendice}. The basic two-point functions necessary for an explicit computation are
%%%%%%%%%
\begin{equation}
\boxed{
\begin{aligned}
&\langle e^{-\phi}(z)e^{-\phi}(\bar z)\rangle_{_{RP_2}} = \frac{1}{(1 + z\bar z)}\,, \quad \langle X^M(z)X^N(\bar{z})\rangle_{_{RP_2}}=-{\alpha' \over 2} \eta^{M N} \log|1 +z\bar{z}|  \\
%%%%%
&\langle X^M(z) X^N(\bar{w})\rangle_{_{RP_2}}=-{\alpha' \over 2} \eta^{M N} \log (1 +z\bar{w})\,,\quad \langle X^M(z) X^N(w)\rangle_{_{RP_2}}=-{\alpha' \over 2} \eta^{M N} \log(z-w)\\
%%%%%
&\langle \psi^M(z)\psi^N(\bar{z})\rangle_{_{RP_2}} = \frac{\eta^{M N}}{( 1+ z\bar{z})}\,,\quad\langle\psi^M(z) \psi^N(\bar{w})\rangle_{_{RP_2}} = \frac{\eta^{M N}}{(1 + z\bar{w})}\,,\quad\langle \psi^M(z) \psi^N(w) \rangle_{_{RP_2}} = \frac{\eta^{M N}}{(z - w)}  \\
%%%%%%
\end{aligned}}\label{eq:correlator10D_RP2}
\end{equation}

%thus and for each $\Lambda_i$ we give explicitly ${\cal A}^{j}_{\Lambda_i}$
%%%%%%%%%%%%\Lambda_1
%\begin{equation}
%\begin{aligned}
%{\cal M}^{(1)}_{_{\Lambda_1}} = &\epsilon^1_{M_1N_1}\epsilon^2_{M_2N_2}D^{N_1}_{P_1}D^{N_2}_{P_2}\langle:\psi^{M_1}e^{ik_1X}(z_1)::\psi^{P_1}e^{ik_1DX}(\bar z_1)::i\DP X^{M_2}e^{ik_2X}(z_2)::i\bar\DP X^{P_2}e^{ik_2D X}(\bar z_2):\rangle\\\\
%%%%%%%%%%%%%%%%%%%
%\end{aligned}
%\end{equation}

Using the definitions of the usual kinematical invariants (see eq. \eqref{KIN}) for the Koba-Nielsen factors, it is convenient to combine the $\Lambda$-subamplitudes in pairs as follows\footnote{No picture changing is needed to combine together the $\Lambda$-subamplitudes. We checked that the amplitude is picture changing invariant, i.e there is no dependence on the picture distribution for the vertex operators, different of what is  argued in \cite{Garousi:2006zh}.}
\begin{equation}
A = \bigg(\Lambda_1 + \Lambda_4\bigg) + \bigg(\Lambda_2 + \Lambda_3\bigg)\,.
\label{eq:Asum1100}
\end{equation}
As is well known, in order to compute the integral in (\ref{eq:RP1100}), one has to fix the invariance under the CKG of $RP_2$. 
The latter is the $SU(2)$ subgroup of $SL(2,\mathcal{C})$ that is identified as the invariant part under the anti-conformal involution $\mathfrak{I}_{{RP_2}}(z)$. For $SU(2)$ the finite transformation reads \begin{equation}
\begin{aligned}
&z \mapsto z'=\frac{uz + v}{-\bar vz + \bar u}\,,\quad L=\begin{pmatrix}u&v\\-\bar v&\bar u\end{pmatrix}\in SU(2)\qquad i.e.\begin{cases}u = 1+i\beta \\ v = \gamma + i \lambda\\\beta,\gamma,\lambda\in\mathbb R \\\abs{u}^2 + \abs{v}^2 = 1\end{cases}\,.
\end{aligned}\label{symRp2}
\end{equation}
In agreement with \cite{Grinstein:1987}
the infinitesimal transformation is
\begin{equation}
\begin{aligned}
\delta(z')= \delta\Big(\frac{uz {+} v}{-\bar vz {+} \bar u}\Big)\Big|_{\beta=\gamma=\lambda=0} &=\frac{iz[(-\gamma {+} i\lambda)z + (1{-}i\beta)] {+} i[(1{+}i\beta)z {+} (\gamma {+} i\lambda)]}{[(-\gamma {+}i\lambda)z + (1 {-} i\beta)]^2}\Big|_{\beta{=}\gamma=\lambda=0}\,\delta\beta \\
&+\frac{[(-\gamma {+} i\lambda)z {+} (1{-}i\beta)] {+} z[(1{+}i\beta)z {+} (\gamma {+} i\lambda)]}{[(-\gamma {+}i\lambda)z {+} (1 {-} i\beta)]^2}\Big|_{\beta=\gamma=\lambda=0}\,\delta\gamma\\
%%%
&+ \frac{i[(-\gamma {+} i\lambda)z {+} (1{-}i\beta)] {-} iz[(1{+}i\beta)z {+} (\gamma {+} i\lambda)]}{[(-\gamma {+}i\lambda)z {+} (1 {-} i\beta)]^2}\Big|_{\beta=\gamma=\lambda=0}\,\delta\lambda
\end{aligned}
\label{infinite_trasRP2}
\end{equation}
yielding
\begin{equation}
\begin{aligned}
2iz\,\delta\beta + (1+z^2)\,\delta\gamma + i(1-z^2)\,\delta\lambda = (f + ig) + i\,e\,z + (f -ig)\,z^2\qquad e,f,g \in \mathbb R
\end{aligned}
\label{infinite_trasRP2_2}
\end{equation}
where $e= 2\,\delta\beta$, $f= \delta\gamma$ and $g=\delta\lambda$.
Setting $z = q + it$, 
\begin{equation}
\begin{aligned}
&\delta(q' + it') = (f +ig) + ie(q + it) + (f -ig)(q + it)^2\rightarrow\begin{cases}\delta q' = [1 + (q^2 - t^2)]f - e\,t + 2\,t\,q\,g\\\delta t' = 2\,t\,q\,f + e\,q+ [1 - (q^2 - t^2)]g\end{cases}
\end{aligned}
\label{infinite_trasRP2_3}
\end{equation}
one can fix three positions choosing the value of, for instance, $ q_1,t_1$ and $q_2$ using the $SU(2)$ symmetry. One gets in this way
\begin{equation}
dz_1\,d\bar z_1\,dz_2 \,d\bar z_2 = 4\,dq_1\,dt_1\,dq_2\,dt_2 = 4\,\abs{J}\,df\,dg\,de\,dt_2
\label{volum_elem_2}
\end{equation}
where the Jacobian $J$ is given by
\begin{equation}
\begin{aligned}
J &= \text{det}\begin{vmatrix} 1 {+} (q^2_1 {-} t^2_1)& 2q_1t_1 & 1 {+} (q^2_2 {-} t^2_2)\\ 2q_1t_1& 1 {-} (q^2_1 {-} t^2_1)& 2q_2t_2\\ -t_1& q_1& -t_2\end{vmatrix}\\
%%%%
&=t_1\Big\{[1 {+} (q^2_1 {+} t^2_1)][1 {+} (q^2_2 {-} t^2_2)]\Big\} -t_2[1 {+} (q^2_1 {+} t^2_1)]\Big\{2 q_2q_1 + [ 1 - (q^2_1 + t^2_1)]\Big\}\label{eq:Jacob3}
\end{aligned}
\end{equation}
The combination of $\langle\abs{ccc}\rangle$ ghost correlators is again
\begin{equation}
\begin{aligned}
\abs{\langle c(z_1)c(\bar z_1)c(z_2)\rangle &{-} \langle c(z_1)c(\bar z_1)c(\bar z_2)\rangle} = \abs{(1 +z_1\bar z_1)(1 + z_2\bar z_1)(z_1 - z_2) - (1 + z_1\bar z_1)(\bar z_1 - \bar z_2)(1 + z_1\bar z_2)}%\\&=|[ 1 + (q_1 +it_1)(q_1 - it_1)][1 + (q_2 + it_2)(q_1 - it_1)][ q_1 + it_1 - q_2 -it_2]\\&\,\,\,\,-[ 1 + (q_1 +it_1)(q_1 - it_1)][q_1 - it_1 -q_2 + it_2][1 + (q_1 + it_1)(q_2 - it_2)]|%\\&=\abs{2i[ 1 + j^2_1 + t^2_1]\Big\{(j_1 - j_2)[j_1t_2 - j_2t_1] + (t_1 - t_2)[1 + j_2j_1 + t_2t_1]\Big\}}\\&= 2[ 1 + j^2_1 + t^2_1][j^2_1t_2 - 2j_2j_1t_2 + j^2_2t_1 + t_1 +t_2t^2_1 - t_2 - t^2_2t_1]\\&=2\Big[ j^2_1t_2 - 2j_2j_1t_2 + j^2_2t_1 + t_1 + t_2t^2_1 - t_2 -t^2_2t_1 + j^4_1t_2 - 2j_2j^3_1t_2 + j^2_1j^2_2t_1 + j^2_1t_1\\& + j^2_1t_2t^2_1 - j^2_1t_2 - j^2_1t^2_2t_1 + j^2_1t^2_1t_2 - 2j_2j_1t^2_1t_2 + j^2_1t^3_1 + t^3_1 + t_2t^4_1 -t_2t^2_1 - t^2_2t^3_1\Big]\\&= 2t_1\Big[1 + j^2_2 - t^2_2 + j^2_1j^2_2 + j^2_1 - j^2_1t^2_2 +j^2_2t^2_1 + t^2_1 - t^2_2t^2_1\Big]\\& -2t_2\Big[ 1 + 2j_2j_1 - j^4_1 + 2j_2j^3_1 - 2j^2_1t^2_1 + j^2_1 + 2j_2j_1t^2_1 - t^4_1\Big]
\\&=\abs{2t_1\Big\{[1 {+} (q^2_1 {+} t^2_1)][1 {+} (q^2_2{ -} t^2_2)]\Big\} -2t_2[1 {+} (q^2_1 {+} t^2_1)]\Big\{2 q_2q_1 {+} [ 1 - (q^2_1 {+} t^2_1)]\Big\}}\label{eq:ghost3}
\end{aligned}
\end{equation}
which allows us to write
\begin{equation}
dz_1\,d\bar z_1\,dz_2\,d\bar z_2 = 4\,\abs{J}\,df\,dg\,de\,dt_2= 2\abs{\langle c(s_1)c(\bar s_1)c(s_2)\rangle {-} \langle c(s_1)c(\bar s_1)c(\bar s_2)\rangle}df\,dg\,de\,dt_2\,.
\label{eq:cambio2}
\end{equation}
In particular, for the specific case of interest a convenient choice is to fix the vertices in
\begin{equation}
\begin{aligned}
&z_1 = 0;\quad z_2 = iy\,\rightarrow q_1 = 0,\quad t_1= 0;\quad q_2 = 0, \quad t_2 = y\\&\bar z_1 = 0;\quad\bar z_2 = -iy\,\rightarrow q_1 = 0,\quad t_1= 0;\quad q_2 = 0,\quad  t_2 = y
\end{aligned}
\label{set_point_rp2}
\end{equation}
with this choice many terms vanish (see Appendix A.4 in ref. \cite{AM:appendice}). From \eqref{eq:cambio2} using \eqref{eq:Jacob3} and \eqref{eq:ghost3} one gets
\begin{equation}
\begin{aligned}
dz_1\,d\bar z_1\,dz_2\,d\bar z_2 = 4\,y\,df\,dg\,de\,dy\,. %&\equiv 2\,\abs{(1 +0)(1+0)(0 - iy) - (1 + 0)(0 +iy)(1 + 0)}\\&\equiv 4y\,df\,dg\,de\,dy
\end{aligned}\label{eq:c-ghostRP2}
\end{equation}
%%%%%%%%%%%%%%%%%%%%%%%%%
%%%%%%%%%%%%%%%%%%%%%%%%%
%%%%%%%%%%%%%%%%%%%%%%%%%
%%%%%%%%%%%
%Fixing the vertices using the $SU(2)$ symmetry, $\{z_1 = 0; \bar z_1 = 0; z_2 = iy; \bar z_2 =-iy\}$ and summing the equal terms we obtain
%
%
%
%%%%%%%%%%%%%%%%%%%%%%
%What happened is that the subamplitudes $\Lambda_i$ are equal two by two. Now one can sums up all the terms that have a common expression that involves the polarization tensors with the help of the  conservation and transversality of the momentum, i.e. $k_i\cdot\epsilon_i = 0$ , with the
%manifestation of the bose symmetry $(1\leftrightarrow 2)$, as reported in the appendix (\ref{termini RP2 10dim})
%
%
%%%%%%%%%%%%%%%%%%
%%
%Now we can put together all the terms, that involve the same denominator, in a function that we will call kinematical function $a_i$ \cite{Garousi:2006zh}. Thus we have
%\\
In this setup, only remain the following two contributions

\begin{equation}
\begin{aligned}
&\Lambda_1 + \Lambda_4 = \frac{2g_c^2C_{_{RP_2}}}{\al}\int_{0}^1dy^2\big(1 + y^2\big)^{-\al s} y^{-2\al \frac{t}{4}}\bigg(\frac{a_1}{y^2} + \frac{a_2}{(1 + y^2)} + \frac{a_3\,(1 + \al s)}{(1 + y^2)^2}\bigg)\\
&\Lambda_2 + \Lambda_3 = \frac{2g_c^2C_{_{RP_2}}}{\al}\int_{0}^1dy^2\big(1 + y^2\big)^{-\al s} y^{-2\al \frac{u}{4}}\bigg(\frac{a_1}{y^2} + \frac{a_2}{y^2(1 + y^2)} + \frac{a_3\,(1 + \al s)}{(1 + y^2)^2}\bigg) 
\end{aligned}
\label{Lambada_couple}
\end{equation}

with 
\begin{equation}
\begin{aligned}
& a_1 = -\frac{\al^2}{4}\bigg\{%-\frac{1}{2}\Tr({E}_2^{_T}{E}_1)k_2{\R}k_2
 -\frac{1}{2}\Tr({E}_1^{_T}{E}_2)k_1{\R}k_1 - \Tr({\R}{E}_1)k_1{E}_2k_1 %- \Tr({\R}{E}_2)k_2{E}_1k_2\,+
 %\\
%%%%
%&\hspace{2cm} %+ k_1{E}_2{\R}{E}_1k_2 
+ k_2{E}_1{\R}{E}_2k_1 -k_2{\R}{E}_1{E}_2^{_T}k_1 -\frac{1}{2}k_2{E}_1{E}_2^{_T}k_1 %-k_1{\R}{E}_2{E}_1^{_T}k_2 -\frac{1}{2}k_1{E}_2{E}_1^{_T}k_2\,+
\\
%%%
&\hspace{2cm} - k_2{\R}{E}_1^{_T}{E}_2k_1 %- k_1{\R}{E}_2^{_T}{E}_1k_2 -\frac{1}{2}k_1{E}_2^{_T}{E}_1k_2 
-\frac{1}{2}k_2{E}_1^{_T}{E}_2k_1 + (1\leftrightarrow2)\bigg\}\\
&a_2 = -\frac{\al^2}{4}\bigg\{%\frac{1}{2}\Tr({E}_2^{_T}{E}_1)k_2{\R}k_2 + 
\frac{1}{2}\Tr({E}_1^{_T}{E}_2)k_1{\R}k_1 -\frac{1}{2}\Tr({E}_1{\R}{E}_2{\R})k_2{\R}k_2
%+\\&\hspace{2cm} -\frac{1}{2}\Tr({E}_2{\R}{E}_1{\R})k_1{\R}k_1
%%%
%-\Tr({\R}{E}_2)k_2{E}_1{\R}k_1
 + \Tr({\R}{E}_1)k_2{\R}{E}_2{\R}k_1
 %+\\&\hspace{2cm} + \Tr({\R}{E}_2)k_1{\R}{E}_1{\R}k_2 - k_2{\R}{E}_2{\R}{E}_1{\R}k_1 - k_1{\R}{E}_1{\R}{E}_2{\R}k_2
%%%
+\frac{1}{2}k_1{\R}{E}_1{E}_2^{_T}{\R}k_2+\\&\hspace{2cm} %+\frac{1}{2}k_2{\R}{E}_2{E}_1^{_T}{\R}k_1 + 
\frac{1}{2}k_1{\R}{E}_1^{_T}{E}_2{\R}k_2
%+ \frac{1}{2}k_2{\R}{E}_2^{_T}{E}_1{\R}k_1
 -\Tr({\R}{E}_1)k_1{E}_2{\R}k_2 + (1\leftrightarrow2)\bigg\}\\
& a_3 = \frac{\al}{2}\Tr({\R}{E}_1)\Tr({\R}{E}_2)
\end{aligned}
\label{ai_RP2}
\end{equation}
are relevant. The details on the terms appearing in the $a_i$ are spelled out in Appendix A.5 of ref. \cite{AM:appendice}. Using the integral representation of the hypergeometric function 
\begin{equation}
{}_2F_{1}(\alpha,\beta;\gamma; z) = \frac{\Gamma(\gamma)}{\Gamma(\beta)\Gamma(\gamma - \beta)}\int_0^1 du\,u^{\beta -1}(1 - u)^{\gamma -\beta -1}(1 - uz)^{-\alpha}
\label{eq:HY21}
\end{equation}
with $\gamma = \beta + 1$ and $z = -1$, one obtains % i.e.
%\begin{equation}
%F(\alpha,\beta,\beta +1; -1) = \beta \int_0^1 du\,u^{\beta -1}(1 + u)^{-\alpha}\label{eq:HY21}
%\end{equation}
%
\begin{equation}
\begin{aligned}
&\Lambda_1 + \Lambda_4 = \frac{2g_c^2C_{_{RP_2}}}{\al}\bigg\{ a_1\frac{_2F_1\big(\al s, -\al t/4; -\al t/4 +1; -1\big)}{(-\al t/4)} + a_2\frac{_2F_1\big(\al s + 1, -\al t/4 + 1; -\al t/4 + 2; -1\big)}{(-\al t/4 + 1\big)}\\&\hspace{1.6cm} + a_3\frac{(1 +\al s){}_2F_1\big(\al s + 2, -\al t/4 + 1; -\al t/4 + 2; -1\big)}{\big(-\al t/4 + 1\big)}\bigg\}\\\\
&\Lambda_2 + \Lambda_3 = \frac{2g_c^2C_{_{RP_2}}}{\al}\bigg\{a_1\frac{_2F_1\big(\al s , - \al u/4; - \al u/4 + 1; -1\big)}{(-\al u/4)} +a_2\frac{_2F_1\big(\al s + 1, -\al u/4; -\al u/4 + 1; -1\big)}{(-\al u/4)} \\&\hspace{1.6cm}+ a_3\frac{(1 +\al s){}_2F_1\big(\al s + 2, -\al u/4 + 1; -\al u/4 + 2; -1\big)}{\big(-\al u/4 + 1\big)}\bigg\}\,.
\end{aligned}
\label{Lambda_couple2}
\end{equation}
With the help of the identity
\begin{equation}
b \,\,_2 F_1(a, a+b; a+1; -1) + a \,\,_2F_1(b, a+b; b+1;-1) = \frac{\Gamma(a+1)\Gamma(b+1)}{\Gamma(a+b)}
\label{HYPid}
\end{equation}
it is straightforward to see that all the subamplitudes $\Lambda_i$ combine together, because for instance
\begin{equation}
\small{\begin{aligned}
&_2F_1(\al s, - \al u/4; - \al u/4 {+} 1; -1) = \frac{\Gamma\left(-\al t/4\right)\Gamma\left(-\al u/4 {+} 1\right)}{\Gamma\big(\al s\big)} - \frac{(-\al u/4)}{(-\al t/4)}{}_2F_1(\al s, -\al t/4; -\al t/4 {+} 1; - 1)
%
%&_2F_1(\al s {+} 1, -\al t/4 {+} 1; -\al t/4 {+} 2; -1) = -\frac{(-\al u/4)}{(-\al t/4 {+}1)}{}_2F_1(\al s {+} 1, -\al t/4{ +} 1; -\al t/4; - 1)\\&\hspace{6.6cm} +\frac{\Gamma(-\al t/4 {+} 1)\Gamma(-\al u/4 {+} 1)}{\Gamma(\al s {+} 1)}\\\\
%
%&_2F_1(\al s {+} 2, -\al u/4 {+} 1; -\al u/4; -1) = -\frac{(-\al u/4 {+}1)}{(-t/4 {+} 1)}{}_2F_1(\al s {+} 2, -\al t/4 {+} 1; -\al t/4 {+} 2; - 1)\\&\hspace{6.4cm}+\frac{\Gamma(-\al t/4 {+} 1)\Gamma(-\al u/4 {+} 2)}{\Gamma(\al s {+} 2)}
\end{aligned}}
\label{HYPexe}
\end{equation}
and exploiting the $\Gamma$ function properties and using eq.\,\eqref{eq:onshellness_2tC}, one arrives at the same conclusion as \cite{Garousi:2006zh}, namely
\begin{equation}
\begin{aligned}
\Lambda_1 + \Lambda_2 + \Lambda_3 + \Lambda_4 =&\frac{2g_c^2C_{_{RP_2}}}{\al}\bigg\{a_1(\al s)\frac{\Gamma(-\al t/4)\Gamma(-\al u/4)}{\Gamma(-\al t/4 - \al u/4 + 1)} +a_2(-\al t/4)\frac{\Gamma(-\al t/4)\Gamma(-\al u/4)}{\Gamma(-\al t/4 - \al u/4 + 1)}\\& + a_3( 1 +\al s)(-\al t/4)(-\al u/4)\frac{\Gamma(-\al t/4)\Gamma(-\al u/4)}{\Gamma(\al s + 2)}\bigg\}\\=
&\frac{2g_c^2C_{_{RP_2}}}{\al}\bigg\{\al s\, a_1 - \al \frac{t}{4}\,\hat{a}_2\bigg\}\frac{\Gamma(-\al t/4)\Gamma(-\al u/4)}{\Gamma(-\al t/4 - \al u/4 + 1)}
\end{aligned}
\label{eq:resOp}
\end{equation}
where $\hat{a}_2 = a_2 - \al\frac{u}{4}a_3$. Taking the limit $\al\rightarrow 0$ as in \eqref{eq:gammaexp} the amplitude \eqref{eq:resOp} exhibits only closed string poles in the t-channel and u-channel due to the presence of  $\Omega_P$-planes where no open strings can be attached. As for the disk case, we can extract the terms proportional to $(\al)^2$ which read
\begin{equation}
g_c^2C_{_{RP_2}}\zeta(2)\left\{-2s\,a_1 + \frac{t}{2}\,a_2 + \al\,\frac{ut}{8}\,a_3\right\}
\label{eq:alpha22}
\end{equation}
in agreement with the results of refs. \cite{Garousi:2006zh, Garousi:2017fbe}. At this point it is also interesting to compare the pole structures of both disk and real projective plane amplitudes for which we obtain
\begin{equation}
\begin{aligned}
{\cal A}_{D_2}&=\frac{4g_c^2C_{_{D_2}}}{\al}\left(-\al s\,a^{_{D_2}}_1 - \al \frac{t}{4}\,\hat{a}^{_{D_2}}_2\right)\frac{\Gamma(-\al s)\Gamma(-\al t/4)}{\Gamma(-\al s - \al t/4 +1)}\\\\
{\cal A}_{_{RP_2}}&=\frac{2g_c^2C_{_{RP_2}}}{\al}\bigg(\al s\, a^{_{RP_2}}_1 - \al \frac{t}{4}\,\hat{a}^{_{RP_2}}_2\bigg)\frac{\Gamma(-\al t/4)\Gamma(-\al u/4)}{\Gamma(-\al t/4 - \al u/4 + 1)}\,.
\end{aligned}
\label{eq:conf.pole-t}
\end{equation} 
One can verify that the factor inside the round bracket is the same in both the amplitudes since $a_{1}^{_{RP_2}}=-a_{1}^{_{D_2}}$ and $\hat{a}_{2}^{_{RP_2}}=\hat{a}_{2}^{_{D_2}}$. We can match the pole expansion in the $t$-channel, using the following representations 
\begin{equation}
\begin{aligned}
{\cal A}_{D_2}
\sim 
%\frac{\Gamma(-\al s)\Gamma\left(-\al \frac{t}{4}\right)}{\Gamma\left(-\al s - \al \frac{t}{4} +1\right)} =
&\frac{\Gamma(\al \frac{u}{4} + \al\frac{t}{4})\Gamma\left(-\al \frac{t}{4}\right)}{\Gamma\left(\al \frac{u}{4} +1\right)}\\
=& -\frac{\text{sin}\left(\pi\left(\al\frac{u}{4}\right)\right)}{\pi}\Gamma\left(-\al \frac{t}{4}\right)\Gamma\left(-\al \frac{u}{4}\right)\Gamma\left(\al \frac{t}{4} +\al \frac{u}{4}\right)\\\\
{\cal A}_{_{RP_2}}&\sim\frac{\Gamma\left(-\al \frac{t}{4}\right)\Gamma\left(-\al \frac{u}{4}\right)}{\Gamma\left(-\al \frac{t}{4} - \al \frac{u}{4} + 1\right)}\\
&= \frac{\text{sin}\left(\pi\left(\al \frac{t}{4} + \al\frac{u}{4}\right)\right)}{\pi}\Gamma\left(-\al \frac{t}{4}\right)\Gamma\left(-\al \frac{u}{4}\right)\Gamma\left(\al \frac{t}{4} +\al \frac{u}{4}\right)
\label{eq:expa-t}
\end{aligned}
\end{equation}
%%%%%%%
where we have used $\al s = -\al t/4 - \al u/4$ and the Gamma function property $\Gamma(z)\Gamma(1 - z)=\pi/\text{sin}(\pi\,z)$. In \eqref{eq:expa-t} the $t$-channel poles are due to $\Gamma(-\al t/4)$ and the matched residue at the pole $\al t/4 = n$ is equal to
%%%%
\begin{equation}
\begin{aligned}
Res({\cal A}_{D_2})\big|_{\al \frac{t}{4} = n}&\sim \frac{(-1)^{n +1}\,\text{sin}\left(\pi\,\al\frac{u}{4}\right)}{\pi\,n!}\Gamma\left(-\al \frac{u}{4}\right)\Gamma\left(n +\al \frac{u}{4}\right)\\
Res({\cal A}_{_{RP_2}})\big|_{\al \frac{t}{4} = n}&\sim \frac{\text{sin}\left(\pi\,\al\frac{u}{4}\right)}{\pi\,n!}\Gamma\left(-\al \frac{u}{4}\right)\Gamma\left(n +\al \frac{u}{4}\right)
\end{aligned}
\label{res_compare}
\end{equation}
%%%%%%%%%%
with $n$ positive integer. At fixed $n$, one can verify that, for $n$-odd the poles in \eqref{eq:expa-t} for ${\cal A}_{D_2}$ and ${\cal A}_{_{RP_2}}$ have the same sign while for $n$-even the sign of the poles are opposite in agreement with \cite{Garousi:2006zh}.\footnote{Our analysis is done on the full amplitudes \eqref{eq:conf.pole-t}, not on partial subamplitudes as in \cite{Garousi:2006zh}.} We end by remarking that the mass-square $M_n^2$, in the $t$-channel, for the closed string takes the form 
\begin{equation}
M_n^2 = \frac{4n}{\al} := \frac{2}{\al}(2n) = \frac{2}{\al}\left((N_{\alpha} + \tilde{N}_{\alpha}) + (N_{b} + \tilde{N}_{b}) - \delta^{(_{NS},_{R})}\right)
\label{level_mathcing}
\end{equation}
where one can verify that only the GSO-projected states are permitted. In eq. \eqref{level_mathcing} $N_{\alpha}$ and $N_b$ are the bosonic and the fermionic number operators, respectively. 
%%%%%%%%%%%%%%%%%%%%%%%%%%%%%%%%%%%
%%%%%%%%%%%%%%%%%%%%%%%%%%%%%%%%%%%
%%%%%%%%%%%%%%%%%%%%%%%%%%%%%%%%%%%
%%%%%%%%%%%%%%%%%%%%%%%%%%%%%%%%%%%
%%%%%%%%%%%%%%%%%%%%%%%%%%%%%%%%%%%
%%%%%%%%%%%%%%%%%%%%%%%%%%%%%%%%%%%
%%%%%%%%%%%%%%%%%%%%%%%%%%%%%%%%%%%
%%%%%%%%%%%%%%%%%%%%%%%%%%%%%%%%%%%
%%%%%%%%%%%%%%%%%%%%%%%%%%%%%%%%%%%%
\section{Type IIB orientifold on $\T^6/\mathbb{Z}_2\times\mathbb{Z}_2$}\label{ch:3}\label{sec3}

Our aim in this section is to compute string two-point scattering amplitude involving closed string moduli, at tree-level in $g_s$ expansion, in particular on the first worldsheet spanned by open string and closed unoriented string, i.e the disk $D_2$ and the real projective plane $RP_2$ respectively. We postponed to Section \ref{scattering_sec} the explicit amplitude calculations. Here our attention is focused on the formulation of the necessary computational tools, like the explicit construction of the relevant vertex operators and two-point functions on these specific worldsheet surfaces. This knowledge is required to bridge the string approach and to the construction of the four dimensional supergravity LEEA.
In the various subsections we wish to summarise the basic aspects of Type IIB orientifolds on $\T^6/\mathbb Z_2\times \mathbb Z_2$. We give an overview of the Type IIB toroidal orbifold $\mathbb Z_2\times Z_2$ spectrum, of the extended objects that the orientifold projection $\Omega\sigma$ induces and the resulting parametrisation of the moduli space at tree-level in a specific situations from the supergravity point of view.

Among the general Toroidal orbifold \eqref{eq:Torbifold}, the Type IIB on $\textbf{T}^6/\mathbb{Z}_2\times\mathbb{Z}_2$ that we want to consider here is the one with Abelian  orbifold group $\Gamma \equiv \mathbb{Z}_2\times\mathbb{Z}_2$ without discrete torsion \cite{Bianchi:1990tb,Bianchi:1990yu,Angelantonj:2002ct,Vafa:1994rv} and satisfying the constraints needed to get $\mathcal N=2$ supersymmetry \cite{BLT,Lust:2005dy,Ibanez:2012zz}, ``before'' the orientifold projection. The root lattice of the underlying $\T^6$ torus is chosen in a way that it factorises as
\begin{equation}
\T^6=\bigotimes_{I=1}^3\T_I^2
\label{torus6}
\end{equation}
where the two-tori $\T^2_I$ are the blocks-diagonal parts of $\T^6$, while the elements of the group orbifold are $\mathcal G_{\mathbb{Z}_2\times\mathbb{Z}_2} = \{1,\theta_1,\theta_2,\theta_3\}$ with
\begin{equation}
\theta_1 = (+,-,-)\quad \theta_2 = (-,+,-)\quad \theta_3 =(-,-,+)
\label{eq:3theta}
\end{equation}
where each element $\theta_I$ of $\mathbb{Z}_2\times\mathbb{Z}_2$ acts as if it were a single $\mathbb Z_2$ thus leaves invariant the $\T^2_I$ lattice base and flips with a minus sign the lattice of the corresponding two-torus $\T^2_{J\ne I}$. The torus partition function of this model is the truncation onto invariant states of the torus partition function of Type IIB on $\T^6$ torus, obtained by the action of a defined orbifold projection operator on the latter \cite{Angelantonj:2002ct} (as the GSO projection operator in superstring). The resulting states of the spectrum are organised in sectors: the \textit{untwisted} sector and the \textit{twisted} sectors. The twisted sectors are characterised by strings that satisfy periodicity condition imposed by the orbifold group elements. Their presence ensures the modular invariance of the torus partition function. As for $CY$ compactification, the spectrum of toroidal orbifold is expressed using the Hodge classes $H^{p,q}$ and their dimensions $h^{p,q}$ are collected into the Hodge diamond that for the $\textbf{T}^6/\mathbb{Z}_2\times\mathbb{Z}_2$ is \cite{Vafa:1994rv}
\begin{equation}
\begin{pmatrix}
1&0&0&1\\
0&3&51&0\\
0&51&3&0\\
1&0&0&1\\
\end{pmatrix}
\label{eq:diamond}
\end{equation}
where $p$ and $q$ span the columns and the rows respectively. The massless spectrum of the Type IIB orbifold is contained in the matrix  \eqref{eq:diamond}, since its entries count the number of super-multiplets in the low-energy effective field theory and we have  
\begin{equation}
\begin{aligned}
&h^{1,1} \equiv {h}_{_{\text{utw}}}^{1,1} + h^{1,1}_{_{\text{tw}}} = 3 + 0\\
&h^{2,1}\equiv h^{2,1}_{_{\text{utw}}} + h^{2,1}_{_{\text{tw}}} = 3 + 48
\end{aligned}
\label{moduli_utw_tw}
\end{equation}
$h^{1,1}$, $h^{2,1}$ are split into the untwisted (utw)  and twisted (tw) sectors respectively. In terms of $\mathcal N=2$ supermultiplets in 4-dimensions the moduli are organised in hypermultiplets and vectormultiplets as follows
\begin{equation}
\begin{aligned}
&\text{Utw} : ({h}_{_{\text{utw}}}^{1,1} + 1) \equiv (3 + 1)\quad\text{Hyper}\,,\quad h^{2,1}_{_{\text{utw}}}\equiv 3\quad \text{Vector} \\  
&\text{Tw} : h^{1,1}_{_{\text{tw}}} \equiv 0\quad \text{Hyper}\,,\quad h^{2,1}_{_{\text{tw}}}\equiv 48 \quad\text{Vector}\,.
\label{eq:N2multipl}
\end{aligned}
\end{equation}
In eq. \eqref{eq:N2multipl} we have also included the axion-dilaton hypermultiplet (or universal hypermultiplet). Our attention is on the untwisted moduli and on the moduli space that they parameterise. Starting from the Type IIB on $\T^6$ in 4-dimensions with $\mathcal N=8$ supersymmetries, one knows that the 70 scalar fields are collected in the gravitational multiplet (unique multiplet) and parametrise a moduli space that is $E_{7(7)}/SU(8)$. The latter can be reduced to a factorised moduli space parameterised only by the geometrical moduli (as in Heterotic case \cite{Schwarz:1992tn})  
\begin{equation}
\frac{SU(1,1)}{U(1)}\times\frac{SO(6,6)}{SO(6)\times SO(6)}
\label{eq:cos-geomT^6}
\end{equation}
when the non-geometrical moduli are frozen \cite{Ferrara:1989ik}. The first coset refers to the axion-dilaton. The reduction of this coset \eqref{eq:cos-geomT^6} due to the $\mathbb{Z}_2\times\mathbb{Z}_2$ action is the moduli space spanned by the untwisted moduli \eqref{eq:N2multipl}(see \cite{Ferrara:1989py})
\begin{equation}
\frac{SU(1,1)}{U(1)}\times\left(\frac{SU(1,1)}{U(1)}\times\frac{SO(2,{h}_{\text{utw}}^{1,1} -1)}{SO(2)\times SO({h}_{\text{utw}}^{1,1} -1)}\right)\times\left(\frac{SU(1,1)}{U(1)}\times\frac{SO(2,{h}_{\text{utw}}^{2,1} -1)}{SO(2)\times SO({h}_{\text{utw}}^{2,1} -1)}\right)
\label{eq:Z2XZ2}
\end{equation}
in agreement (when only geometrical moduli are dynamical fields) with the moduli space factorisation 
\begin{equation}
\frac{SU(1,1)}{U(1)}\times\mathcal {\hat S}_K^{h^{1,1}}\times\mathcal {S}_K^{h^{2,1}}
\end{equation}
valid for Type IIB compactification on $CY_3$ with $\mathcal{N}=2$ superymmetries \eqref{eq:modulispace} with both $\mathcal {\hat S}_K$ and $\mathcal {S}_K$ special K\"ahler manifold parametrised by the hypermultiplets and vectormultiplets, respectively. Considering $h_{\text{utw}}^{1,1} = 3$ in eq.\eqref{eq:Z2XZ2} the factor $SO(2,{h}_{\text{utw}}^{1,1} -1)/SO(2)\times SO({h}_{\text{utw}}^{1,1} -1)$ are equal to $(SU(1,1)/U(1))^2$. The same holds for $h_{\text{utw}}^{2,1}=3$. In the supergravity language the scalar fields in the hypermultiplets and vectormultiplets are the complex K\"ahler moduli $t^I$ and the complex structure $u^I$, respectively, while the universal complex axion-dilaton is denoted by $s$. The K\"ahler potential \eqref{eq:Kmanifold} associated to the space \eqref{eq:Z2XZ2} can be written as 
\begin{equation}
\kappa_4^2\,\mathcal K = - \ln(s + \bar s) - \ln\prod_{I=1}^3(t^I +\bar t^I) - \ln\prod_{I=1}^3(u^I +\bar u^I)
\label{eq:kahlerP}
\end{equation}
with $\kappa_4$ the physical gravitational coupling constant in 4-dimensions introduced in \eqref{eq:effective.act}. 

$\mathcal N=1$ models in 4-dimensions can be obtained by taking the orientifold projection $\Omega\sigma$ of Type II theories. To make more clear how one can obtain such models we wish to explain the key steps that lead to their construction. \footnote{There is no dictated order between the orientifold projection and the orbifold projection, thus one can take, for instance, both projections at the same time referring to the full orientifold group, i.e. $\mathcal G= \mathcal G_{\Gamma}\cup \Omega\sigma\mathcal G_{\Gamma}$.} The main feature that orientifold projection $\Omega\sigma$ introduces is the presence of non-dynamical ${\Omega}_{P}$-planes, with $P$ the dimensions of the worldvolume \cite{Polchinski:1995mt,Angelantonj:2002ct}. Moreover different kinds of orientifold projections $\Omega\sigma$ on Type IIB induce specific ${\Omega}_{P}$-planes summarised in Table \ref{tab:tabOM},
\begin{table}[h!]
\begin{center}
\begin{tabular}{|c|c|c|c|c|}
\hline
$\Omega\sigma$&$\Omega I_0$&$\Omega I_2(-1)^{F_L}$&$\Omega I_4$&$\Omega I_6(-1)^{F_L}$\\
\hline
${\Omega}_P$&${\Omega}_9$&${\Omega}_7$&${\Omega}_5$&${\Omega}_3$\\
\hline
\end{tabular}
\end{center}
\caption{}
\label{tab:tabOM}
\end{table}

The discrete involution $I_n (n=0,2,4,6)$ acts on the internal compact space as a reflection that fixs also the positions of the $\Omega_{P}$-planes in the target space, while $(-1)^{F_L}$ ensures that the $\Omega\sigma$ operator square to unity, as required for an involution. 

Additional ${\Omega}$-planes can appear when the orbifold group $\mathcal G_{\Gamma}$ contains ${\mathbb Z}_2$ elements because they mix with $\Omega I_n$ giving either ${\Omega}_{9 - (4 -n)}$-planes for $I_n = I_0,I_4$ or ${\Omega}_{3 + (n - 2)}$-planes for $I_n= I_2,I_6$. To each kind of ${\Omega}_{P}$-plane a stack of $D_{P}$-branes needs to be added for consistency. The truncation of the spectrum of the Type IIB theories onto $\Omega\sigma$ invariant states takes into account that worldsheet parity projection $\Omega$ exchanges the left moving part with right moving part of the string. Furthermore the $(-1)^{F_L}$   
assigns a (+)-eigenvalue and (-)-eigenvalue to the NS-NS and R-R states, respectively, while the pullback of $I_n$ acts on the $H^{p,q}$ Hodge classes of the toroidal orbifold (or $CY$ manifold) dividing each class into two subclasses characterised by (+)-eigenvalue $H^{p,q}_+$ and (-)-eigenvalue $H^{p,q}_-$, while the unique 3-form ${\cal\omega}_3$ has (+)-eigenvalue or (-)-eigenvalue respectively for $I_n=I_0,I_4$ or $I_n=I_2,I_6$ \cite{Blumenhagen:2006ci}.

The states in the spectrum that survive the $\Omega\sigma$ projection are those with overall (+)-eigenvalue. All the moduli fields fall into chiral multiples of $\mathcal N=1$ in 4-dimensions, but due to the orientifold projection $\Omega\sigma$, when one has to complexify the K\"ahler moduli $t$, the net distinction between geometrical and non-geometrical moduli for the $\mathcal N =2$ case is lost because they can mix. As a result the linear combinations allowed in general are those with the real K\"ahler moduli coming from the same Hodge subclasses, i.e both in $H^{p,q}_+$ or  $H^{p,q}_-$ \cite{Blumenhagen:2006ci}. Complex structure moduli $u^I$ are complex by definition. 

We continue to explore the Type IIB orientifold on $\T^6/ \mathbb{Z}_2\times\mathbb{Z}_2$ with a focus on the closed untwisted sector, because the moduli in the twisted sectors can be considered as frozen. We choose as orientifold $\Omega\sigma$ operator the worldsheet parity operator $\Omega$ ($\sigma = I_0$) which by itself introduces ${\Omega_9}$-plane (see Table \ref{tab:tabOM}).\footnote{We want to stress that in that case Type IIB orientifold $\equiv$ Type I.} The mixing of worldsheet parity operator $\Omega$ with the three $\theta_I={\mathbb Z}^I_2$ elements \eqref{eq:3theta} of $\mathcal G_{ \mathbb{Z}_2\times\mathbb{Z}_2}$ orbifold group add three different ${\Omega}_5^I$-planes fixed at the ${\mathbb Z}^I_2$ invariant loci respectively. Both ${\Omega_9}$ and ${\Omega}_5^I$-planes are balanced by stacks of $D_9$- and $D^{I}_5$-branes, respectively, where the specific number of branes in each stack and the gauge group localised on the $D$-branes worldvolume, can be determinated by computing the partition function of the orientifold model.  In this calculation one has to include all the one-loop oriented and unoriented worldsheet surfaces torus ($\mathscr T$) and Klein bottle ($\mathscr K$) from closed string, as well as the annulus ($\mathscr A$) and M\"obius strip ($\mathscr M$) from open string. We do not discuss this point in a detailed way, see for instance \cite{Angelantonj:2002ct}, but for completeness the model admits coincident 32 $D_9$-branes on top of the $ {\Omega_9}$-plane that wrap the full internal space, three sets of coincident 32 $ D_5^I $ -branes on top of the $ {\Omega} _5 ^ I $ -planes wrapped along the $\T^2_I$-torus with $(Usp (16))^4$ gauge group \cite{Berg:2005ja,Berkooz:1996km}. The orientifold $\Omega I_0$ action splits the Hodge classes $H^{1,1}$ and $H^{2,1}$, as said before, but does not reduce the number of untwisted moduli for which we find 
\begin{itemize}
\item Utw :
\begin{equation}
\begin{aligned}
&h^{1,1}_{_{\text{utw}}} = h^{1,1}_{_{\text{utw}}+} + h^{1,1}_{_{\text{utw}}-} \equiv 3 + 0\leftrightarrow \text{(compl.) K\"ahler moduli}\,\,t^I\,\,,\quad I = 1,{\dots}, h^{1,1}_{_{\text{utw}}+}\\
&h^{2,1}_{_{\text{utw}}} = h^{2,1}_{_{\text{utw}}+} + h^{2,1}_{_{\text{utw}}-} \equiv 3 + 0 \leftrightarrow \text{complex structure}\,\,u^I\,\,,\quad I = 1,{\dots}, h^{2,1}_{_{\text{utw}}+}\,\,.
\end{aligned}
\label{eq:moduliSugra}
\end{equation} 
\end{itemize}

In this situation, however, the complex K\"ahler moduli $t^I$ are no longer purely geometric because their imaginary parts come from the moduli of $C_2$-RR field,  while the real parts involve the moduli of the NS-NS graviton field, all of them in the same Hodge subclass $H^{1,1}_+$. The complex axion-dilaton $s$ contains the scalar dual to the $C_2$-RR field as imaginary part and the four dimensional dilaton $\phi$ as real part. The complex structures $u^I$ remain purely geometrical moduli. The orientifold projection doesn't modify the structure of the moduli space parametrised by untwisted moduli
\eqref{eq:Z2XZ2} at tree level. Thus we are left with the direct product of moduli space for the $s$ axion-dilaton $SU(1,1)/U(1)$, the complex K\"ahler $t^I$ $(SU(1,1)/U(1))^3$ and complex structure $u^I$ $(SU(1,1)/U(1))^3$ \cite{Blumenhagen:2006ci,Ferrara:1989ik,Lust:2005dy,Ibanez:2012zz}. As a consequence also the K\"ahler potential $\mathcal K$ is the same, i.e. \eqref{eq:kahlerP}. Some modifications can occur when open string moduli (Wilson lines) coming from $D$-branes wrapping the internal space, are taken into account. The moduli space is no longer factorisable (see for instance \cite{Blumenhagen:2006ci}) and the definitions of complex K\"ahler moduli $t$ (by open string moduli from $D_9$-branes) and of axion-dilaton $s$ (by open string moduli from $D_5^{I}$-branes) become more involved  \cite{Blumenhagen:2006ci,Berg:2005ja,Antoniadis:1996vw}. Furthermore the gauge group $(Usp (16))^4$ can be broken when some open string moduli take non-trivial vev. A discussion of this situation is beyond the scope of this paper \cite{Berg:2005ja,Berkooz:1996dw}. 
%%%%%%%%%%%%
%%%%%%%%%%%%
\subsection{Derivation of the compactified vertex operators and their properties}
\label{Toolkit}
In this section we describe the main steps needed to build the vertex operators and the two-point functions for the closed untwisted moduli of Type IIB orientifold on $\T^6/\mathbb{Z}_2\times\mathbb{Z}_2$ (and \textit{mutatis mutandis} for each models that have similar behaviour). The compactifications of Type II string theories with $\mathcal N=2$ supersymmetries in $d=4$ allow a two-dimensional superconformal field theory (SCFT) description on string worldsheet that is locally $N=(1,1)$ invariant in the external space and globally $N=(2,2)$ symmetric in the internal space with central charges $\left(c^{\text{ext}},\bar c^{\text{ext}}\right) = (-9,-9)$ and $\left(c^{\text{int}},\bar c^{\text{int}}\right) = (9,9)$ respectively \cite{Blumenhagen:2006ci,BLT}.\,\footnote{In $c^{\text{ext}}$ the ghost contribution to the central charge is included.} %\footnote{This global $N=(2,2)$ internal SCFT can be obtained enlarging the local $N=(1,1)$ internal SCFT in a way that there will be no additional ghost systems which would appear for local $N=(2,2)$ and affect the value of external space-time dimension. How obtain a global $N=(2,2)$ SCFT from a local $N=(1,1)$ SCFT is beyond our scopes thus more details can be find in \cite{BLT,Blumenhagen:2006ci}.}
%Each resulting $N=2$ superconformal algebra (SCA), mutually for the left and right moving superstring, is generated by: the two-dimesional energy momentum tensor $T({\bm\bar T})$, a global $U(1)$ current $J({\bm\bar J})$ and two fermionic supercurrents $G_F^{\pm}({\bm\bar G_F^{\pm}})$ charged under the global $U(1)$ with charge $Q=\pm 1({\bm\bar Q=\pm 1})$ respectively.
Before the orientifold projection $\Omega\sigma$, the string spectrum originating from the global $N=(2,2)$ SCFT can be put in correspondence with the fields content of the Type II string theories with $\mathcal N=2$ supersymmetries in $d=4$ \cite{Blumenhagen:2006ci}. For instance looking at the massless moduli, this correspondence can occur because at this stage there is no mixing between geometrical and non geometrical moduli. Thus one can start from the vertex operators for the geometrical moduli coming from the NSNS fields (graviton $g$ and Kalb-Ramond $B_2$) because their explicit form can be deduced by the $d=2$ non-linear sigma model. Then using the sypersymmetry transformations the vertex operators for the non-geometrical moduli originating from the RR fields (2-form $C_2$ and 4-form $C_4$) can be obtained.\footnote{Also the fermionic vertex operators can be obtained by supersymmetry transformations.} The closed string complex moduli coming from the CFT point of view, can be considered as the supergravity scalar fields (or K\"ahler coordinates sometimes) of the $\mathcal N=2$ space-time supermultiplets when a flip of imaginary and real part is made because, the supergravity construction, imposes certain constraints on the structure of the LEEA.%\footnote{And in all functions related to it as K\"ahler potential $\mathcal K$, gauge kinetic function $f$.} 
When the orientifold projection $\Omega\sigma$ is considered, the correspondence between vertex operators and supergravity moduli fields does not exist anymore because the $N=(2,2)$ string states do not represent scalars of $\mathcal N=1$ chiral supermultiplets in $d=4$. A mixing between NSNS moduli and RR moduli due to the action of $\Omega\sigma$ occurs. As a consequence, when one calculates string scattering amplitudes, one can extract the specific contribution of the real or imaginary part of a given complex modulus, taking for instance a linear combination of the $N=(2,2)$ SCFT vertex operators in order to mimic the scattering between the NSNS or RR moduli that survive the $\Omega\sigma$ projection.
\\
\\
The two-dimensional non-linear sigma model (closed bosonic part) that describes the string propagation in the background of Type IIB orientifold on $\T^6/{\mathbb Z}_2\times{\mathbb Z}_2$, with orientifold projection performed by the worldsheet parity operator $\Omega$ only, is 
\\
\begin{equation}
S = \frac{1}{4\pi}\int d^2z\,\sum_{I=1}^3\DP X^m(z)\bar\DP \bar X^n(\bar z)(g^I)_{mn}(X)\\
%
%&=\frac{1}{4\pi}\int d^2z\,\big[\DP X^4\bar\DP \bar X^4(g^1)_{44} + \DP X^4\bar\DP \bar X^5(g^1)_{45} + \DP X^5\bar\DP \bar X^4(g^1)_{54} + \DP X^5\bar\DP \bar X^5(g^1)_{55}\big] + \sum_{K=2}^3\DP X^m(z)\bar\DP \bar X^n(\bar z)(g^K)_{mn}(X)\\
%
\label{eq:sigmamodel}
\end{equation}
where due to the $\Omega$ projection, the metric $g$ of the factorisable $\T^6=\bigotimes_{I=1}^3\T_I^2$ torus survives the projection while the Kalb-Ramond $B_2$ is projected out.\footnote{As known also the dilaton $\phi$ survives.} The metric and its inverse for each $\T^2_I$ torus are respectively
\begin{equation}
(g^I)_{mn} = \frac{T^I_2}{U^I_2}\begin{pmatrix}1 & U^I_1\\ U^I_1& \abs{U^I}^2\end{pmatrix}\quad\quad (g^I)^{mn} = \frac{1}{T^I_2U^I_2}\begin{pmatrix}\abs{U^I}^2 & -U^I_1\\ -U^I_1& 1\end{pmatrix}
\label{eq:torusmetric}
\end{equation}
with $I\in\{1,2,3\}$ and $[m,n]\in\{[4,5];[6,7];[8,9]\}$.\footnote{For instance $I=1 \rightarrow [m,n]\in[4,5]$.}From the CFT point of view, the geometric untwisted moduli fields \eqref{eq:moduliSugra} describe the deformations of the underlying sigma-model, i.e. they are the real parameters of the complex $\T^2_I$ tori. Thus $U_1^I$ and $U_2^I$ are the real and imaginary part of the complex structure moduli $U^I$(${\bm\bar U}^I$) that parametrise the shape of the $\T^2_I$ torus, while the size of the $\T^2_I$ torus is parametrised by $T_2^I$, namely by the imaginary part of the complex K\"ahler moduli $T^I$(${\bm\bar T}^I$) (the real part $T_1^I$ comes form the RR $C_2$-form and no longer from the NSNS $B_2$-form). The definition of the complex moduli in the CFT description is flipped with respect to the supergravity description of eq. \eqref{eq:moduliSugra}. Expanding \eqref{eq:sigmamodel}, for instance, along the first torus $\T^2_1$ (the same would hold for the other two $\T^2$ tori) new functions of internal bosonic field $X^m$ can be defined 
%%%%%%%%%%%%%%%%
\begin{equation}
\begin{aligned}
S&=\frac{1}{4\pi}\int d^2z\,\bigg[\DP X^4\bar\DP\bar X^4\frac{T^1_{_2}}{U^1_{_2}} + \DP X^4\bar\DP{\bar X}^5\frac{T^1_{_2}}{U^1_{_2}}U^1_1 + \DP X^5{\bar\DP}{\bar X}^4\frac{T^1_{_2}}{U^1_{_2}}U^1_1 + \DP X^5{\bar\DP}{\bar X}^5\frac{T^1_{_2}}{U^1_{_2}}\abs{U^1}^2\bigg]\\ %+\sum_{K=2}^3\DP X^m(z)\bar\DP\bar X^n(\bar z)(g^K)_{mn}(X)\\
&=\frac{1}{4\pi}\int d^2z\,2\bigg[\sqrt{\frac{T^1_{_2}}{2U^1_{_2}}}\DP(X^4 + {\bar U}^1 X^5)\sqrt{\frac{T^1_{_2}}{2U^1_{_2}}}{\bar\DP}({\bar X}^4 + U^1 {\bar X}^5)\bigg]\\ %+\sum_{K=2}^3\DP X^m(z)\bar\DP\bar X^n(\bar z)(g^K)_{mn}(X)\\
&=\frac{1}{2\pi}\int d^2z\DP Z^1(z){\bar\DP}\bar{\tilde Z}^1(\bar z)\,.
\end{aligned}
\label{eq:NLST2}
\end{equation}
Here the new internal bosonic fields $Z^I, \tilde Z^I$ are identified as\,\cite{Lust:2004cx}\footnote{we have changed a bit the original definitions}
\begin{equation}
\begin{aligned}
Z^I(z) &= \sqrt{\frac{T^I_{_2}}{2U^I_{_2}}}(X^{2I +2} + \bar U^I X^{2I +3})(z)\,,\quad \tilde Z^I(z) = \sqrt{\frac{T^I_{_2}}{2U^I_{_2}}}(X^{2I +2} + U^I X^{2I +3})(z)\\ \bar Z^I(\bar z) &= \sqrt{\frac{T^I_{_2}}{2U^I_{_2}}}(\bar X^{2I +2} + \bar U^I \bar X^{2I +3})(\bar z)\,,\quad \bar{\tilde Z}^I(\bar z) = \sqrt{\frac{T^I_{_2}}{2U^I_{_2}}}(\bar X^{2I +2} + U^I \bar X^{2I +3})(\bar z)
\end{aligned}
\label{eq:InternalB}
\end{equation}
while the supersymmetric partners, i.e the internal fermionic fields $\Psi^I,\tilde\Psi^I$ are \cite{Lust:2004cx}
\begin{equation}
\begin{aligned}
\Psi^I(z) &= \sqrt{\frac{T^I_{_2}}{2U^I_{_2}}}(\psi^{2I +2} + \bar U^I \psi^{2I +3})(z)\,,\quad \tilde \Psi^I(z) = \sqrt{\frac{T^I_{_2}}{2U^I_{_2}}}(\psi^{2I +2} + U^I \psi^{2I +3})(z)\\ \bar \Psi^I(\bar z) &= \sqrt{\frac{T^I_{_2}}{2U^I_{_2}}}(\bar\psi^{2I +2} + \bar U^I \bar\psi^{2I +3})(\bar z)\,,\quad \bar{\tilde \Psi}^I(\bar z) = \sqrt{\frac{T^I_{_2}}{2U^I_{_2}}}(\bar\psi^{2I +2} + U^I \bar\psi^{2I +3})(\bar z)\,.
\end{aligned}
\label{eq:InternalF}
\end{equation}
%%%%%%%%%%%%%%%%%%%%%%
%%%%%%%%%%%%%%%%%%%%%%
\subsubsection{Compactified vertex operators on $S_2$ and $D_2$}\label{subsec1}
Using the previous definitions of internal bosonic \eqref{eq:InternalB} and fermionic fields \eqref{eq:InternalF}, one can construct the NS-NS vertex operators for the holomorphic (antiholomorphic) untwisted complex K\"ahler moduli $T^I({\bm\bar T}^I)$  and the untwisted complex structure moduli $U^I({\bm\bar U}^I)$. The known holomorphic building blocks for the NS sector for uncompactified states are 
\begin{equation}
\begin{aligned}
&\mathcal V^\mu_{(-1)}(k,z) = e^{-\phi}\psi^\mu(z)e^{ikX}(z)\\&\mathcal V^\mu_{(0)}(k,z) = \sqrt{\frac{2}{\al}}\Big(i\DP X^\mu + \frac{\al}{2}(k\psi)\psi^\mu\Big)e^{ikX}(z)
\label{eq:BulBlok}
\end{aligned}
\end{equation}
where (-1) and (0) are the ghost pictures. The tensor product with the antiholomorphic part provides the closed vertex operators in the NS-NS sector which read \cite{BLT}
\begin{equation}
\begin{aligned}
&\mathcal W(k,E)_{(-1,-1)} = E_{MN}:e^{-\phi}\psi^M e^{ikX}(z)e^{-\bar\phi}\bar\psi^N e^{ik\bar X}(\bar z):\\
&\mathcal W(k,E)_{(0,0)} = E_{MN}\frac{2}{\al}:\Big(i\DP X^M + \frac{\al}{2}(k\psi)\psi^M\Big)e^{ikX}(z)\Big(i\bar\DP \bar X^N + \frac{\al}{2}(k\bar\psi)\bar\psi^N\Big)e^{ik\bar X}(\bar z):
\end{aligned}
\label{eq:vertexuncop}
\end{equation}
where the polarisation tensor $E_{MN}$ encodes the properties of the state that the vertex has to represent. Taking the compactification of the vertex operators \eqref{eq:vertexuncop} on the underlying background $\T^6=\bigotimes_{I=1}^3\T_I^2$ torus, the NS-NS vertex operator for the untwisted complex K\"ahler modulus $T^I$ in the canonical ghost picture (-1,-1) is 
\begin{equation}
\small{\begin{aligned}
&\boxed{{\cal W}_{_{T^I({-}1,{-}1)}}(E,k,z,\bar z)}=\,E_{mn}[T^I]:{\cal V}_{({-1})}(k,z){\cal V}_{({-}1)}(k,\bar z): = %\epsilon_{mn}[T^I]:e^{-\phi}\psi^me^{ikX}(z)e^{-\bar \phi}\bar\psi^ne^{ik\bar X}(\bar z):\\&
\bigg(\frac{\DP(g^I)_{mn}}{\DP T^I}\bigg):e^{-\phi}\psi^me^{ikX}(z)e^{-\bar \phi}\bar\psi^ne^{ik\bar X}(\bar z):\\
&=\bigg\{\bigg(\frac{\DP(g^I)}{\DP T^I}\bigg)_{[2I{+}2][2I{+}2]}:\psi^{[2I{+}2]}\bar\psi^{[2I{+}2]} + \bigg(\frac{\DP(g^I)}{\DP T^I}\bigg)_{[2I{+}2][2I{+}3]}:\psi^{[2I{+}2]}\bar\psi^{[2I{+}3]} \\&\,+ \bigg(\frac{\DP(g^I)}{\DP T^I}\bigg)_{[2I{+}3][2I{+}2]}:\psi^{[2I{+}3]}\bar\psi^{[2I{+}2]} + \bigg(\frac{\DP(g^I)}{\DP T^I}\bigg)_{[2I{+}3][2I{+}3]}:\psi^{[2I{+}3]}\bar\psi^{[2I{+}3]}\bigg\}e^{-\phi}e^{ikX}(z)e^{-\bar\phi}e^{ik\bar{X}}(\bar z):\\
%&=\Big[\Big(\frac{\DP}{\DP T^I}\frac{(T^I - \bar T^I)}{2iU^I_2}\Big)\psi_{_L}^{[2I+2]}(z)\bar\psi_{_R}^{[2I+2]}(\bar z) + \Big(\frac{\DP}{\DP T^I}\frac{(T^I - \bar T^I)U^I_1}{2iU^I_2}\Big)\psi_{_L}^{[2I+2]}(z)\bar\psi_{_R}^{[2I+3]}(\bar z)\\&+ \Big(\frac{\DP}{\DP T^I}\frac{(T^I - \bar T^I)U^I_1}{2iU^I_2}\Big)\psi_{_L}^{[2I+3]}(z)\bar\psi_{_R}^{[2I+2]}(\bar z) + \Big(\frac{\DP}{\DP T^I}\frac{(T^I - \bar T^I)\abs{U^I}^2}{2iU^I_2}\Big)\psi_{_L}^{[2I+3]}(z)\bar\psi_{_R}^{[2I+3]}(\bar z)\Big]e^{-\phi_{_L}(z)}e^{-\bar\phi_{_R}(\bar z)}e^{ik\cdot X(z,\bar z)}\\
%&=\Big[\frac{1}{2iU^I_2}\psi_{_L}^{[2I+2]}(z)\bar\psi_{_R}^{[2I+2]}(\bar z) + \frac{U^I_1}{2iU^I_2}\psi_{_L}^{[2I+2]}(z)\bar\psi_{_R}^{[2I+3]}(\bar z)\\&+ \frac{U^I_1}{2iU^I_2}\psi_{_L}^{[2I+3]}(z)\bar\psi_{_R}^{[2I+2]}(\bar z) + \frac{\abs{U^I}^2}{2iU^I_2}\psi_{_L}^{[2I+3]}(z)\bar\psi_{_R}^{[2I+3]}(\bar z)\Big]e^{-\phi_{_L}(z)}e^{-\bar\phi_{_R}(\bar z)}e^{ik\cdot X(z,\bar z)}
&= \frac{1}{(T^I {-} \bar T^I)}\bigg\{\frac{T^I_2}{U^I_2}:\psi^{[2I{+}2]}\bar\psi^{[2I{+}2]} + \frac{T^I_2U^I_1}{U^I_2}:\psi^{[2I{+}2]}\bar\psi^{[2I{+}3]}
+ \frac{T^I_2U^I_1}{U^I_2}:\psi^{[2I{+}3]}\bar\psi^{[2I{+}2]}\\&+ \frac{T^I_2\abs{U^I}^2}{U^I_2}:\psi^{[2I{+}3]}\bar\psi^{[2I{+}3]}\bigg\}e^{-\phi}e^{ikX}(z)e^{-\bar\phi}e^{ik\bar{X}}(\bar z):\\
&=\boxed{\frac{2}{(T^I {-}\bar T^I)}:\tilde{\Psi}^Ie^{-\phi}e^{ikX}(z)\bar\Psi^Ie^{-\bar\phi}e^{ik\bar{X}}(\bar z):}
%\\&\Big[(Q,\bar Q) = (1,-1)\Big]
\end{aligned}}
\label{eq:VT}
\end{equation}
\\
where the polarisation tensor $E_{mn}[T^I]$ can be determined taking the variation of the sigma-model \eqref{eq:sigmamodel} with respect to the complex K\"ahler modulus $T^I$. The NS-NS vertex operator for the untwisted complex structure modulus $U^I$ in the same ghost picture (-1,-1) reads
\begin{equation}
\small{\begin{aligned}
&\boxed{{\cal W}_{_{U^I({-}1,{-}1)}}(E,k,z,\bar z)}=\,E_{mn}[U^I]:{\cal V}_{({-1})}(k,z){\cal V}_{({-}1)}(k,\bar z): = %\epsilon_{mn}[U^I]:e^{-\phi}\psi^me^{ikX}(z)e^{-\bar \phi}\bar\psi^ne^{ik\bar X}(\bar z):\\=
\bigg(\frac{\DP(g^I)_{mn}}{\DP U^I}\bigg):e^{-\phi}\psi^me^{ikX}(z)e^{-\bar \phi}\bar\psi^ne^{ik\bar X}(\bar z):\\
%&=\Big[\Big(\frac{\DP}{\DP U^I}\frac{2iT^I_2}{(U^I - \bar U^I)}\Big)\psi_{_L}^{[2I+2]}(z)\bar\psi_{_R}^{[2I+2]}(\bar z) + \Big(\frac{\DP}{\DP U^I}\frac{iT^I_2(U^I + \bar U^I)}{(U^I - \bar U^I)}\Big)\psi_{_L}^{[2I+2]}(z)\bar\psi_{_R}^{[2I+3]}(\bar z)\\&+ \Big(\frac{\DP}{\DP U^I}\frac{iT^I_2(U^I + \bar U^I)}{(U^I - \bar U^I)}\Big)\psi_{_L}^{[2I+3]}(z)\bar\psi_{_R}^{[2I+2]}(\bar z) + \Big(\frac{\DP}{\DP U^I}\frac{2iT^I_2\abs{U^I}^2}{(U^I - \bar U^I)}\Big)\psi_{_L}^{[2I+3]}(z)\bar\psi_{_R}^{[2I+3]}(\bar z)\Big]e^{-\phi_{_L}(z)}e^{-\bar\phi_{_R}(\bar z)}e^{ik\cdot X(z,\bar z)}\\
%&= \Big[-\frac{2iT_2^I}{(U^I - \bar U^I)^2}\psi_{_L}^{[2I+2]}(z)\bar\psi_{_R}^{[2I+2]}(\bar z) + \Big(\frac{iT^I_2(U^I - \bar U^I) - iT^I_2(U^I +\bar U^I)}{(U^I -\bar U^I)^2}\Big)\psi_{_L}^{[2I+2]}(z)\bar\psi_{_R}^{[2I+3]}(\bar z)\\& + \Big(\frac{iT^I_2(U^I - \bar U^I) - iT^I_2(U^I +\bar U^I)}{(U^I -\bar U^I)^2}\Big)\psi_{_L}^{[2I+3]}(z)\bar\psi_{_R}^{[2I+2]}(\bar z)\\& +\Big(\frac{2iT^I_2\bar U^I(U^I - \bar U^I) - 2iT^I_2\abs{U^I}^2}{(U^I - \bar U^I)^2}\Big)\psi_{_L}^{[2I+3]}(z)\bar\psi_{_R}^{[2I+3]}(\bar z)\Big]e^{-\phi_{_L}(z)}e^{-\bar\phi_{_R}(\bar z)}e^{ik\cdot X(z,\bar z)}\\
&=\bigg\{-\frac{T^I_2}{(U^I {-}\bar U^I)U^I_2}:\psi^{[2I{+}2]}\bar\psi^{[2I{+}2]} - \frac{T^I_2\bar U^I}{(U^I {-} \bar U^I)U^I_2}:\psi^{[2I{+}2]}\bar\psi^{[2I{+}3]}-\frac{T^I_2\bar U^I}{(U^I {-} \bar U^I)U^I_2}\psi^{[2I{+}3]}\bar\psi^{[2I{+}2]}\\& - \frac{T^I_2(\bar U^I)^2}{(U^I {-} \bar U^I)U^I_2}\psi^{[2I{+}3]}\bar\psi^{[2I{+}3]}\bigg\}e^{-\phi}e^{ikX}(z)e^{-\bar\phi}e^{ik\bar{X}}(\bar z):\\
&=\boxed{-\frac{2}{(U^I {-} {\bm\bar U}^I)}:\Psi^Ie^{-\phi}e^{ikX}(z)\bar\Psi^Ie^{-\bar\phi}e^{ik\bar{X}}(\bar z):}\,\,.
%\\&\Big[\text{R-symmetry charges}\,\,(Q,\bar Q) = (-1,-1)\Big]
\end{aligned}}
\label{eq:VU}
\end{equation}
From the $N=(2,2)$ SCFT point of view \eqref{eq:VT} and \eqref{eq:VU} can be written in a more general form \cite{Blumenhagen:2006ci}
\begin{equation}
\begin{aligned}
&{\cal W}_{_{T^I({-}1,{-}1)}}(\mathcal E,k,z,\bar z) = {\mathcal E}_I\left[T^I\right]:\Delta^I(z,\bar z)e^{-\phi}e^{ikX}(z)e^{-\bar\phi}e^{ik\bar{X}}(\bar z):\quad,\quad I = 1,{\dots}, h^{1,1}_{\text{utw}+}\\
&{\cal W}_{_{U^I({-}1,{-}1)}}(\mathcal E,k,z,\bar z) = {\mathcal E}_I\left[U^I\right]:\Sigma^I(z,\bar z)e^{-\phi}e^{ikX}(z)e^{-\bar\phi}e^{ik\bar{X}}(\bar z):\quad,\quad I = 1,{\dots}, h^{2,1}_{\text{utw}+}
\end{aligned}
\label{SCFT_vertXmoduli_def}
\end{equation}
where $\Delta^I(z,\bar z)$ and $\Sigma^I(z,\bar z)$ are conformal field with conformal dimensions $(h,\bar h) = (1/2,1/2)$ respect to the internal $N=(2,2)$ SCFT and charged under the pair of $U(1)$ currents $(J,\bar J)$: (1,-1) for the field $\Delta^I(z,\bar z)$ and (-1,-1) for the field $\Sigma^I(z,\bar z)$. The complex conjugate conformal fields ${\bm\bar\Delta}^I(z,\bar z)$ and ${\bm\bar\Sigma}^I$ have the same conformal dimensions and opposite charges. Thus the $U(1)$ charges of the fields $Z(\Psi)$ and $\tilde Z(\tilde\Psi)$ are -1 and 1, respectively.\,\footnote{The same holds for the $\bar Z(\bar\Psi)$ and $\bar{\tilde Z}(\bar{\tilde\Psi})$}
%%%%%%%
Here, the vertex operators for the antiholomorphic complex K\"ahler modulus ${\bm\bar T}^I$ and  the complex structure modulus ${\bm\bar U}^I$ in the (-1,-1) ghost pictures and in the (0,0) ghost pictures are collected
%%%%%%%%%%%%%%%%%%%%%%%%%%
%%%%%%%%%%%%%%%%%%%%%%%%%%
\begin{equation*}
%\begin{aligned}
%\begin{equation*}
\begin{aligned}
{\cal W}_{_{{\bm\bar{T}}^I({-}1,{-}1)}}(E,k,z,\bar z)&=%&\,\_{mn}[T^I]:{\cal V}_{({-1})}(k,z){\cal V}_{({-}1)}(k,\bar z): = 
E_{mn}[{\bm\bar T}^I]:e^{-\phi}\psi^me^{ikX}(z)e^{-\bar \phi}\bar\psi^ne^{ik\bar X}(\bar z):\\&= - \frac{2}{(T^I {-}{\bm\bar T}^I)} :\Psi^Ie^{-\phi}e^{ikX}(z)\bar{\tilde\Psi}^Ie^{-\bar\phi}e^{ik\bar{X}}(\bar z):
%\\&\Big[\text{R-symmetry charges}\,\,(Q,\bar Q) = (-1,1)\Big]
\end{aligned}
\end{equation*}
\begin{equation}
\begin{aligned}
%\label{eq:VbarT}
%\end{equation*}
%%%%%%%%%%%%%%%%%%
%%%%%%%%%%%%%%%%%%
%%%%%%%%%%%%
%%%%%%%%%%%%
%\begin{equation*}
&\begin{aligned}
{\cal W}_{_{{\bm\bar U}^I({-}1,{-}1)}}(E,k,z,\bar z)&=%\,\epsilon_{mn}[\bar {U}^I]:{\cal V}_{({-1})}(k,z){\cal V}_{({-}1)}(k,\bar z): =
 E_{mn}[{\bm\bar U}^I]:e^{-\phi}\psi^me^{ikX}(z)e^{-\bar \phi}\bar\psi^ne^{ik\bar X}(\bar z):\\
%&= \Big[\frac{T^I_2}{(U^I - \bar U^I)U^I_2}\bar\psi^{[2I+2]}(\bar z)\psi^{[2I+2]}(z) +\frac{T^I_2 U^I}{(U^I - \bar U^I)U^I_2}\bar\psi^{[2I+2]}(\bar z)\psi^{[2I+3]}(z)\\&+\frac{T^I_2 U^I}{(U^I - \bar U^I)U^I_2}\bar\psi^{[2I+3]}(\bar z)\psi^{[2I+2]}(z) + \frac{T^I_2(U^I)^2}{(U^I - \bar U^I)U^I_2}\bar\psi^{[2I+3]}(\bar z)\psi^{[2I+3]}(z)\Big]e^{-\bar\phi(\bar z)}e^{-\phi(z)}e^{ik\cdot X(z,\bar z)}\\
&= \frac{2}{(U^I {-}{\bm\bar U}^I)}\tilde\Psi^Ie^{-\phi}e^{ikX}(z)\bar{\tilde\Psi}^Ie^{-\bar \phi}e^{ik\bar X}(\bar z):
%\\&\Big[\text{R-symmetry charges}\,\,(Q,\bar Q) = (1,1)\Big]
\end{aligned}\\
%\label{eq:VbarU}
%\end{equation*}
%%%%%%%%%%%%
%%%%%%%%%%%%
%\begin{equation*}
&\begin{aligned}
{\cal W}_{_{T^I(0,0)}}(E,k,z,\bar z)&=%\,\epsilon_{mn}[T^I]:{\cal V}_{(0)}(k,z){\cal V}_{(0)}(k,\bar z):=
\frac{2}{\al}E_{mn}[T^I]:\Big(i\DP X^n {+} \frac{\al}{2}(k\psi)\psi^n\Big)e^{ikX}(z)\Big(i\bar\DP \bar X^m {+} \frac{\al}{2}(k \bar\psi)\bar\psi^m\Big)e^{ik\bar X}(\bar z):\\
&=\frac{4}{\al(T^I {-}{\bm\bar T}^I)}:\Big(i\DP \tilde Z^I + \frac{\al}{2}(k\psi)\tilde\Psi^I\Big)e^{ikX}(z)\Big(i\bar\DP \bar Z^I + \frac{\al}{2}(k \bar\psi)\bar \Psi^I\Big)e^{ik\bar X}(\bar z):
%\\&\Big[\text{R-symmetry charges}\,\,(Q,\bar Q) = (-1,1)\Big]
\end{aligned}\\\\
%\label{eq:VT0}
%\end{equation*}
%%%%%%%%%%%
%%%%%%%%%%%
%\begin{equation*}
&\begin{aligned}
{\cal W}_{_{{\bm\bar T}^I(0,0)}}(E,k,z,\bar z)&=%\,\epsilon_{mn}[\bar{T}^I]:{\cal V}_{(0)}(k,z){\cal V}_{(0)}(k,\bar z):
\frac{2}{\al}E_{mn}[{\bm\bar T}^I]\Big(i\DP X^m + \frac{\al}{2}(k\psi)\psi^m\Big)e^{ikX}(z)\Big(i\bar\DP \bar X^n + \frac{\al}{2}(k\bar \psi)\bar \psi^n\Big)e^{ik\bar X}(\bar z):
\\&=-\frac{4}{\al(T^I {-} {\bm\bar T}^I)}:\Big(i\DP Z^I + \frac{\al}{2}(k\psi)\Psi^I\Big)e^{ikX}(z)\Big(i\bar\DP\bar{\tilde Z}^I + \frac{\al}{2}(k\bar\psi)\bar{\tilde\Psi}^I\Big)e^{ik\bar X}(\bar z):
%\\&\Big[\text{R-symmetry charges}\,\,(Q, \bar Q) = (1,-1)\Big]
\end{aligned}\\\\
%\label{eq:VbarT0}
%\end{equation*}
%%%%%%%%%%%%
%%%%%%%%%%%%
%\begin{equation*}
&\begin{aligned}
{\cal W}_{_{U^I(0,0)}}(E,k,z,\bar z)&=%\,E_{mn}[U^I]:{\cal V}_{(0)}(k,z){\cal V}_{(0)}(k,\bar z):
\frac{2}{\al}E_{mn}[U^I]:\Big(i\DP X^m + \frac{\al}{2}(k\psi)\psi^m\Big)e^{ikX}(z)\Big(i\bar\DP \bar X^n + \frac{\al}{2}(k\bar \psi)\bar \psi^n\Big)e^{ik\bar X}(\bar z):
\\&=-\frac{4}{\al(U^I {-} {\bm\bar U}^I)}:\Big(i\DP Z^I + \frac{\al}{2}(k\psi)\Psi^I\Big)e^{ikX}(z)\Big(i\bar\DP \bar Z^I + \frac{\al}{2}(k\bar\psi)\bar\Psi^I\Big)e^{ik\bar X}(\bar z):
%\\&\Big[\text{R-symmetry charges}\,\,( Q,\bar Q) = (-1,-1)\Big]
\end{aligned}\\\\
%\label{eq:VU0}
%\end{equation*}
%%%%%%%%%%%%
%%%%%%%%%%%%
%\begin{equation*}
&\begin{aligned}
{\cal W}_{_{{\bm\bar U}^I(0,0)}}(E,k,z,\bar z)&=%\,[\bar{U}^I]:{\cal V}_{(0)}(k,z){\cal V}_{(0)}(k,\bar z):
\frac{2}{\al}E_{mn}[{\bm\bar U}^I]:\Big(i\DP X^m + \frac{\al}{2}(k\psi)\psi^m
\Big)e^{ikX}(z)\Big(i\bar\DP \bar X^n + \frac{\al}{2}(k\bar \psi)\bar \psi^n\Big)e^{ik\bar X}(\bar z):
\\&=\frac{4}{\al(U^I {-} {\bm\bar U}^I)}:\Big(i\DP \tilde Z^I + \frac{\al}{2}(k \psi)\tilde\Psi^I\Big)e^{ikX}(z)\Big(i\bar\DP \bar{\tilde Z}^I + \frac{\al}{2}(k\bar\psi)\bar {\tilde\Psi}^I\Big)e^{ik\bar X}(\bar z):
%\\&\Big[\text{R-symmetry charges}\,\,(Q,\bar Q) = (1,1)\Big]
\label{eq:vertexD2}
\end{aligned}
\end{aligned}
\end{equation}
\\
Such vertex operators can be of interest when calculation of string scattering amplitudes involve oriented surfaces like the $S_2$ sphere and the $D_2$ disk. To be more precisely, we are interested on the imaginary part $T_2$ of the K\"ahler modulus $T$ and its vertex operator is given by the following combination
\begin{equation}
\mathcal W_{{T_2^I}{(q,\bar q)}}(E,z,\bar z, k) = -\frac{i}{2}\left(\mathcal W_{{T^I}{(q,\bar q)}}(E,z,\bar z, k) - \mathcal W_{{\bar T^I}{(q,\bar q)}}(E,z,\bar z, k)\right)\,.
\label{eq:VT2}
\end{equation}

%%%%%%%%%%%%%
%%%%%%%%%%%%% 
\subsubsection{Compactified vertex operators on $RP_2$}\label{subsec2}
When an unoriented surface as the real projective plane $RP_2$ is considered, the usual string vertex operators can not be used due to the presence of $\Omega_P$-planes induced by the action of $\Omega\sigma$ as indicated in Table \ref{tab:tabOM} \cite{Garousi:2006zh, BLT}.
Closed vertex operators in a generic picture $(q,\bar q)$ for a generic $\Omega_P$-plane, represent string states that are invariant under the $\Omega\sigma$ action and manifestly invariant under the involution $\mathfrak {I}_{_{}RP_2}(z)= - 1/\bar z$
\begin{equation}
\mathcal W^{\otimes}_{(q,\bar q)}(k,{E}) = \frac{1}{2}\Big(E_{MN}\mathcal V^{M}_{(q)}(k,z)\mathcal V^{N}_{(\bar q)}(k,\bar z) + E_{MN}\mathcal R^{M}_{P}\mathcal R^{N}_{S}\mathcal V_{(\bar q)}^{P}(k\mathcal R,\bar z)\mathcal V^{S}_{(q)}(k\mathcal R,z)\Big)\,.
\label{eq:VRP2gen}
\end{equation}
This vertex is a symmetric combination of holomorphic and antiholomorphic vertices, since $\Omega$ operator exchanges the left-movers with the right-movers of the string while the reflection matrix $\mathcal R$, defined in the next section (see eq. \eqref{eq:matrixRint}), manifests the action of $\sigma=I_n$, that acts as a reflection in the $n$ directions perpendicular to the $\Omega_P$-plane. Moreover the operator $I_n$ can be viewed as the result of $n\equiv(9-p)$ $T$-dualities of Type I theory.\footnote{For instance a single $T$-duality acts on the worldsheet parity operator $\Omega$ as $T^{-1}\Omega T = \Omega I_1$.}
%
%\begin{equation}
%\begin{aligned}
%&\Omega\sigma = \Omega I_n \rightarrow O_{[9 - n]}\\
%&I_0\equiv 1 : O_9\,; \quad I_2\equiv R_i : O_7\,;\quad I_4\equiv R_iR_j : O_5\,;\quad I_6\equiv R_iR_jR_k : O_3
%\end{aligned}\label{eq:Orientifold_Op}
%\end{equation}
% 
%Now in our compactified theory we have to take into account the presence of the $U(1)$-charges $Q(\bar Q)$, so how the capitol field $Z, \Psi$ works under the $\Omega$ actions. We argue that the actions of $\Omega$ on the capitol field preserves the eigenvalue of the $U(1)$-charge, thus 
%
%\\
%\begin{equation}
%\begin{aligned}
%&\tilde{Z}^I(z)\,\,(Q=1) \xleftrightarrow[]{\Omega} \bar{\tilde{Z}}^I(\bar z)\,\,(\bar Q = 1)\,;\quad Z^I(z)\,\,(Q= -1)  \xleftrightarrow[]{\Omega} \bar{Z}^I(\bar z)\,\,(\bar Q = -1)\\\\
%&\tilde{\Psi}^I(z)\,\,(Q=1) \xleftrightarrow[]{\Omega} \bar{\tilde{\Psi}}^I(\bar z)\,\,(\bar Q = 1)\,;\quad \Psi^I(z)\,\,(Q= -1)  \xleftrightarrow[]{\Omega} \bar{\Psi}^I_{_R}(\bar z)\,\,(\bar Q = -1)
%\end{aligned}
%\end{equation}
% 
 For the untwisted moduli $T^I$ and $U^I$ the vertex operators ${\cal W}^{\otimes}_{_{(q,\bar q)}}$ in the $(-1,-1)$, $(0,0)$ picture read%The explicit definition of the reflection matrix $\mathcal R$ is left for the next section. %Using the properties $(\mathcal R)^2 =1$ one we can write 
\begin{equation*}
\begin{aligned}
&{\cal W}^{\otimes}_{_{T^I({-}1,{-}1)}}(E,k,z,\bar z)=%\,\frac{1}{2}E_{mn}[T^I]\Big\{:{\cal V}_{({-1})}(k,z){\cal V}_{({-}1)}(k,\bar z): +\mathcal R^{m}_{p}\mathcal R^{n}_{s} :{\cal V}^p_{({-1})}(k\R,\bar z){\cal V}^s_{({-}1)}(\kt\R,z):\Big\} \\&= %\frac{1}{2}\Bigg[\epsilon_{mn}[T^I]e^{-\bar\phi(\bar z)}\bar\psi^m(\bar z)e^{-\phi(z)}\psi^n(z)e^{ik\cdot X(z,\bar z)} + (D\epsilon^t[T^I]D)_{mn}e^{-\bar\phi(\bar z)}\bar\psi^m(\bar z)e^{-\phi(z)}\psi^n(z)e^{ik\cdot D X(z,\bar z)}\Bigg]\\&= \frac{1}{2}\Bigg[ \frac{1}{(T^I - \bar T^I)}\bigg(\frac{T^I_2}{U^I_2}\bar\psi^{[2I+2]}(\bar z)\psi^{[2I+2]}(z) + \frac{T^I_2U^I_1}{U^I_2}\bar\psi^{[2I+2]}(\bar z)\psi^{[2I+3]}(z)+ \frac{T^I_2U^I_1}{U^I_2}\bar\psi^{[2I+3]}(\bar z)\psi^{[2I+2]}(z)\\& + \frac{T^I_2\abs{U^I}^2}{U^I_2}\bar\psi^{[2I+3]}(\bar z)\psi^{[2I+3]}(z)\bigg)e^{-\bar\phi(\bar z)}e^{-\phi(z)}e^{ik\cdot X(z,\bar z)} +\frac{1}{(T^I - \bar T^I)}\bigg(\frac{T^I_2}{U^I_2}\bar\psi^{[2I+2]}(\bar z)\psi^{[2I+2]}(z) \\&+ \frac{T^I_2U^I_1}{U^I_2}\bar\psi^{[2I+2]}(\bar z)\psi^{[2I+3]}(z)+ \frac{T^I_2U^I_1}{U^I_2}\bar\psi^{[2I+3]}(\bar z)\psi^{[2I+2]}(z) + \frac{T^I_2\abs{U^I}^2}{U^I_2}\bar\psi^{[2I+3]}(\bar z)\psi^{[2I+3]}(z)\bigg)e^{-\bar\phi(\bar z)}e^{-\phi(z)}e^{ik\cdot D X(z,\bar z)}\Bigg]\\&=
\frac{1}{(T^I {-}\bar T^I)}\bigg\{{:}\tilde\Psi^Ie^{-\phi}e^{ikX}(z)\bar\Psi^Ie^{-\bar\phi}e^{ik\bar X}(\bar z){:} {+}{:}\left(\{L,k,z\}\leftrightarrow\{R,k\R,\bar z\}\right)
%\bar{\tilde\Psi}^Ie^{-\bar\phi}e^{ik {\R}\bar X}(\bar z)\Psi^Ie^{-\phi}e^{ik {\R}X}(z)
{:}\bigg\}
%\\&\Big[\text{R-symmetry charges}\,\,(Q,\bar Q) = (1, -1)\Big]
\end{aligned}
\end{equation*}
%%%%%%%%%%%%
\begin{equation*}
\begin{aligned}
&{\cal W}^{\otimes}_{_{\bar{T}^I({-}1,{-}1)}}(E,k,z,\bar z)=%\,E_{mn}[\bar{T}^I]\frac{1}{2}\Big\{:{\cal V}^m_{({-1})}(k,z){\cal V}^n_{({-}1)}(k,\bar z): + \mathcal R^{m}_{p}\mathcal R^{n}_{s} :{\cal V}^p_{({-1})}(k\R,\bar z){\cal V}^s_{({-}1)}(k\R,z):\Big\} \\&= %\frac{1}{2}\Bigg[\epsilon_{mn}[\bar T^I]e^{-\bar\phi(\bar z)}\bar\psi^m(\bar z)e^{-\phi(z)}\psi^n(z)e^{ik\cdot X(z,\bar z)} + (D\epsilon^t[\bar T^I]D)_{mn}e^{-\bar\phi(\bar z)}\bar\psi^m(\bar z)e^{-\phi(z)}\psi^n(z)e^{ik\cdot D X(z,\bar z)}\Bigg]\\&=\frac{1}{2}\Bigg[ -\frac{1}{(T^I - \bar T^I)}\bigg(\frac{T^I_2}{U^I_2}\bar\psi^{[2I+2]}(\bar z)\psi^{[2I+2]}(z) + \frac{T^I_2U^I_1}{U^I_2}\bar\psi^{[2I+2]}(\bar z)\psi^{[2I+3]}(z)+ \frac{T^I_2U^I_1}{U^I_2}\bar\psi^{[2I+3]}(\bar z)\psi^{[2I+2]}(z) \\&+ \frac{T^I_2\abs{U^I}^2}{U^I_2}\bar\psi^{[2I+3]}(\bar z)\psi^{[2I+3]}(z)\bigg)e^{-\bar\phi(\bar z)}e^{-\phi(z)}e^{ik\cdot X(z,\bar z)} -\frac{1}{(T^I - \bar T^I)}\bigg(\frac{T^I_2}{U^I_2}\bar\psi^{[2I+2]}(\bar z)\psi^{[2I+2]}(z) \\&+ \frac{T^I_2U^I_1}{U^I_2}\bar\psi^{[2I+2]}(\bar z)\psi^{[2I+3]}(z)+ \frac{T^I_2U^I_1}{U^I_2}\bar\psi^{[2I+3]}(\bar z)\psi^{[2I+2]}(z) + \frac{T^I_2\abs{U^I}^2}{U^I_2}\bar\psi^{[2I+3]}(\bar z)\psi^{[2I+3]}(z)\bigg)e^{-\bar\phi(\bar z)}e^{-\phi(z)}e^{ik\cdot D X(z,\bar z)}\Bigg]\\&=
-\frac{1}{(T^I{-}\bar T^I)}\bigg\{{:}\Psi^Ie^{-\phi}e^{ikX}(z)\bar{\tilde\Psi}^Ie^{-\bar\phi}e^{ik\bar X}(\bar z) {:}{+}{:}\left(\{L,k,z\}\leftrightarrow\{R,k\R,\bar z\}\right)
%\bar{\Psi}^Ie^{-\bar\phi}e^{ik {\R}\bar X}(\bar z)\tilde\Psi^Ie^{-\phi}e^{ik {\R}X}(z)
{:}\bigg\}
%\\&\Big[\text{R-symmetry charges}\,\,(Q,\bar Q) = (-1,1)\Big]
\end{aligned}
\end{equation*}
%\\&\\
%%%%%%%%%%%%%%
%%%%%%%%%%%%%%
\begin{equation}
\begin{aligned}
&\begin{aligned}
&{\cal W}^{\otimes}_{_{U^I({-}1,{-}1)}}(E,k,z,\bar z)=%\,E_{mn}[U^I]\frac{1}{2}\Big\{:{\cal V}_{({-1})}(k,z){\cal V}_{({-}1)}(k,\bar z): + \mathcal R^{m}_{p}\mathcal R^{n}_{s} :{\cal V}^p_{({-1})}(k\R,\bar z){\cal V}^s_{({-}1)}(k\R,z):\Big\} \\&=%\frac{1}{2}\Bigg[\epsilon_{mn}[U^I]e^{-\bar\phi(\bar z)}\bar\psi^m(\bar z)e^{-\phi(z)}\psi^n(z)e^{ik\cdot X(z,\bar z)} + (D\epsilon^t[U^I]D)_{mn}e^{-\bar\phi(\bar z)}\bar\psi^m(\bar z)e^{-\phi(z)}\psi^n(z)e^{ik\cdot D X(z,\bar z)}\Bigg]\\&=\frac{1}{2}\Bigg[-\frac{1}{(U^I - \bar U^I)}\bigg(\frac{T^I_2}{U^I_2}\bar\psi^{[2I+2]}(\bar z)\psi^{[2I+2]}(z) + \frac{T^I_2\bar U^I}{U^I_2}\bar\psi^{[2I+2]}(\bar z)\psi^{[2I+3]}(z)+\frac{T^I_2\bar U^I}{U^I_2}\bar\psi^{[2I+3]}(\bar z)\psi^{[2I+2]}(z)\\& + \frac{T^I_2(\bar U^I)^2}{U^I_2}\bar\psi^{[2I+3]}(\bar z)\psi^{[2I+3]}(z)\bigg)e^{-\bar\phi(\bar z)}e^{-\phi(z)}e^{ik\cdot X(z,\bar z)}- \frac{1}{(U^I - \bar U^I)}\bigg(\frac{T^I_2}{U^I_2}\bar\psi^{[2I+2]}(\bar z)\psi^{[2I+2]}(z)\\& + \frac{T^I_2\bar U^I}{U^I_2}\bar\psi^{[2I+2]}(\bar z)\psi^{[2I+3]}(z)+\frac{T^I_2\bar U^I}{U^I_2}\bar\psi^{[2I+3]}(\bar z)\psi^{[2I+2]}(z) + \frac{T^I_2(\bar U^I)^2}{U^I_2}\bar\psi^{[2I+3]}(\bar z)\psi^{[2I+3]}(z)\bigg)e^{-\bar\phi(\bar z)}e^{-\phi(z)}e^{ik\cdot D X(z,\bar z)}\Bigg]\\&= 
-\frac{1}{(U^I {-} \bar U^I)}\bigg\{{:}\Psi^Ie^{-\phi}e^{ikX}(z)\bar\Psi^Ie^{-\bar\phi}e^{ik\bar X}(\bar z){:} +{:} \left(\{L,k,z\}\leftrightarrow\{R,k\R,\bar z\}\right)
%\bar\Psi^Ie^{-\bar\phi}e^{ik {\R}\bar X}(\bar z)\Psi^Ie^{-\phi}e^{ik {\R}X}(z)
{:}\bigg\}
%\\&\Big[\text{R-symmetry charges}\,\,(Q,\bar Q) = (-1,-1)\Big]
\end{aligned}
%\end{equation}
\\&\\
%%%%%%%%%%%%
%%%%%%%%%%%%
%\begin{equation}
&\begin{aligned}
&{\cal W}^{\otimes}_{_{\bar{U}^I({-}1,{-}1)}}(E,k,z,\bar z)=%\,E_{mn}[\bar{U}^I]\frac{1}{2}\Big\{:{\cal V}^m_{({-1})}(k,z){\cal V}^n_{({-}1)}(k,\bar z): + \mathcal R^{m}_{p}\mathcal R^{n}_{s} :{\cal V}^p_{({-1})}(k\R,\bar z){\cal V}^s_{({-}1)}(k\R,z):\Big\} \\&= %\frac{1}{2}\Bigg[\epsilon_{mn}[\bar U^I]e^{-\bar\phi(\bar z)}\bar\psi^m(\bar z)e^{-\phi(z)}\psi^n(z)e^{ik\cdot X(z,\bar z)} + (D\epsilon^t[\bar U^I]D)_{mn}e^{-\bar\phi(\bar z)}\bar\psi^m(\bar z)e^{-\phi(z)}\psi^n(z)e^{ik\cdot D X(z,\bar z)}\Bigg]\\&=\frac{1}{2}\Bigg[\frac{1}{(U^I - \bar U^I)}\bigg(\frac{T^I_2}{U^I_2}\bar\psi^{[2I+2]}(\bar z)\psi^{[2I+2]}(z) + \frac{T^I_2U^I}{U^I_2}\bar\psi^{[2I+2]}(\bar z)\psi^{[2I+3]}(z) + \frac{T^I_2U^I}{U^I_2}\bar\psi^{[2I+3]}(\bar z)\psi^{[2I+2]}(z)\\& + \frac{T^I_2(U^I)^2}{U^I_2}\bar\psi^{[2I+3]}(\bar z)\psi^{[2I+3]}(z)\bigg)e^{\bar\phi(\bar z)}e^{-\phi(z)}e^{ik\cdot X(z,\bar z)} + \frac{1}{(U^I - \bar U^I)}\bigg(\frac{T^I_2}{U^I_2}\bar\psi^{[2I+2]}(\bar z)\psi^{[2I+2]}(z)\\& + \frac{T^I_2U^I}{U^I_2}\bar\psi^{[2I+2]}(\bar z)\psi^{[2I+3]}(z)+\frac{T^I_2U^I}{U^I_2}\bar\psi^{[2I+3]}(\bar z)\psi^{[2I+2]}(z) + \frac{T^I_2(U^I)^2}{U^I_2}\bar\psi^{[2I+3]}(\bar z)\psi^{[2I+3]}(z)\bigg)e^{-\bar\phi(\bar z)}e^{-\phi(z)}e^{ik\cdot D X(z,\bar z)}\Bigg]\\&= 
\frac{1}{(U^I {-} \bar U^I)}\bigg\{{:}\tilde\Psi^Ie^{-\phi}e^{ikX}(z)\bar{\tilde\Psi}^Ie^{-\bar\phi}e^{ik\bar X}(\bar z){:} + {:}\left(\{L,k,z\}\leftrightarrow\{R,k\R,\bar z\}\right)
%\bar{\tilde\Psi}^Ie^{-\bar\phi}e^{ik {\R}\bar X}(\bar z)\tilde\Psi^Ie^{-\phi}e^{ik {\R}X}(z)
{:}\bigg\}
%\\&\Big[\text{R-symmetry charges}\,\,(Q,\bar Q) = (1,1)\Big]
\end{aligned}
%\end{equation}
\\&\\
%%%%%%%%%%%%
%%%%%%%%%%%%
%\begin{equation}
&\begin{aligned}
{\cal W}^{\otimes}_{_{T^I(0,0)}}(E,k,z,\bar z)&=%\,E_{mn}[T^I]\frac{1}{2}\Big\{:{\cal V}^m_{(0)}(k,z){\cal V}^n_{(0)}(k,\bar z): +\mathcal R^{m}_{p}\mathcal R^{n}_{s} :{\cal V}^p_{(0)}(k\R,\bar z){\cal V}^s_{(0)}(k\R,z):\Big\} \\=&%\frac{1}{2}\Bigg[\epsilon_{mn}[T^I]\frac{2}{\al}\bigg(i\bar\DP \bar X^m(\bar z) + \frac{\al}{2}(k\cdot \bar \psi(\bar z))\bar \psi^m(\bar z)\bigg)\bigg(i\DP X^n(z) + \frac{\al}{2}(k\cdot \psi(z))\psi^n(z)\bigg)e^{ik\cdot X(z,\bar z)}\\& +(D\epsilon^t[T^I]D)_{mn}\frac{2}{\al}\bigg(i\bar\DP \bar X^m(\bar z) + \frac{\al}{2}(k\cdot D \bar \psi(\bar z))\bar \psi^m(\bar z)\bigg)\bigg(i\DP X^n(z) + \frac{\al}{2}(k\cdot D \psi(z))\psi^n(z)\bigg)e^{ik\cdot D X(z,\bar z)}\Bigg]\\&=
\frac{2}{\al(T^I {-} \bar T^I)}\bigg\{:\bigg(i\DP \tilde Z^I + \frac{\al}{2}(k\psi)\tilde\Psi^I\bigg)e^{ikX}(z)\bigg(i\bar\DP \bar Z^I + \frac{\al}{2}(k\bar \psi)\bar \Psi^I\bigg)e^{ik\bar X}(\bar z):\\&\hspace{2.7cm} + :\left(\{L,k,z\}\leftrightarrow\{R,k\R,\bar z\}\right)
%\bigg(i\bar\DP \bar {\tilde Z}^I + \frac{\al}{2}(k {\R}\bar \psi)\bar{\tilde\Psi}^I\bigg)e^{ik {\R}\bar X}(\bar z)\bigg(i\DP Z^I + \frac{\al}{2}(k {\R}\psi)\Psi^I\bigg)e^{ik {\R}X}(z)
:\bigg\}
%\\&\Big[\text{R-symmetry charges}\,\,(Q,\bar Q) = (1,-1)\Big]
\end{aligned}
%\end{equation}
\\&\\
%%%%%%%%%%%
%%%%%%%%%%%
%\begin{equation}
&\begin{aligned}
{\cal W}^{\otimes}_{_{\bar{T}^I(0,0)}}(E,k,z,\bar z)&=%\,E_{mn}[\bar{T}^I]\frac{1}{2}\Big\{:{\cal V}_{(0)}(k,z){\cal V}_{(0)}(k,\bar z): + \mathcal R^{m}_{p}\mathcal R^{n}_{s} :{\cal V}^p_{(0)}(k\R,\bar z){\cal V}^s_{(0)}(k\R,z):\Big\} \\&= %\frac{1}{2}\Bigg[\epsilon_{mn}[\bar T^I]\frac{2}{\al}\bigg(i\bar\DP \bar X^m(\bar z) + \frac{\al}{2}(k\cdot \bar \psi(\bar z))\bar \psi^m(\bar z)\bigg)\bigg(i\DP X^n(z) + \frac{\al}{2}(k\cdot \psi(z))\psi^n(z)\bigg)e^{ik\cdot X(z,\bar z)}\\& +(D\epsilon^t[\bar T^I]D)_{mn}\frac{2}{\al}\bigg(i\bar\DP \bar X^m(\bar z) + \frac{\al}{2}(k\cdot D \bar \psi(\bar z))\bar \psi^m(\bar z)\bigg)\bigg(i\DP X^n(z) + \frac{\al}{2}(k\cdot D \psi(z))\psi^n(z)\bigg)e^{ik\cdot D X(z,\bar z)}\Bigg]\\&=
-\frac{2}{\al(T^I {-} \bar T^I)}\bigg\{:\bigg(i\DP Z^I + \frac{\al}{2}(k\psi)\Psi^I\bigg)e^{ikX}(z)\bigg(i\bar\DP \bar{\tilde Z}^I + \frac{\al}{2}(k\bar \psi)\bar{\tilde \Psi}^I\bigg)e^{ik\bar X}(\bar z):\\&\hspace{2.7cm} +:
\left(\{L,k,z\}\leftrightarrow\{R,k\R,\bar z\}\right)
%\bigg(i\bar\DP \bar{Z}^I + \frac{\al}{2}(k {\R}\bar \psi)\bar{\Psi}^I\bigg)e^{ik {\R}\bar X}(\bar z)\bigg(i\DP \tilde{Z}^I + \frac{\al}{2}(k {\R} \psi)\tilde{\Psi}^I\bigg)e^{ik {\R}X}(z)
:\bigg\}
%\\&\Big[\text{R-symmetry charges}\,\,(Q,\bar Q) = (-1,1)\Big]
\end{aligned}
%\end{equation}
\\&\\
%\begin{equation}
&\begin{aligned}
{\cal W}^{\otimes}_{_{U^I(0,0)}}(E,k,z,\bar z)&=%\,E_{mn}[U^I]\frac{1}{2}\Big\{:{\cal V}^m_{(0)}(k,z){\cal V}^n_{(0)}(k,\bar z): + \mathcal R^{m}_{p}\mathcal R^{n}_{s} :{\cal V}^p_{({-1})}(k\R,\bar z){\cal V}^s_{({-}1)}(k\R,z):\Big\} \\&= %\frac{1}{2}\Bigg[\epsilon_{mn}[U^I]\frac{2}{\al}\bigg(i\bar\DP \bar X^m(\bar z) + \frac{\al}{2}(k\cdot \bar \psi(\bar z))\bar \psi^m(\bar z)\bigg)\bigg(i\DP X^n(z) + \frac{\al}{2}(k\cdot \psi(z))\psi^n(z)\bigg)e^{ik\cdot X(z,\bar z)}\\& +(D\epsilon^t[U^I]D)_{mn}\frac{2}{\al}\bigg(i\bar\DP \bar X^m(\bar z) + \frac{\al}{2}(k\cdot D \bar \psi(\bar z))\bar \psi^m(\bar z)\bigg)\bigg(i\DP X^n(z) + \frac{\al}{2}(k\cdot D \psi(z))\psi^n(z)\bigg)e^{ik\cdot D X(z,\bar z)}\Bigg]\\&=
-\frac{2}{\al(U^I - \bar U^I)}\bigg\{:\bigg(i\DP Z^I + \frac{\al}{2}(k\psi)\Psi^I\bigg)e^{ikX}(z)\bigg(i\bar\DP \bar Z^I + \frac{\al}{2}(k\bar \psi)\bar \Psi^I\bigg)e^{ik\bar X}(\bar z)\\&\hspace{2.7cm}  +:\left(\{L,k,z\}\leftrightarrow\{R,k\R,\bar z\}\right)
%\bigg(i\bar\DP \bar Z^I + \frac{\al}{2}(k {\R}\bar \psi)\bar \Psi^I\bigg)e^{ik {\R}\bar X}\bigg(i\DP Z^I + \frac{\al}{2}(k {\R}\psi)\Psi^I\bigg)e^{ik {\R}X}(z)
:\bigg\}
%\\&\Big[\text{R-symmetry charges}\,\,(Q,\bar Q) = (-1,-1)\Big]
\end{aligned}
%\end{equation}
\\&\\
%%%%%%%%%%%
%%%%%%%%%%%
%\begin{equation}
&\begin{aligned}
{\cal W}^{\otimes}_{_{\bar{U}^I(0,0)}}(E,k,z,\bar z)&=%\,E_{mn}[\bar{U}^I]\frac{1}{2}\Big\{:{\cal V}^m_{(0)}(k,z){\cal V}^n_{(0)}(k,\bar z): + \mathcal R^{m}_{p}\mathcal R^{n}_{s} :{\cal V}^p_{(0)}(k\R,\bar z){\cal V}^s_{(0)}(k\R,z):\Big\} \\&=%\frac{1}{2}\Bigg[\epsilon_{mn}[\bar U^I]\frac{2}{\al}\bigg(i\bar\DP \bar X^m(\bar z) + \frac{\al}{2}(k\cdot \bar \psi(\bar z))\bar \psi^m(\bar z)\bigg)\bigg(i\DP X^n(z) + \frac{\al}{2}(k\cdot \psi(z))\psi^n(z)\bigg)e^{ik\cdot X(z,\bar z)}\\& +(D\epsilon^t[\bar U^I]D)_{mn}\frac{2}{\al}\bigg(i\bar\DP \bar X^m(\bar z) + \frac{\al}{2}(k\cdot D \bar \psi(\bar z))\bar \psi^m(\bar z)\bigg)\bigg(i\DP X^n(z) + \frac{\al}{2}(k\cdot D \psi(z))\psi^n(z)\bigg)e^{ik\cdot D X(z,\bar z)}\Bigg]\\&=
\frac{2}{\al(U^I {-} \bar U^I)}\bigg\{:\bigg(i\DP \tilde Z^I + \frac{\al}{2}(k\psi)\tilde\Psi^I\bigg)e^{ikX}(z)\bigg(i\bar\DP \bar{\tilde Z}^I + \frac{\al}{2}(k\bar \psi)\bar{\tilde \Psi}^I\bigg)e^{ik\bar X}(\bar z):\\&\hspace{2.7cm} +:\left(\{L,k,z\}\leftrightarrow\{R,k\R,\bar z\}\right)
%\bigg(i\bar\DP \bar{\tilde Z}^I + \frac{\al}{2}(k {\R}\bar \psi)\bar{\tilde \Psi}^I\bigg)e^{ik {\R}\bar X}\bigg(i\DP\tilde Z^I + \frac{\al}{2}(k {\R}\psi)\tilde\Psi^I\bigg)e^{ik {\R}X}(z)
:\bigg\}
%\\&\Big[\text{R-symmetry charges}\,\,(Q,\bar Q) = (1,1)\Big]
\end{aligned}
\end{aligned}
\end{equation}
where the building block \eqref{eq:BulBlok} and the polarisation tensor $E_{mn}\left[T^I(U^I)\right]$ coming from the variation of the sigma model \eqref{eq:sigmamodel} with respect to $T^I (U^I)$ have been used. As for the  $S_2$ sphere and $D_2$ disk, the vertex operator for the $T_2$ fields here is given by the following liner combination
\begin{equation}
\mathcal W^{\otimes}_{{T_2^I}{(q,\bar q)}}(E,z,\bar z, k) = \frac{i}{2}\left(\mathcal W^{\otimes}_{{T^I}{(q,\bar q)}}(E,z,\bar z, k) - \mathcal W^{\otimes}_{{\bar T^I}{(q,\bar q)}}(E,z,\bar z, k)\right)\,.
\label{eq:VT22}
\end{equation}
%%%%%%%%%%%%%%%%%%%%%%%%%%%%%%%%%%
\subsubsection{Two-point functions for the $Z,\Psi$ system}
It is known that, when string scattering amplitudes on surfaces with boundaries as the $D_2$ disk are considered, the two-point functions need to be modified owing to the $\bf{Z}_2$ involution $\mathfrak{I}_{D_2}(z)= \bar z$ that acts on the complex plane $\mathbb C$ ($S_2$) giving the upper half plane $\mathcal H_+ (D_2)$. This implies an interaction between the left- and right-moving parts of closed string fields \cite{BLT,Garousi:1996ad}, giving
\begin{equation}
\begin{aligned}
\langle\DP X^M(z_1)\bar\DP\bar X^N(\bar z_2)\rangle_{_{D_2}}&= - \frac{\alpha'}{2}\frac{\R^{MN}}{(z_1 - \bar z_2)^2}\\
\big<\psi^M(z_1)\bar\psi^N(\bar z_2)\big>_{_{D_2}}&=\frac{\R^{MN}}{(z_1- \bar z_2)}\\
\big<\phi(z_1)\bar\phi(\bar z_2)\big>_{_{D_2}}&=- \ln(z_1 - \bar z_2)
\end{aligned}
\label{eq:diskcorre_stan}
\end{equation}
where the non trivial interaction between holomorphic and antiholomorphic part can be obtained using the doubling trick.\footnote{Extend the fields to the entire complex plane.} In this way each right-moving field of the closed string vertex operator is replaced by 
\begin{equation}
\bar X^M(\bar z)\rightarrow \R^M_{\,\,\,N}X^N(\bar z)\,,\quad \bar\psi^M(\bar z)\rightarrow \R^M_{\,\,\,N}\psi^N(\bar z)\,,\quad \bar\phi(\bar z)\,\rightarrow\phi(\bar z)\,.
\label{doublig_trik}
\end{equation}
With the help of the standard two-point functions on the sphere \cite{BLT,Garousi:1996ad}, obtaining \eqref{eq:diskcorre_stan} on the disk is straightforward. The presence of both closed and open strings, especially for the latter, involves the presence of $D$-brane on which one can impose Neumann or Dirichlet (or mixed ones) boundary conditions on the directions parallel and transverse to the brane, respectively. The reflection matrix $\R$ allows imposing the above conditions.
\begin{equation}
\R^{MN} = \begin{cases} g^{ab}\,,\quad a,b= 0,\dots ,p\,\,\,&(NN)\\-g^{ij}\, ,\quad i,j= p + 1,\dots, 9\,\,\,&(DD) \end{cases}\quad M=0,\dots,9
\label{eq:Dmatrix}
\end{equation}  
Neumann boundary conditions are imposed on coordinates $X^a(\psi^a)$ for $0\le a\le p$ and Dirichlet boundary conditions on coordinates $X^i(\psi^i)$ for $p+1\le i \le 9$. In this paper we don't consider the mixed boundary conditions case, i.e ND (DN).\footnote{That enters when twisted fields are considered.}

When orbifolds compactifications are considered, one has to specify both vertex operators (sections \ref{subsec1} and \ref{subsec2}) and two-point functions associated to the internal part that locally, aside singular points, looks like a $\T^6$ torus. In our case the internal six-torus factorizes as $\T^6=\bigotimes_{I=1}^3\T^2_I$ and the matrix $\mathcal{R}$ \eqref{eq:Dmatrix} in the internal directions takes the form
\begin{equation}
\R^{mn} = \begin{cases} (g^I)^{mn}\,\,\,&(NN)\\-(g^I)^{mn}\,\,\,&(DD)\,\,.\end{cases}
\label{eq:matrixRint}
\end{equation}
In \eqref{eq:matrixRint} the metric and the associated boundary conditions refer to the $\T^2$-torus that the specific $D$-brane wraps (as discussed at the beginning of Section \ref{sec3}), with the index $\{I, m, n\}$ the same as \eqref{eq:torusmetric}. In the case of $D9$-branes that are characterised only by Neumann boundary conditions as they wrap the full $\T^6$, eq. \eqref{eq:matrixRint}  reads
\begin{equation}
\R^{mn}_{D9} = \bigotimes_{I=1}^3 (g^I)^{mn}\,\,\,(NN)\,,\quad[m,n]\in\{[4,5];[6,7];[8,9]\}\,.
\label{RM9}
\end{equation}
Taking for instance the set of $D5_1$ out of the three sets of $D5$-branes, one has 
 \begin{equation}
 \R^{mn}_{D5_1} =\begin{cases} (g^1)^{mn}\,\,\,&(NN)\\-(g^1)^{mn}\,\,\,&(DD)\,\end{cases}\quad [m,n] \in [4,5]
\label{RM5}
\end{equation}
where the Neumann boundary conditions refer to the $\T^2_1$ torus that $D5_1$ branes wrap, while the Dirichlet boundary conditions refer to the $\T^2_2\otimes \T^2_3$ torus transverse to the $D5_1$ branes.\footnote{Dirichlet boundary conditions refer to the internal directions where the $D5_1$ branes are fixed.}
%%%%%%%%%%
%%%%%%%%%%
Concerning the correlators on the $S_2$ sphere, one has for the internal bosonic fields $Z$ \eqref{eq:InternalB} \cite{Lust:2004cx}
\begin{equation}
\begin{aligned}
&\langle\DP Z^I(z_1)\DP \tilde Z^J(z_2)\rangle_{_{S_2}} = %\frac{T^I_{_2}}{2U^I_{_2}}\langle\DP(X^{2I +2} + \bar U^I X^{2I +3})(z_1)\DP(X^{2I +2} + U^I X^{2I +3})(z_2)\rangle\\
\sqrt{\frac{T^I_{_2}}{2U^I_{_2}}}\sqrt{\frac{T^J_{_2}}{2U_{_2}^J}}\bigg[\langle\DP X^{2I +2}(z_1)\DP X^{2J +2}(z_2)\rangle + \bar U^I\langle\DP X^{2I +3}(z_1)\DP X^{2J +2}(z_2)\rangle\\&\hspace{3cm}+ U^J\langle\DP X^{2I +2}(z_1)\DP X^{2J +3}(z_2)\rangle + \bar U^IU^J\langle\DP X^{2I +3}(z_1)\DP X^{2J +3}(z_2)\rangle\bigg]\\
&= -\sqrt{\frac{T^I_{_2}}{2U^I_{_2}}}\sqrt{\frac{T^J_{_2}}{2U_{_2}^J}}\frac{\al}{2(z_1 - z_2)^2}\bigg[ g^{[2I+2][2J+2]} +\bar U^Ig^{[2I+3][2J+2]} + U^Jg^{[2I +2][2J+3]} + \bar U^IU^Jg^{[2I+3][2J+3]}\bigg]\\
%&= -\frac{T^I_{_2}}{2U^I_{_2}}\frac{\al\delta_{I,J}}{2(z_1 - z_2)^2}\bigg[\frac{\abs{U^I}^2}{T^I_{_2}U^I_{_2}} - \frac{\bar U^IU^I_{_1}}{T^I_{_2}U^I_{_2}} - \frac{U^IU^I_{_1}}{T^I_{_2}U^I_{_2}} + \frac{\abs{U^I}^2}{T^I_{_2}U^I_{_2}}\bigg]\\
%&= -\frac{T^I_{_2}}{2U^I_{_2}}\frac{\al}{2(z_1 - z_2)^2}\bigg[2\frac{(U_{_1}^I)^2 + (U^I_{_2})^2 - U^I_{_1}U^I_{_1}}{T^I_{_2}U^I_{_2}}\bigg]\\
&= -\frac{\al\delta_{I,J}}{2(z_1 - z_2)^2}
\end{aligned}
\label{eq:ZtildeZ}
\end{equation}
%%%%%%%%%%%%
%%%%%%%%%%%%
\begin{equation}
\begin{aligned}
&\langle\DP Z^I(z_1)\DP Z^J(z_2)\rangle_{_{S_2}}%= \frac{T^I_{_2}}{2U^I_{_2}}\big<\DP(X^{2I +2} + \bar U^I X^{2I +3})(z_1)\DP(X^{2I +2} + \bar U^I X^{2I +3})(z_2)\big>_{\mathbb C}\\
= \sqrt{\frac{T^I_{_2}}{2U^I_{_2}}}\sqrt{\frac{T^J_{_2}}{2U_{_2}^J}}\bigg[\langle\DP X^{2I +2}(z_1)\DP X^{2J +2}(z_2)\rangle + \bar U^I\langle\DP X^{2I +3}(z_1)\DP X^{2J +2}(z_2)\rangle\\&\hspace{3cm}+ \bar U^J\langle\DP X^{2I +2}(z_1)\DP X^{2J +3}(z_2)\rangle + \bar U^I\bar U^J\langle\DP X^{2I +3}(z_1)\DP X^{2J +3}(z_2)\rangle\bigg]\\
&= -\sqrt{\frac{T^I_{_2}}{2U^I_{_2}}}\sqrt{\frac{T^J_{_2}}{2U_{_2}^J}}\frac{\al}{2(z_1 - z_2)^2}\bigg[ g^{[2I+2][2J+2]} +\bar U^Ig^{[2I+3][2J+2]} + \bar U^Jg^{[2I +2][2J+3]} + \bar U^I\bar U^Jg^{[2I+3][2J+3]}\bigg]\\
%&= -\frac{T^I_{_2}}{2U^I_{_2}}\frac{\al\delta_{I,J}}{2(z_1 - z_2)^2}\bigg[\frac{\abs{U^I}^2}{T^I_{_2}U^I_{_2}} - \frac{\bar U^IU^I_{_1}}{T^I_{_2}U^I_{_2}} - \frac{\bar U^IU^I_{_1}}{T^I_{_2}U^I_{_2}} + \frac{(\bar U^I)^2}{T^I_{_2}U^I_{_2}}\bigg]\\
%&= -\frac{T^I_{_2}}{2U^I_{_2}}\frac{\al}{2(z_1 - z_2)^2}\bigg[\frac{(U_{_1}^I)^2 + (U^I_{_2})^2 -2 (U^I_{_1})^2 +2iU^I_{_2}U^I_{_1} +(U^I_{_1})^2 - (U^I_{_2})^2 - 2iU^I_{_1}U^I_{_2}}{T^I_{_2}U^I_{_2}}\bigg]\\
&=0
\label{eq:ZZ}
\end{aligned}
\end{equation}
where in \eqref{eq:ZtildeZ} the $\delta_{I,J}$ is due to the vanishing of the off diagonal block matrix $g^{mn}$ when $I\ne J$, while \eqref{eq:ZZ} vanishes also in the case $I=J$.\,\footnote{Following our definitions the vanishing correlators that doesn't contain one $Z$ and one $\tilde Z$ occur when the $U(1)$ charge is not conserved. The same will holds for the correlators of the fermionic $\Psi$ fields.}
%%%%%%%%%%%%%
The two-point functions for the compactified bosons $Z$ on the sphere are 
\begin{equation}
\boxed{
\begin{aligned}
&\langle\DP Z^I(z_1)\DP Z^J(z_2)\rangle_{_{S_2}}=\langle\DP \tilde Z^I(z_1)\DP \tilde Z^J(z_2)\rangle_{_{S_2}}= \langle\bar\DP \bar Z^I(\bar z_1)\bar\DP \bar Z^J(\bar z_2)\rangle_{_{S_2}}=\langle\bar\DP \bar{\tilde Z}^I(\bar z_1)\bar\DP \bar{\tilde Z}^J(\bar z_2)\rangle_{_{S_2}} =0
\\&\langle\DP Z^I(z_1)\DP \tilde Z^J(z_2)\rangle_{_{S_2}}= -\frac{\al\delta_{I,J}}{2(z_1 -z_2)^2}\,\,,\quad\langle\bar\DP \bar Z^I(\bar z_1)\bar\DP \bar{\tilde Z}^J(\bar z_2)\rangle_{_{S_2}} =  -\frac{\al\delta_{I,J}}{2(\bar z_1 - \bar z_2)^2}%\\& \langle\DP Z^I( z_1)\bar\DP \bar{\tilde Z}^I(\bar z_2)\rangle = 0\,\,,\hspace{2.6cm} \langle\bar\DP \bar Z^I(\bar z_1)\DP \tilde Z^I(z_2)\rangle=0\,\,,
\end{aligned}}
\label{eq:boscomSfera}
\end{equation}
\\
%%%%%%%%%%
%%%%%%%%%%
while for the compactified fermions $\Psi$ \eqref{eq:InternalF} one has 
\begin{equation}
\boxed{
\begin{aligned}
&\langle\Psi^I(z_1)\Psi^J(z_2)\rangle_{_{S_2}} = \langle\tilde\Psi^I(z_1)\tilde\Psi^J(z_2)\rangle_{_{S_2}}=\langle\bar\Psi^I(\bar z_1)\bar\Psi^J(\bar z_2)\rangle_{_{S_2}}=\langle\bar{\tilde\Psi}^I(\bar z_1)\bar{\tilde\Psi}^J(\bar z_2)\rangle_{_{S_2}} = 0\\
&\langle\Psi^I(z_1)\tilde\Psi^J(z_2)\rangle_{_{S_2}} = \frac{\delta_{I,J}}{(z_1 - z_2)}\,\,,\quad\langle\bar\Psi^I(\bar z_1)\bar{\tilde\Psi}^J(\bar z_2)\rangle_{_{S_2}}= \frac{\delta_{I,J}}{(\bar z_1 - \bar z_2)}\\
%&\big<\Psi^I(z_1)\bar{\tilde\Psi}^I(\bar z_2)\big>_{\mathbb C} = 0\,\,,\hspace{2cm}\big<\bar{\tilde\Psi}^I(\bar z_1)\tilde\Psi^I(z_2)\big>_{\mathbb C} = 0
\end{aligned}}\label{eq:fermcomSfera}
\end{equation}
\\
where in the fermionic analog of \eqref{eq:ZtildeZ}, \eqref{eq:ZZ} the standard fermonic correlator on the sphere is used \cite{BLT,Garousi:1996ad}. To generalize the result to the $D_2$ disk, one has to calculate the correlators on the upper-half plane $\mathcal H^+$. First of all using the doubling trick one is able to replace the right-moving part according to the formulae
\begin{equation}
\bar Z^I(\bar z)\rightarrow \R^IZ^I(\bar z)\,\,,\quad\bar{\tilde Z}^I(\bar z)\rightarrow \R^I\tilde Z^I(\bar z)\,\,,\quad \bar\Psi^I(\bar z)\rightarrow \R^I\Psi^I(\bar z)\,\,,\quad\bar{\tilde\Psi}^I(\bar z)\rightarrow \R^I\tilde\Psi^I(\bar z)
\end{equation}
where the reflection matrix $\R$ expressed in a non compact form is $\R^I = (\R^I)^m_{\,\,\,\,n}$ with $\{I,m,n\}$ as \eqref{eq:torusmetric}. More specifically, one can use $(\R^I)^{mn} =\mathscr R_{a}^I (g^I)^{mn}$ with $\mathscr R_{a}^I$ equal to $+1(-1)$ for $NN(DD)$-directions, $(g^I)^{mn}$ the internal metric \eqref{eq:torusmetric} and $a$ labels the type of $D_P$-brane. Examples of correlators for the boson fields $Z$ on $\mathcal H^+$ are 
\begin{equation}
\begin{aligned}
&\langle\DP Z^I(z_1)\bar\DP\bar{\tilde Z}^J(\bar z_2)\rangle_{_{D_2}} = \langle\DP Z^I(z_1)\bar\DP\tilde Z^J(\bar z_2)\rangle_{_{S_2}} \R^J\\
&= \sqrt{\frac{T^I_{_2}}{2U^I_{_2}}}\sqrt{\frac{T^J_{_2}}{2U_{_2}^J}}\bigg[(\R^J)^{[2J{+}2]}_{\,\,[2J{+}2]}\langle\DP X^{2I {+}2}(z_1)\bar\DP X^{2J {+}2}(\bar z_2)\rangle + \bar U^I(\R^J)^{[2J{+}2]}_{\,\,[2J{+}2]}\langle\DP X^{2I {+}3}(z_1)\bar\DP X^{2J{ +}2}(\bar z_2)\rangle\\&\,\,\,+ U^J(\R^J)^{[2J{+}3]}_{\,\,[2J{+}3]}\langle\DP X^{2I {+}2}(z_1)\bar\DP X^{2I {+}3}(\bar z_2)\rangle + \bar U^IU^J(\R^J)^{[2J{+}3]}_{\,\,[2J{+}3]}\langle\DP X^{2I {+}3}(z_1)\bar\DP X^{2J {+}3}(\bar z_2)\rangle\bigg]\\
%&=-\frac{T^I_{_2}}{2U^I_{_2}}\frac{\al}{2(z_1 - \bar z_2)^2}\bigg[ (D^I)^{[2I+2]}_{\,\,[2I+2]}g^{[2I+2][2I+2]} +\bar U^I(D^I)^{[2I+3]}_{\,\,[2I+3]}g^{[2I+3][2I+2]}\\&\,\,\, + U^I(D^I)^{[2I+2]}_{\,\,[2I+2]}g^{[2I +2][2I+3]} + \abs{U^I}^2(D^I)^{[2I+3]}_{\,\,[2I+3]}g^{[2I+3][2I+3]}\bigg]\\
&=-\sqrt{\frac{T^I_{_2}}{2U^I_{_2}}}\sqrt{\frac{T^J_{_2}}{2U_{_2}^J}}\frac{\al}{2(z_1 - \bar z_2)^2}\bigg[(\R^J)^{[2J{+}2][2I{+}2]} +\bar U^I(\R^J)^{[2J{+}2][2I{+}3]} + U^J(\R^J)^{[2J{+}3][2I{+}2]} \\&\,\,\,+ \bar U^IU^J(\R^J)^{[2J{+}3][2I{+}3]}\bigg]\\
%&=-\frac{T^I_{_2}}{2U^I_{_2}}\frac{\al}{2(z_1 - \bar z_2)^2}\bigg[\mathcal D^Ig^{[2I+2][2I+2]} +\bar U^I\mathcal D^Ig^{[2I+3][2I+2]}\\&\,\,\, + U^I\mathcal D^Ig^{[2I+2][2I+3]} + \abs{U^I}^2\mathcal D^Ig^{[2I+3][2I+3]}\bigg]
%&= -\frac{T^I_{_2}}{2U^I_{_2}}\frac{\al\mathscr R^J_a\,\delta_{I,J}}{2(z_1 - \bar z_2)^2}\bigg[2\frac{(U_{_1}^I)^2 + (U^I_{_2})^2 - U^I_{_1}U^I_{_1}}{T^I_{_2}U^I_{_2}}\bigg]\\
&= -\frac{\al\mathscr R^J_a\,\delta_{I,J}}{2(z_1 - \bar z_2)^2}
\end{aligned}
\label{corre1ZD2}
\end{equation}

%%%%%%%%%%%
%%%%%%%%%%%
\begin{equation}
\begin{aligned}
&\langle\DP Z^I(z_1)\bar\DP\bar Z^J(\bar z_2)\rangle_{_{D_2}}= \langle\DP Z^I(z_1)\bar\DP Z^J(\bar z_2)\rangle_{_{S_2}} \R^J\\&= \sqrt{\frac{T^I_{_2}}{2U^I_{_2}}}\sqrt{\frac{T^J_{_2}}{2U_{_2}^J}}\bigg[(\R^J)^{[2J+2]}_{\,\,[2J+2]}\langle\DP X^{2I +2}(z_1)\bar\DP X^{2J +2}(\bar z_2)\rangle + \bar U^I(\R^J)^{[2J+2]}_{\,\,[2J+2]}\langle\DP X^{2I +3}(z_1)\bar\DP X^{2J +2}(\bar z_2)\rangle\\&\,\,\,+ \bar U^J(\R^J)^{[2J+3]}_{\,\,[2J+3]}\langle\DP X^{2I +2}(z_1)\bar\DP X^{2J +3}(\bar z_2)\rangle + \bar U^I\bar U^J(\R^J)^{[2J+3]}_{\,\,[2J+3]}\langle\DP X^{2I +3}(z_1)\bar\DP X^{2J +3}(\bar z_2)\rangle\bigg]\\
%&= -\frac{T^I_{_2}}{2U^I_{_2}}\frac{\al}{2(z_1 - \bar z_2)^2}\bigg[ (D^I)^{[2I+2]}_{\,\,[2I+2]}g^{[2I+2][2I+2]} +\bar U^I(D^I)^{[2I+3]}_{\,\,[2I+3]}g^{[2I+3][2I+2]}\\&\,\,\, + \bar U^I(D^I)^{[2I+2]}_{\,\,[2I+2]}g^{[2I +2][2I+3]} + (\bar U^I)^2(D^I)^{[2I+3]}_{\,\,[2I+3]}g^{[2I+3][2I+3]}\bigg]\\
& =-\sqrt{\frac{T^I_{_2}}{2U^I_{_2}}}\sqrt{\frac{T^J_{_2}}{2U_{_2}^J}}\frac{\al}{2(z_1 - \bar z_2)^2}\bigg[(\R^J)^{[2J+2][2I+2]} +\bar U^I(\R^J)^{[2J+2][2I+3]} + \bar U^J(\R^J)^{[2J+3][2I+2]}\\&\,\,\, + \bar U^I\bar U^J(\R^J)^{[2J+3][2I+3]}\bigg]\\
%&= -\frac{T^I_{_2}}{2U^I_{_2}}\frac{\al}{2(z_1 - \bar z_2)^2}\bigg[\mathcal D^Ig^{[2I+2][2I+2]} +\bar U^I\mathcal D^Ig^{[2I+3][2I+2]}\\&\,\,\, + \bar U^I\mathcal D^Ig^{[2I+2][2I+3]} +( \bar U^I)^2\mathcal D^Ig^{[2I+3][2I+3]}\bigg]\\
%&= -\frac{T^I_{_2}}{2U^I_{_2}}\frac{\al\mathscr R_a^J\delta_{I,J}}{2(z_1 - \bar z_2)^2}\bigg[\frac{(U_{_1}^I)^2 + (U^I_{_2})^2 -2 (U^I_{_1})^2 +2iU^I_{_2}U^I_{_1} +(U^I_{_1})^2 - (U^I_{_2})^2 - 2iU^I_{_1}U^I_{_2}}{T^I_{_2}U^I_{_2}}\bigg]\\
&= 0
\end{aligned}
\label{corre2zD2}
\end{equation}
where $(\mathcal R^J)^m_{\,\,\,n} = \mathscr R_{a}^J\,(\delta^J)^{m}_{\,\,\,n}$ and the $\delta_{I,J}$ has been used as in the sphere case. Thus, besides \eqref{eq:boscomSfera} and \eqref{eq:fermcomSfera}, one has  
\begin{equation}
\boxed{
\begin{aligned}
&\langle\DP Z^I(z_1)\bar\DP\bar Z^J(\bar z_2)\rangle_{_{D_2}}= 0\,\,\quad\langle\DP \tilde Z^I(z_1)\bar\DP\bar{\tilde Z}^J(\bar z_2)\rangle_{_{D_2}} = 0\,\,,\\
&\langle\DP Z^I(z_1)\bar\DP\bar{\tilde Z}^J(\bar z_2)\rangle_{_{D_2}}=-\frac{\al\mathscr R_a^J\delta_{I,J}}{2(z_1 - \bar z_2)^2} \,\,,\quad \langle\DP\tilde Z^I(z_1)\bar\DP\bar Z^J(\bar z_2)\rangle_{_{D_2}}= -\frac{\al\mathscr R_a^J\delta_{I,J}}{2(z_1 - \bar z_2)^2}\\
\end{aligned}}\label{eq:boscomUp}
\end{equation} 
for the compactified bosons $Z$ and 
\\
%%%%%%%%%%%%%%%%
\begin{equation}
\boxed{
\begin{aligned}
&\langle\Psi^I(z_1)\bar\Psi^J(\bar z_2)\rangle_{_{D_2}} = 0\,\,,\quad\langle\tilde\Psi^I(z_1)\bar{\tilde\Psi}^J(\bar z_2)\rangle_{_{D_2}} = 0\,\,,\\
&\langle\Psi^I(z_1)\bar{\tilde\Psi}^J(\bar z_2)\rangle_{_{D_2}} = \frac{\mathscr R_a^J\delta_{I,J}}{(z_1 - \bar z_2)}\,\,,\quad\langle\tilde\Psi^I(z_1)\bar\Psi^J(\bar z_2)\rangle_{_{D_2}} = \frac{\mathscr R_a^J\delta_{I,J}}{(z_1 - \bar z_2)}
\end{aligned}}\label{eq:fermcomUp}
\end{equation}
\\
%%%%%%%%%
for the compactified fermions $\Psi$.\footnote{
Since we are not taking fluxes the matrix $\mathcal R$ has only block diagonal components and one is in the simplified case of \cite{Lust:2004cx,Bertolini:2005qh}, in which one has to send the fluxes \textit f$^I$ to zero.}
At tree level the presence of $\Omega_P$-planes in orientifold models suggests that one has to consider the internal bosonic and fermionic correlators on the real projective plane $RP_2$. Taking the action of the $\bm Z_2$ involution to be $\mathfrak{I}_{RP_2}(z) = -1/\bar z$ on $S_2$, one obtains the $RP_2$, which is a disk $D_2$ with antipodal points on the boundary identified\,\cite{BLT,Garousi:2006zh,Burgess:1986ah}. 
The basic two-point functions on the $RP_2$, employing the method of images and the doubling trick, are \cite{BLT}
\begin{equation}
\begin{aligned}
&\langle X^M(z)\bar X^N(\bar w)\rangle_{_{RP_2}}= - \frac{\al}{2}\R^{MN}\ln(1 + z\bar w)
\\&\langle\psi(z)^M\bar\psi^N(\bar w)\rangle_{_{RP_2}} = \frac{\R^{MN}}{(1 + z\bar w)}
\\
&\langle\phi(z)\bar\phi(\bar w)\rangle_{_{RP_2}} = -\ln(1 + z\bar w)
\label{eq:RP2coor}
\end{aligned}
\end{equation}
with $\R^{MN}$ the reflection matrix of eq. \eqref{eq:Dmatrix}. The two-point functions for the compactified bosons $Z$ and fermions $\Psi$ can be computed using the building block \eqref{eq:RP2coor}, the correlators on the sphere $S_2$ \eqref{eq:boscomSfera} and \eqref{eq:fermcomSfera}, respectively. They differ from those on  $\mathcal H^+$ only by the $z\bar w$ dependence and for the compactified bosons and for the compactified fermions read
%%%%%%%%%%%%%%%%%%%%
\\
\begin{equation}
\boxed{
\begin{aligned}
&\langle\DP Z^I(z_1)\bar\DP\bar Z^J(\bar z_2)\rangle_{_{RP_2}}= 0\,\,\quad\langle\DP \tilde Z^I(z_1)\bar\DP\bar{\tilde Z}^J(\bar z_2)\rangle_{_{RP_2}} = 0\,\,,\\
&\langle\DP Z^I(z_1)\bar\DP\bar{\tilde Z}^J(\bar z_2)\rangle_{_{RP_2}}=-\frac{\al\mathscr R_{\alpha}^J\delta_{I,J}}{2(1 +z_1\bar z_2)^2} \,\,,\quad \langle\DP\tilde Z^I(z_1)\bar\DP\bar Z^J(\bar z_2)\rangle_{_{RP_2}}= -\frac{\al\mathscr R_{\alpha}^J\delta_{I,J}}{2(1 + z_1\bar z_2)^2}\\
\end{aligned}}\label{eq:boscomRp}
\end{equation} 

\begin{equation}
\boxed{
\begin{aligned}
&\langle\Psi^I(z_1)\bar\Psi^J(\bar z_2)\rangle_{_{RP_2}} = 0\,\,,\quad\langle\tilde\Psi^I(z_1)\bar{\tilde\Psi}^J(\bar z_2)\rangle_{_{RP_2}} = 0\,\,,\\
&\langle\Psi^I(z_1)\bar{\tilde\Psi}^J(\bar z_2)\rangle_{_{RP_2}} = \frac{\mathscr R_{\alpha}^J\delta_{I,J}}{(1 +z_1\bar z_2)}\,\,,\quad\langle\tilde\Psi^I(z_1)\bar\Psi^J\bar z_2)\rangle_{_{RP_2}}= \frac{\mathscr R_{\alpha}^J\delta_{I,J}}{(1 +z_1\bar z_2)}
\end{aligned}}\label{eq:fermcomRp}
\end{equation}
%%%%%%%%%%%%%%%
where $\mathscr R_{\alpha}^I$ is $+1(-1)$ for $NN(DD)$-directions and $\alpha$ labels the type of $\Omega_P$-planes one is considering. 
%%%%%%%%%%%%%%%%%%%%%%%%%%%%%%%%%%%
%%%%%%%%%%%%%%%%%%%%%%%%%%%%%%%%%%%
%%%%%%%%%%%%%%%%%%%%%%%%%%%%%%%%%%%
%%%%%%%%%%%%%%%%%%%%%%%%%%%%%%%%%%%
%%%%%%%%%%%%%%%%%%%%%%%%%%%%%%%%%%%
%%%%%%%%%%%%%%%%%%%%%%%%%%%%%%%%%%%
%%%%%%%%%%%%%%%%%%%%%%%%%%%%%%%%%%%
%%%%%%%%%%%%%%%%%%%%%%%%%%%%%%%%%%%
%%%%%%%%%%%%%%%%%%%%%%%%%%%%%%%%%%%
%%%%%%%%%%%%%%%%%%%%%%%%%%%%%%%%%%%
%%%%%%%%%%%%%%%%%%%%%%%%%%%%%%%%%%%

\section{Scattering Amplitudes of the closed untwisted moduli \\ in Type IIB $T^6/\mathbb{Z}_2\times\mathbb{Z}_2$ orientifold}\label{scattering_sec}\label{sec4}

In this section string scattering amplitudes with two untwisted closed string moduli on the $D_2$ disk \cite{Lust:2004cx} are reviewed and extended to the real projective plane $RP_2$ worldsheet surface, as needed at tree-level when unoriented string models are considered. String scattering amplitudes on the $D_2$ disk with fundamental region the upper-half plane $\mathcal H_+$ that we will compute are
\\
\begin{equation}
\small{
\begin{aligned}
{\cal A}_{a}\left(T^{I},\bar{T}^{J}\right) + {\cal A}_{a}\left(T^{I},{T}^{J}\right)&= g_c^2C_{_{D_2}}\int_{\Hp}\frac{d^2z_1d^2z_2}{V_{CKG}}\Big(\langle{\cal W}_{T^I}(z_1, \bar z_1){\cal W}_{\bar T^J}(z_2, \bar z_2)\rangle +\langle{\cal W}_{T^I}(z_1, \bar z_1){\cal W}_{T^J}(z_2, \bar z_2)\rangle\Big)\\
%%%%%
{\cal A}_{a}\left(U^{I},\bar{U}^{J}\right) + {\cal A}_{a}\left(U^{I},{U}^{J}\right)&=g_c^2C_{_{D_2}}\int_{\Hp}\frac{d^2z_1d^2z_2}{V_{CKG}}\Big(\langle{\cal W}_{U^I}(z_1, \bar z_1){\cal W}_{\bar U^J}(z_2, \bar z_2)\rangle + \langle{\cal W}_{U^I}(z_1, \bar z_1){\cal W}_{U^J}(z_2, \bar z_2)\rangle\Big)\\
%%%%%
{\cal A}_{a}\left(T^{I},\bar{U}^{J}\right) + {\cal A}_{a}\left(T^{I},{U}^{J}\right)&=g_c^2C_{_{D_2}}\int_{\Hp}\frac{d^2z_1d^2z_2}{V_{CKG}}\Big(\langle{\cal W}_{T^I}(z_1, \bar z_1){\cal W}_{\bar U^J}(z_2, \bar z_2)\rangle + \langle{\cal W}_{T^I}(z_1, \bar z_1){\cal W}_{U^J}(z_2, \bar z_2)\rangle \Big)\\&\quad+\,(T\leftrightarrow U)\\
\end{aligned}\label{eq:2untwmoduli_disk}}
\end{equation}
where the different kinds of $D$-branes are labelled by $a \in\{9,5^I\}$. On the real projective plane $RP_2$ the string scattering amplitudes involving the same states taking the unit disk $\abs{z}\leq 1$ as fundamental region, 
reads
\begin{equation}
\small{
\begin{aligned}
{\cal A}_{\alpha}\left(T^{I},\bar{T}^{J}\right) + {\cal A}_{\alpha}\left(T^{I},{T}^{J}\right)&= g_c^2C_{_{RP_2}}\int_{\abs{z}\leq 1}\frac{d^2z_1d^2z_2}{V_{CKG}}\Big(\langle{\cal W}^{\otimes}_{T^I}(z_1, \bar z_1){\cal W}^{\otimes}_{\bar T^J}(z_2, \bar z_2)\rangle +\langle{\cal W}^{\otimes}_{T^I}(z_1, \bar z_1){\cal W}^{\otimes}_{T^J}(z_2, \bar z_2)\rangle\Big)\\
%%%%%
{\cal A}_{\alpha}\left(U^{I},\bar{U}^{J}\right) + {\cal A}_{\alpha}\left(U^{I},{U}^{J}\right)&=g_c^2C_{_{RP_2}}\int_{\abs{z}\leq 1}\frac{d^2z_1d^2z_2}{V_{CKG}}\Big(\langle{\cal W}^{\otimes}_{U^I}(z_1, \bar z_1){\cal W}^{\otimes}_{\bar U^J}(z_2, \bar z_2)\rangle + \langle{\cal W}^{\otimes}_{U^I}(z_1, \bar z_1){\cal W}^{\otimes}_{U^J}(z_2, \bar z_2)\rangle\Big)\\
%%%%%
{\cal A}_{\alpha}\left(T^{I},\bar{U}^{J}\right) + {\cal A}_{\alpha}\left(T^{I},{U}^{J}\right)&=g_c^2C_{_{RP_2}}\int_{\abs{z}\leq 1}\frac{d^2z_1d^2z_2}{V_{CKG}}\Big(\langle{\cal W}^{\otimes}_{T^I}(z_1, \bar z_1){\cal W}^{\otimes}_{\bar U^J}(z_2, \bar z_2)\rangle +\langle{\cal W}^{\otimes}_{T^I}(z_1, \bar z_1){\cal W}^{\otimes}_{U^J}(z_2, \bar z_2)\rangle \Big)\\&\quad+\,(T\leftrightarrow U) \\
\end{aligned}\label{eq:AMPLRP2_New}}
\end{equation}
with $\alpha\in\{9,5^I\}$ that label the different kinds of $\Omega$-planes. In Section \ref{Toolkit} the derivation of the compactified vertex operators and the structure that enters in the specific scattering amplitudes were provided. Moreover the doubling trick needed to make the correlation among the left and the right field, for convenience, will not be manifest in the definition of the vertex operators, as in Section \ref{sec2}. 
%%%%%%%%%%%%%%%%%%%%%%%%%%%%%%%%%%%%%%%%%%%%%%%%%%%%%%%%%%%%%%%%%%%%%%%%%%%%%%%%%%%%%%%%%%%%%%%%%%%%%%%%%%%%%%%%%%%%%%%%%%%%%%%%%%%%%%%%%%%%%%%%%%%%%%%%%%%%%%%%%%%%%%%%%%%%%%%%%%%%%%%%%%%%%%%%%%%%%%%%%%%%%%%%%%%%%%%%%%%%%%%%%%%%%%%%%%%%%%%%%%%%%%%%%%%%%%%%%%%%%%%%%%%%%%%%%%%%%%%%%%%%%%%%%%%%%%%%%%%%%%%%%%%%%%%%%%%%%%%%%%%%%%%%%%%%%
\subsection{$\mathcal A^{D_2}_{a}\left({T^{I}},\bar{T}^{J} \right)$ and $\mathcal A^{D_2}_{a}\left(T^{I},{T}^{J}\right)$}
Let start with the first set of string scattering amplitudes in \eqref{eq:2untwmoduli_disk} that involves two untwisted K\"ahler moduli $T^I$.
\subsubsection*{$\boxed{{\cal A}_{a}\left(T^{I},\bar{T}^{J}\right)}$}
The amplitude which mixes $T$ and $\bar T$ K\"ahler moduli is given by
\begin{equation}
\begin{aligned}
&{\cal A}_{a}\left(T^{I},\bar{T}^{J}\right)=\,g_c^2C_{_{D_2}}\int_{\Hp}\frac{d^2z_1d^2z_2}{V_{CKG}}\langle{\cal W}_{T^{I}(-1,-1)}(E_1,k_1,z_1, \bar z_1)\,{\cal W}_{\bar{T}^{J}(0,0)}(E_2,k_2,z_2, \bar z_2)\rangle\\
%%%%%%%%%%%%%%%%%%%
&=-\frac{8g_c^2C_{_{D_2}}}{\al(T^I -\bar T^I)(T^J - \bar T^J)}\int_{\Hp}\frac{d^2z_1d^2z_2}{V_{CKG}}\,\langle:e^{-\phi}\tilde\Psi^I e^{ik_1X}(z_1)e^{-\bar\phi}\bar\Psi^I e^{ik_1\bar X}(\bar z_1):\\
%%%%%%%%%%%
&\hspace{5cm}
\Big(i\DP Z^J {+} \frac{\al}{2}\big(k_2\psi\big)\Psi^J\Big)e^{ik_2X}(z_2)\Big(i\bar\DP \bar{\tilde Z}^J {+} \frac{\al}{2}\big(k_2\bar\psi\big)\bar{\tilde\Psi}^J\Big)e^{ik_2\bar X}(\bar z_2):\rangle\\
%%%%%%%%%%%%%%%%%%%
&= -\frac{8g_c^2C_{_{D_2}}}{\al(T^I -\bar T^I)(T^J - \bar T^J)}\int_{\Hp}\frac{d^2z_1d^2z_2}{V_{CKG}}\,\langle :e^{-\phi}(z_1)::e^{-\bar\phi}(\bar{z}_{1}): \rangle   \left({\cal M}^{(1)}+{\cal M}^{(2)}+{\cal M}^{(3)}+{\cal M}^{(4)} \right)\\
\end{aligned}\label{eq:disk_TT_-1-100}
\end{equation}
where with ${\cal M}^{(i)}$'s $(i=1,2,3,4)$ we denote all the possible different contractions. For instance, for ${\cal M}^{(1)}$ one gets
\begin{equation}
{\cal M}^{(1)}=\langle:\tilde\Psi^I e^{ik_1X}(z_1)\bar\Psi^I e^{ik_1\bar X}(\bar z_1)::i\DP Z^J e^{ik_2X}(z_2)i\bar\DP\bar{\tilde Z}^J e^{ik_2\bar X}(\bar z_2):\rangle\,\,.
\label{M1_TbarT}
\end{equation}
For the uncompactified fields the relevant two-point functions are given in \eqref{eq:correlator10D_disk}\,\footnote{Where in \eqref{eq:correlator10D_disk} one has to reinsert the reflection matrix $\R$ on the two points correlation function which involves left and right field.}. For the compactified fields the two-point functions can be found in equations \eqref{eq:boscomSfera}, \eqref{eq:fermcomSfera}, \eqref{eq:boscomUp} and \eqref{eq:fermcomUp}.
\\
%\begin{equation}
%\small{
%\boxed{
%\begin{aligned}
%\\&\langle\DP Z^I(z_1)\DP \tilde Z^J(z_2)\rangle= -\frac{\al\delta_{I,J}}{2(z_1 -z_2)^2}\,\,,\quad\langle\bar\DP \bar Z^I(\bar z_1)\bar\DP \bar{\tilde Z}^J(\bar z_2)\rangle =  -\frac{\al\delta_{I,J}}{2(\bar z_1 - \bar z_2)^2}\\\\
%&\langle\DP Z^I(z_1)\bar\DP\bar{\tilde Z}^J(\bar z_2)\rangle=-\frac{\al{\mathscr R}^J_a\delta_{I,J}}{2(z_1- \bar z_2)^2} \,\,,\quad \langle\DP\tilde Z^J(z_1)\bar\DP\bar Z^J(\bar z_2)\rangle= -\frac{\al{\mathscr R}^J_a\delta_{I,J}}{2(z_1-\bar z_2)^2}\\\\
%&\langle\Psi^I(z_1)\tilde\Psi^J(z_2)\rangle = \frac{\delta_{I,J}}{(z_1 - z_2)}\,\,,\quad\langle\bar\Psi^I(\bar z_1)\bar{\tilde\Psi}^J(\bar z_2)\rangle= \frac{\delta_{I,J}}{(\bar z_1 - \bar z_2)}\\\\
%&\langle\Psi^I(z_1)\bar{\tilde\Psi}^J(\bar z_2)\rangle = \frac{{\mathscr R}_a^J\delta_{I,J}}{(z_1-\bar z_2)}\,\,,\quad\langle\tilde\Psi^I(z_1)\bar\Psi^J(\bar z_2)\rangle = \frac{{\mathscr R}_a^J\delta_{I,J}}{(z_1-\bar z_2)}
%\end{aligned}}\label{eq:FermBoseComMIX}}
%\end{equation}
The explicit derivation of  the form of ${\cal M}^{(i)}$'s and the explanation of why some combinations vanish are given in Appendix A.6 of ref. \cite{AM:appendice}.
%%%%%%%%%%%%%%%%%%%%%%%%%%%%%%%%%%%%%%%%
Putting all together one gets
\begin{equation}
\begin{aligned}
{\cal A}_{a}\left(T^{I},\bar{T}^{J}\right) =&-\frac{8g_c^2C_{_{D_2}}}{\al(T^I {-}\bar T^I)(T^J {-} \bar T^J)}\int_{\Hp}\frac{d^2z_1d^2z_2}{V_{CKG}}\bigg(\frac{\abs{z_1 {-}\bar z_1}\abs{z_2 {-} \bar z_2}}{\abs{z_1 {-} \bar z_2}^2}\bigg)^{-\al s}\bigg(\frac{\abs{z_1 {-}z_2}^2}{\abs{z_1 {-} \bar z_2}^2}\bigg)^{-\al \frac{t}{4}}\frac{1}{(z_1 {-} \bar z_1)}\\&\bigg(\frac{\al{\mathscr R}^I_a{\mathscr R}^J_a}{2(z_1 {-} \bar z_1)(z_2 {-} \bar z_2)^2} +\frac{\al^2s\,\,{\mathscr R}^I_a{\mathscr R}^J_a}{2(z_2 {-}\bar z_2)(z_1 {-} \bar z_1)(z_2 {-}\bar z_2)} -\frac{\al^2s\,\delta_{I,J}}{2(z_2 {-}\bar z_2)(z_1 {-} z_2)(\bar z_1 {-} \bar z_2)}\bigg)\\&
\end{aligned}\label{eq:disk_TT_1234_-1-100}
\end{equation}
Using the $PSL(2,R)$ symmetry (see \eqref{eq:PSL2t} and refs. \cite{Garousi:1996ad,Hashimoto:1996kf,Hashimoto:1996bf}) in order to fix  vertex operators at the points 
\begin{equation}
z_1= i,\quad\bar z_1= -i,\quad z_2= iy,\quad\bar z_2= -iy
\label{eq:fixpoints}
\end{equation}
and inserting the c-ghost determinant of eq. \eqref{eq:cghostD2}, one obtains \cite{Lust:2004cx}
\begin{equation}
\begin{aligned}
 &\frac{8g_c^2C_{_{D_2}}}{(T^I {-}\bar T^I)(T^J {-} \bar T^J)}\int_0^1\,dy\,\left(\frac{4y}{(1 {+} y)^2}\right)^{{-}\al s}\left(\frac{(1{-}y)^2}{(1{+}y)^2}\right)^{-\al \frac{t}{4}}4(1 {-} y^2)\bigg(\frac{(-1-\al s)\,\mathscr{R}^I_a\mathscr{R}^J_a}{16y^2} -\frac{\al s\,\delta_{I,J}}{4y(1 {-} y)^2}\bigg)\,.\\&
%%%%%%%%%%%%
%&=g_s\frac{8\mathcal{D}^I\mathcal{D}^J}{\al(T^I -\bar T^I)(T^J - \bar T^J)}\,2^3 \frac{\al}{2}\int_0^1\,dy \bigg(\frac{4y}{(1 + y)^2}\bigg)^{-\al s}\bigg(\frac{(1-y)^2}{(1+y)^2}\bigg)^{-\al \frac{t}{4}}\bigg(\big(-1 {-} \al s\big)\frac{1 - y^2}{16y^2} -\al s\frac{1+y}{4y(1 - y)}\delta^{IJ}\bigg)
\end{aligned}
\label{eq:disk_TT_1234_-1-100_fix}
\end{equation}
Exploiting the substitution \eqref{sub}
%\begin{equation}
%y = \frac{1 - \sqrt{x}}{1 + \sqrt{x}}\,\,\,,\quad dy = -\frac{1}{\sqrt x}\frac{1}{(1 + \sqrt x)^2}dx\,\,\,,\quad \{y=0, y= 1\}\rightarrow\{x=1, x=0\} 
%\end{equation}\label{eq:changeVar}
\cite{Garousi:1996ad,Hashimoto:1996kf,Hashimoto:1996bf} and the $\Gamma$ function properties eq. \eqref{eq:disk_TT_1234_-1-100_fix} becomes
%
%\begin{equation}
%\begin{aligned}
%&{\cal A}_{a}\left(T^{I},\bar{T}^{J}\right) = %\frac{8g_c^2C_{_{D_2}}}{(T^I {-}\bar T^I)(T^J {-} \bar T^J)} \int_0^1\,dx\,(1 {-} x)^{-\al s} x^{-\al \frac{t}{4}}\bigg(\frac{(-1 {-} \al s)\mathscr{R}^I_a\mathscr{R}^J_a}{(1 {-} x)^2} {-}\frac{\al s\,\delta_{I,J}}{x(1 {-} x)}\bigg)\\
%%%%
%&=\frac{8g_c^2C_{_{D2}}}{(T^I {-}\bar T^I)(T^J {-} \bar T^J)} \bigg((-1 {-} \al s)\mathscr{R}^I_a\mathscr{R}^J_a\int_0^1 dx\,(1 {-} x)^{-\al s{-}2}  x^{-\al \frac{t}{4}} - \al s\,\delta_{I,J}\int_0^1 dx\,(1 {-} x)^{-\al s{-}1} x^{{-}\al \frac{t}{4}{-}1}\bigg)
%%%%%
%\frac{8g_c^2C_{_{D2}}}{(T^I {-}\bar T^I)(T^J {-} \bar T^J)}\bigg(\mathscr{R}^I_a\mathscr{R}^J_a\frac{(-1 {-} \al s)\Gamma(-\al s {-}1)\Gamma(-\al t/4 {+} 1)}{\Gamma(-\al s {-} \al t/4)} - \al s\,\delta_{I,J} \frac{\Gamma(-\al s)\Gamma( -\al t/4)}{\Gamma(-\al s {-} \al t/4)}\bigg)\\&
%\end{aligned}\label{eq:disk_TT_subyx_-1-100}
%\end{equation}
%
%which using  becomes

%\begin{equation}
%\begin{aligned}
%\Gamma(-\al s {-}1) = \frac{\Gamma(-\al s)}{(-\al s {-} 1)}\,\,\,,\quad\Gamma(-\al t/4 {+}1)=(-\al t/4)\,\Gamma(-\al t/4)\,\,\,,\quad\Gamma(-\al s {-}\al t/4) = \frac{\Gamma(-\al s {-}\al t/4 + 1)}{(-\al s {-}\al t/4)}\\&
%\end{aligned}
%\label{eq:Gammafun}
%\end{equation}
\begin{equation}\boxed{
\small{\begin{aligned}
{\cal A}_{a}\left(T^{I},\bar{T}^{J}\right)=&\frac{8g_c^2C_{_{D_2}}}{(T^I {-}\bar T^I)(T^J {-} \bar T^J)} \bigg\{\al \frac{t}{4}\mathscr{R}^I_a\mathscr{R}^J_a\Big(\al s {+} \al \frac{t}{4}\Big) {+}\al s\,\delta_{I,J}\Big(\al s {+}\al \frac{t}{4}\Big)\bigg\}\frac{\Gamma({-}\al s)\Gamma({-}\al t/4)}{\Gamma({-}\al s {-} \al t/4 {+} 1)}
\end{aligned}}}\,.\label{eq:disk_TT_res_-1-100}
\end{equation}
Naturally two different kinds of contributions are present, the diagonal one with $({\mathscr R}^J_a)^2 = 1$
\begin{itemize}
\item{$I=J$}
\begin{equation}
\begin{aligned}
&\frac{8g_c^2C_{_{D_2}}}{(T^I {-}\bar T^I)^2} \Big(\al s{+} \frac{\al t}{4}\Big)^2  \frac{\Gamma(-\al s)\Gamma(-\al t/4)}{\Gamma(-\al s - \al t/4 {+} 1)}=  \\
&\boxed{\mathcal A_a(T^I,\bar T^I)=\frac{8g_c^2C_{_{D_2}}}{(T^I {-}\bar T^I)^2}\bigg\{\frac{4s}{t} {+} \frac{t}{4s} {+} 2 {+}\frac{\al^2\,u^2}{16}\left(- \zeta(2) + O(\al)\right)\bigg\}}
\end{aligned}\label{eq:disk-TbarT-res1}
\end{equation}
\end{itemize}

and the off-diagonal one
\begin{itemize}
\item{$I\ne J$}
\begin{equation}
\begin{aligned}
&\frac{8g_c^2C_{_{D_2}}\mathscr{R}^I_a\mathscr{R}^J_a}{(T^I -\bar T^I)(T^J - \bar T^J)} \frac{ \al t}{4}\Big(\al s {+}  \frac{\al t}{4}\Big) \frac{\Gamma(-\al s)\Gamma(-\al t/4)}{\Gamma(-\al s - \al t/4 {+} 1)}=\\
&\boxed{\mathcal A_a(T^I,\bar T^J)=\frac{8\,g_c^2\mathscr{R}^I_a\mathscr{R}^J_a}{(T^I {-}\bar T^I)(T^J {-} \bar T^J)}\bigg\{1 {+}\frac{t}{4s} {-} \frac{\al^2 tu}{16} \left(- \zeta(2) + O(\al)\right)\bigg\}}
\end{aligned}\label{eq:disk-TbarT-res2}
\end{equation}
\end{itemize}
\vspace{2mm}
where, as in \cite{Garousi:2006zh,Lust:2004cx}, we have used the gamma function expansion
\begin{equation}
\frac{\Gamma(-\al s)\Gamma(-\al t/4)}{\Gamma(-\al s - \al t/4 {+} 1)} = \frac{1}{\al^2}\frac{4}{st} - \zeta(2) + O(\al)\,\,\,.
\label{eq:betaexp}
\end{equation}
\subsubsection*{$\boxed{{\cal A}_{a}\left(T^{I},{T}^{J}\right)}$}
Considering the $(T,T)$ pair, i.e  the pair of two $T$ K\"ahler moduli (the same would hold for a pair of $\bar T$) rather than the K\"ahler moduli pair $(T,\bar T)$, the resulting scattering, to which the amplitude \eqref{eq:disk_TT_res_-1-100} needs to be added, provides information on the geometrical modulus $T_2^I$ (imaginary part of $T$), as we will see at the end of this section. As before, giving all the details on our computation, we get 
\begin{equation}
\begin{aligned}
&{\cal A}_{a}\left(T^{I},{T}^{J}\right)=\,g_c^2C_{_{D_2}}\int_{\Hp}\frac{d^2z_1d^2z_2}{V_{CKG}}\langle:{\cal W}_{_{T^I(-1,-1)}}(E_1,k_1,z_1, \bar z_1) \,{\cal W}_{_{T^J(0,0)}}(E_2,k_2,z_2, \bar z_2)\rangle\\
&=\frac{8g_c^2C_{_{D_2}}}{\al(T^I -\bar T^I)(T^J - \bar T^J)}\int_{\Hp}\frac{d^2z_1d^2z_2}{V_{CKG}}\langle:\tilde{\Psi}^Ie^{-\phi}e^{ik_1X}(z_1)\bar\Psi^Ie^{-\bar\phi}e^{ik_1\bar X}(\bar z_1):\\
%%%%%%
&\hspace{5cm}:\Big(i\DP \tilde Z^J + \frac{\al}{2}\big(k_2\psi\big)\tilde\Psi^J\Big)e^{ik_2X
}(z_2)\Big(i\bar\DP\bar{Z}^J + \frac{\al}{2}\big(k_2\bar\psi\big)\bar{\Psi}^J\Big)e^{ik_2\bar X}(\bar z_2):\rangle\\
%%%%%%
&=\frac{8g_c^2C_{_{D_2}}}{\al(U^I -\bar U^I)(U^J - \bar U^J)}\int_{\Hp}\frac{d^2z_1d^2z_2}{V_{CKG}} \langle :e^{-\phi}(z_1): :e^{-\bar\phi}(\bar{z}_1):\rangle\left( {\cal M}^{(1)}+{\cal M}^{(2)}+{\cal M}^{(3)}+{\cal M}^{(4)}\right)
\end{aligned}
\label{eq:disk_TnobarT_1234_-1-100}
\end{equation}
where the appropriate vertex operators from Section\,\ref{Toolkit} have been taken. We report here the expression of $\mathcal M_1$ as an example of the several $\mathcal M^{(i)}$'s terms entering the previous formula
\begin{equation}
\mathcal M_1 = \langle:\tilde{\Psi}^Ie^{ik_1X}(z_1)\bar\Psi^Ie^{ik_1\bar X}(\bar z_1)::i\DP\tilde{Z}^Je^{ik_2X}(z_2)i\bar\DP\bar Z^Je^{ik_2\bar X}(\bar z_2):\rangle\,.
\end{equation}
The other $\mathcal M^{(i)}$'s (that can be obtained using  \eqref{eq:correlator10D_disk} and \eqref{eq:boscomSfera}, \eqref{eq:fermcomSfera}, \eqref{eq:boscomUp} as well as \eqref{eq:fermcomUp}) can be found in the Appendix A.6 of ref. \cite{AM:appendice}. Exploiting the $PSL(2,R)$ symmetry to fix the vertex operators at  the points \eqref{eq:fixpoints} gives 
\begin{equation}
\begin{aligned}
&\frac{8g_c^2C_{_{D_2}}}{(T^I {-} \bar T^I)(T^J {-}\bar T^J)}\int^1_0dy\,\left(\frac{4y}{(1{+}y)^2}\right)^{-\al s}\left(\frac{(1 {-} y)^2}{(1 {+} y)^2}\right)^{-\al \frac{t}{4}}4(1 {-} y^2)\left(\frac{(1 {+} \al s){\mathscr R}_a^I{\mathscr R}_a^J}{16y^2} - \frac{\al s\,\,\delta_{I,J}({\mathscr R}_a^I)^2}{4y(1 {+} y)^2}\right)
\end{aligned}
\label{eq:disk_TnobarT_1234_-1-100_fix}
\end{equation}
that after the change of variable \eqref{sub} leads to
\begin{equation}
\begin{aligned}
%&\frac{8g_c^2C_{_{D2}}}{(T^I{-}\bar T^I)(T^J {-} \bar T^J)}\left((1 + \al s)\mathscr R^I_a\mathscr R^J_a\int^1_0dx\,(1 {-} x)^{-\al s {-} 2}x^{-\al\frac{t}{4}} - \al s\,\,\delta_{I,J}(\mathscr R^I_a)^2\int^1_0dx\,(1 {-} x)^{-\al s {-}1}x^{-\al\frac{t}{4}}\right)\\
%
%&=\frac{8g_c^2C_{_{D2}}}{(T^I{-}\bar T^I)(T^J {-} \bar T^J)}\left(\mathscr R^I_a\mathscr R^J_a\frac{(1 {+} \al s)\Gamma(-\al s {-} 1)\Gamma(-\al t/4 {+}1)}{\Gamma(-\al s {-} \al t/4)} - \delta_{I,J}(\mathscr R^I_a)^2\frac{\al s\,\Gamma(-\al s)\Gamma(-\al t/4 {+}1)}{\Gamma(-\al s {-} \al t/4 {+}1)}\right)\\\\
%
&\small{\boxed{{\cal A}_{a}\left(T^{I}{,}{T}^{J}\right) {=} \frac{8g_c^2C_{_{D_2}}}{(T^I {-} \bar T^I)(T^J {-} \bar T^J)}\bigg\{{\mathscr R}^I_a{\mathscr R}^J_a\frac{\al t}{4}\left(-\al s {-} \al\frac{t}{4}\right) {+} \delta_{I,J}({\mathscr R}^I_a)^2\frac{\al^2t\,s}{4}\bigg\}\frac{\Gamma(-\al s)\Gamma(-\al t/4)}{\Gamma(-\al s {-} \al t/4 {+} 1)}}}\,.
\end{aligned}
\label{eq:TTresultD2}
\end{equation}
The diagonal and off-diagonal cases are, respectively
\begin{itemize}
\item{$I=J$}
\end{itemize}
\begin{equation}
\begin{aligned}
&\frac{8g_c^2C_{_{D_2}}}{(T^I - \bar T^I)^2}\left(-\al^2\frac{t^2}{16}\right)\frac{\Gamma(-\al s)\Gamma(-\al t/4)}{\Gamma(-\al s - \al t/4 + 1)}=\\
&\boxed{{\cal A}_{a}\left(T^{I},{T}^{I}\right)=\frac{8g_c^2C_{_{D_2}}}{(T^I - \bar T^I)^2}\left\{-\frac{t}{4s} - \al^2\frac{t^2}{16}\left(- \zeta(2) + O(\al)\right)\right\}}
\end{aligned}
\label{TT,IugJ}
\end{equation}
\begin{itemize}
\item{$I\ne J$}
\end{itemize}
\begin{equation}
\begin{aligned}
&\frac{8g_c^2C_{_{D_2}}\mathscr R^I_a\mathscr  R^J_a}{(T^I - \bar T^I)(T^J - \bar T^J)}\left(\al\frac{t}{4}\right)\left(-\al s - \al\frac{t}{4}\right)\frac{\Gamma(-\al s)\Gamma(-\al t/4)}{\Gamma(-\al s - \al t/4 + 1)}=\\
&\boxed{{\cal A}_{a}\left(T^{I},{T}^{J}\right)=\,-\frac{8g_c^2C_{_{D_2}}\mathscr R^I_a\mathscr R^J_a}{(T^I - \bar T^I)(T^J - \bar T^J)}\bigg\{1 {+}\frac{t}{4s} {-} \frac{\al^2 tu}{16}\left(- \zeta(2) + O(\al)\right)\bigg\}}\,\,.
\end{aligned}
\label{TT,IneJ}
\end{equation}
As we said in Section \ref{Toolkit}, the vertex operator associated to the NS-NS untwisted modulus field $T^I_2$ (the imaginary part of the complex K\"ahler moduli $T^I$) is given by \eqref{eq:VT2}, so the true string scattering amplitude that involves two NS-NS untwisted moduli $T^I_2$ is 
\begin{equation}
\mathcal A_{a}(T_2^I,T^J_2) = \frac{1}{4}\left({\cal A}_{a}\left(\bar T^{I}, {T}^{J}\right) -{\cal A}_{a}\left(T^{I},{T}^{J}\right) - {\cal A}_{a}\left(\bar T^{I},\bar {T}^{J}\right) + {\cal A}_{a}\left(T^{I},\bar{T}^{J}\right)\right)\,.
\label{eq:amplitudedef}
\end{equation}
The results for the two distinct cases using \eqref{eq:disk_TT_res_-1-100}, \eqref{eq:TTresultD2} are\,\footnote{The same results hold for $(\bar T,\bar T)$ and $(\bar T, T)$ amplitudes}
\begin{itemize}
\item{$I=J$}
\end{itemize}
\begin{equation}
\boxed{\mathcal A_{a}(T_2^I,T^I_2)
%\frac{1}{4}\left({\cal A}_{a}\left(\bar T^{I}, {T}^{I}\right) -{\cal A}_{a}\left(T^{I},\bar{T}^{I}\right) - {\cal A}_{a}\left(\bar T^{I},\bar {T}^{I}\right) + {\cal A}_{a}\left(T^{I},{T}^{I}\right)\right)
= \frac{4g_c^2C_{_{D_2}}}{(T^I - \bar T^I)^2}\left\{\frac{4s}{t} + \frac{t}{2s} + 2+ \al^2\frac{\left(u^2 + t^2\right)}{16}\left(- \zeta(2) + O(\al)\right)\right\}}\\
\label{eq:TT}
\end{equation}
\begin{itemize}
\item{$I\ne J$}
\end{itemize}
\begin{equation}
\boxed{\mathcal A_{a}(T_2^I,T^J_2)
%\frac{1}{4}\left({\cal A}_{a}\left(\bar T^{I}, {T}^{J}\right) -{\cal A}_{a}\left(T^{I},\bar{T}^{J}\right) - {\cal A}_{a}\left(\bar T^{I},\bar {T}^{J}\right) + {\cal A}_{a}\left(T^{I},{T}^{J}\right)\right) 
= \frac{8g_c^2C_{_{D_2}}\mathscr R^I_a\mathscr R^J_a}{(T^I - \bar T^I)(T^J - \bar T^J)}\left\{1 + \frac{t}{4s} - \al^2\frac{tu}{16}\left(- \zeta(2) + O(\al)\right)\right\}}
\label{eq:TbarT}
\end{equation}
where on the $D9$-branes $a = 9$ and $\mathscr R^I_9 = +1$, while on the $D5_I$-branes $a=5_I$ one has $\mathscr R^I_{5_I} = +1$ and $\mathscr R^J_{5_I} = -1$.
As expected, there is no off-diagonal mixing at tree level for the kinetic terms between different K\"ahler moduli $T^I_2$ because the closed $t$-pole channel is absent in \eqref{eq:TbarT}, while \eqref{eq:TT} suggests that the K\"ahler potential has the expected structure of eq. \eqref{eq:kahlerP} \cite{Blumenhagen:2006ci, Berg:2005ja,Ibanez:2012zz,Lust:2004cx}.

%%%%%%%%%%%%%%%%%%%%%%%%%%%%%%%%%%%%%%%%
%%%%%%%%%%%%%%%%%%%%%%%%%%%%%%%%%%%%%%%%
%%%%%%%%%%%%%%%%%%%%%%%%%%%%%%%%%%%%%%%%
%%%%%%%%%%%%%%%%%%%%%%%%%%%%%%%%%%%%%%%%
\subsection{$\mathcal A^{D_2}_{a}\left(U^{I},\bar{U}^{J}\right)$ and $\mathcal A^{D_2}_{a}\left(U^{I},{U}^{J}\right)$}
The next set of amplitudes in \eqref{eq:2untwmoduli_disk} involves two complex structure moduli $U^I$. Apart from the different vertex operator definition for the complex structure $U^I$, the main steps leading the calculation of the relevant scattering amplitudes are the same as for the K\"ahler modulus $T^I$.
\subsubsection*{$\boxed{{\cal A}_{a}\left(U^{I},\bar{U}^{J}\right)}$}
\begin{equation}
\begin{aligned}
&{\cal A}_{a}\left(U^{I},\bar{U}^{J}\right)=\,g_c^2C_{_{D_2}}\int_{\Hp}\frac{d^2z_1d^2z_2}{V_{CKG}}\langle:{\cal W}_{_{U^I(-1,-1)}}(E_1,k_1,z_1, \bar z_1) \,{\cal W}_{_{\bar U^J(0,0)}}(E_2,k_2,z_2, \bar z_2)\rangle\\
%%%%%%
&=-\frac{8g_c^2C_{_{D_2}}}{\al(U^I -\bar U^I)(U^J - \bar U^J)}\int_{\Hp}\frac{d^2z_1d^2z_2}{V_{CKG}}\langle:\Psi^Ie^{-\phi}e^{ik_1X}(z_1)\bar\Psi^Ie^{-\bar\phi}e^{ik_1\bar X}(\bar z_1):\\
%%%%%%
&\hspace{5cm}:\Big(i\DP \tilde Z^J + \frac{\al}{2}\big(k_2\psi\big)\tilde\Psi^J\Big)e^{ik_2X
}(z_2)\Big(i\bar\DP\bar{\tilde Z}^J + \frac{\al}{2}\big(k_2\bar\psi\big)\bar{\tilde \Psi}^J\Big)e^{ik_2\bar X}(\bar z_2):\rangle\\
%%%%%%
&=-\frac{8g_c^2C_{_{D_2}}}{\al(U^I -\bar U^I)(U^J - \bar U^J)}\int_{\Hp}\frac{d^2z_1d^2z_2}{V_{CKG}} \langle :e^{-\phi}(z_1): :e^{-\bar\phi}(\bar{z}_1):\rangle\left( {\cal M}^{(1)}+{\cal M}^{(2)}+{\cal M}^{(3)}+{\cal M}^{(4)}\right)
\end{aligned}\label{eq:disk_UU_-1-100}
\end{equation}
that with the help of \eqref{eq:correlator10D_disk}, \eqref{eq:boscomSfera}, \eqref{eq:fermcomSfera}, \eqref{eq:boscomUp} and \eqref{eq:fermcomUp}, lead to ${\cal M}$'s contractions of the kind  
\begin{equation}
{\cal M}^{(1)}=\langle:\Psi^I e^{ik_1X}(z_1)\bar\Psi^I e^{ik_1\bar X}(\bar z_1)::i\DP \tilde Z^Je^{ik_2X}(z_2)i\bar\DP\bar{\tilde Z}^J e^{ik_2\bar X}(\bar z_2):\rangle\,\,.
\label{M1UbarU}
\end{equation}
Details on all $\mathcal M^{(i)}$'s are collected in Appendix A.6 of ref. \cite{AM:appendice}. The net result for the amplitude for the pair $(U^I,\bar U^J)$ is
%%%%%%%%%%%%%%%%%%%%%%%%%%%%%%%%%%
%%%%%%%%%%%%%%%%%%%%%%%%%%%%%%%%%%
\begin{equation}
\begin{aligned}
{\cal A}_{a}\left(U^{I},\bar{U}^{J}\right) =&-\frac{8g_c^2C_{_{D_2}}}{\al(U^I {-}\bar U^I)(U^J{-} \bar U^J)}\int_{\Hp}\frac{d^2z_1d^2z_2}{V_{CKG}}\bigg(\frac{\abs{z_1 {-}\bar z_1}\abs{z_2 {-} \bar z_2}}{\abs{z_1 {-} \bar z_2}^2}\bigg)^{-\al s}\bigg(\frac{\abs{z_1 {-} z_2}^2}{\abs{z_1 {-} \bar z_2}^2}\bigg)^{-\al t}\frac{1}{(z_1 - \bar z_1)}\\&\bigg\{-\frac{\al^2s\,\,(\mathscr{R}^I_a)^2\delta_{I,J}}{2(z_2 - \bar z_2)(z_1-\bar z_2)(z_2 - \bar z_1)} - \frac{\al^2s\,\,\delta_{I,J}}{(z_2 -\bar z_2)(z_1 - z_2)(\bar z_1 - \bar z_2)}\bigg\}\,.
\end{aligned}\label{eq:disk_UU_1234_-1-100}
\end{equation}
Exploiting \eqref{eq:fixpoints} to fix the vertex operators and inserting the c-ghost determinant \eqref{eq:cghostD2} the previous expression becomes 
\begin{equation}
\begin{aligned}
&-\delta_{I,J}\frac{8g_c^2C_{_{D_2}}}{(U^I {-}\bar U^I)(U^J {-} \bar U^J)}(\al\,s)\int_0^1\,dy\,\bigg(\frac{4y}{(1 {+} y)^2}\bigg)^{{-}\al s}\bigg(\frac{(1{-}y)^2}{(1{+}y)^2}\bigg)^{{-}\al \frac{t}{4}}\bigg(\frac{(1{+}y)}{y(1 {-} y)}{-}\frac{(\mathscr{R}^I_a)^2(1{-}y)}{y(1{+}y)}\bigg)\,.\\&
%%%
\end{aligned}\label{eq:disk_UU_fix_-1-100}
\end{equation}

Finally after the change of variable \eqref{sub}, using $(\mathscr R^I_a)^2 = 1$ one gets \cite{Lust:2004cx}

\begin{equation}
\begin{aligned}
%&-\delta_{I,J}\frac{8g_sc_{_{D2}}}{(U^I {-}\bar U^I)(U^J {-} \bar U^J)} (\al\,s)\, \bigg(\int_0^1 dx\,(1 {-} x)^{-\al s{-}1} x^{{-}\al \frac{t}{4}{-}1}{-}\int_0^1dx\,(1 {-} x)^{{-}\al s{-}1} x^{{-}\al \frac{t}{4}} \bigg)\\
\boxed{{\cal A}_{a}\left(U^{I},\bar{U}^{J}\right) %=-\delta_{I,J}\frac{8g_sC_{_{D2}}}{(U^I {-}\bar U^I)(U^J {-} \bar U^J)} (\al\,s) \,\bigg(\frac{\Gamma(-\al s)\Gamma(-\al t/4)}{\Gamma(-\al s - \al t/4)}-\frac{\Gamma(-\al s)\Gamma(-\al t/4 +1)}{\Gamma(-\al s - \al t/4 {+} 1)} \bigg)\\
%%%%%
=\delta_{I,J}\,\frac{8g_c^2C_{_{D_2}}}{(U^I {-}\bar U^I)(U^J {-} \bar U^J)}(\al^2 s^2)\frac{\Gamma(-\al s)\Gamma(-\al t/4)}{\Gamma(-\al s - \al t/4 {+}1)}}\,\,\,.
\end{aligned}\label{eq:disk_UU_yxsub_-1-100}
\end{equation}

Separating the diagonal from the off-diagonal case and using the gamma function expansion \eqref{eq:betaexp}, one finds
\begin{itemize}
\item{$I = J$}
\begin{equation}
\begin{aligned}
&%\frac{8g_sC_{_{D_2}}}{(U^I {-}\bar U^I)^2}\big(\al^2 s^2\big)\frac{\Gamma(-\al s)\Gamma(-\al t/4)}{\Gamma(-\al s - \al t/4 {+}1)} =
\boxed{\mathcal A_a(U^I,\bar U^I) = \frac{8g_c^2C_{_{D_2}}}{(U^I {-} \bar U^I)^2}\bigg\{\frac{4s}{t} +\al^2s^2\left(- \zeta(2) + O(\al)\right)\bigg\}}\end{aligned}
 \label{eq:Disk-UbarU-res1}
\end{equation}
\end{itemize}
\begin{itemize}
\item{$I \ne J$}
\begin{equation}
\boxed{\mathcal A_a(U^I,\bar U^J) =\,0}
\label{eq:Disk-UbarU-res2}
\end{equation}
\end{itemize}
giving the same results in all cases $a\in \{9,5_I\}$.
%%%%%%%%%%%%%%%%%%%%%%%%%%%%%%%%%%%%%
%%%%%%%%%%%%%%%%%%%%%%%%%%%%%%%%%%%%
%\subsection{${\cal A}_{a}\left(U^{I},{U}^{J}\right)$}
Upon computing the amplitude which involves the complex structure moduli pair ($U^I,U^J$) we anticipate that one arrives at a vanishing result.
\subsubsection*{$\boxed{{\cal A}_{a}\left(U^{I},{U}^{J}\right)}$}
Considering the vertex operators in Section \ref{Toolkit} one has 
\begin{equation}
\begin{aligned}
&{\cal A}_{a}\left(U^{I},{U}^{J}\right)=\,g_c^2C_{_{D_2}}\int_{\Hp}\frac{d^2z_1d^2z_2}{V_{CKG}}\langle:{\cal W}_{_{U^I(-1,-1)}}(E_1,k_1,z_1, \bar z_1) \,{\cal W}_{_{U^J(0,0)}}(E_2,k_2,z_2, \bar z_2)\rangle\\
%%%%%%
&=\frac{8g_c^2C_{_{D_2}}}{\al(U^I -\bar U^I)(U^J - \bar U^J)}\int_{\Hp}\frac{d^2z_1d^2z_2}{V_{CKG}}\langle:\Psi^Ie^{-\phi}e^{ik_1X}(z_1)\bar\Psi^Ie^{-\bar\phi}e^{ik_1\bar X}(\bar z_1):\\
%%%%%%
&\hspace{5cm}:\Big(i\DP Z^J + \frac{\al}{2}\big(k_2\psi\big)\Psi^J\Big)e^{ik_2X
}(z_2)\Big(i\bar\DP\bar{Z}^J + \frac{\al}{2}\big(k_2\bar\psi\big)\bar{\Psi}^J\Big)e^{ik_2\bar X}(\bar z_2):\rangle\\
%%%%%%
&=\frac{8g_c^2C_{_{D_2}}}{\al(U^I -\bar U^I)(U^J - \bar U^J)}\int_{\Hp}\frac{d^2z_1d^2z_2}{V_{CKG}} \langle :e^{-\phi}(z_1): :e^{-\bar\phi}(\bar{z}_1):\rangle\left( {\cal M}^{(1)}+{\cal M}^{(2)}+{\cal M}^{(3)}+{\cal M}^{(4)}\right)
\end{aligned}\label{eq:disk_UU_-1-100}
\end{equation}
where we notice that the specific combinations of two-point functions \eqref{eq:boscomSfera}, \eqref{eq:fermcomSfera}, \eqref{eq:boscomUp} and \eqref{eq:fermcomUp}, occuring in the $\mathcal M^{(i)}$'s for the computation of the compactified field, are responsible for the vanishing of all the $\mathcal M^{(i)}$ terms. As an example we report the expression of ${\cal M}^{(1)}$
\begin{equation}
{\cal M}^{(1)}=\langle:\Psi^I e^{ik_1X}(z_1)\bar\Psi^I e^{ik_1\bar X}(\bar z_1)::i\DP Z^Je^{ik_2X}(z_2)i\bar\DP\bar{Z}^J e^{ik_2\bar X}(\bar z_2):\rangle
\end{equation}
while for the others see Appendix A.6 of ref.\,\cite{AM:appendice}. Thus owing to the vanishing of some two-point correlation functions, the final result is
\begin{equation}
\boxed{{\cal A}_{a}\left(U^{I},{U}^{J}\right) = 0}
\label{eq:UU}
\end{equation}
independently of whether $I=J$ or $I\ne J$. The complex structure $U^I$ moduli are purely geometrical moduli therefore no vertex redefinition is needed in contrast to the K\"ahler moduli $T^I_2$. Thus \eqref{eq:UU} and \eqref{eq:Disk-UbarU-res2} show that there is no mixing between different complex structure $U$ moduli, while \eqref{eq:Disk-UbarU-res1} is in agreement with the form of the tree-level K\"ahler potential for this model \eqref{eq:kahlerP} \cite{Blumenhagen:2006ci, Berg:2005ja,Ibanez:2012zz,Lust:2004cx}.
%%%%%%%%%%%%%%%%%%%%%%%%%%%%%%%%%%%%%%
%%%%%%%%%%%%%%%%%%%%%%%%%%%%%%%%%%%%%%
%%%%%%%%%%%%%%%%%%%%%%%%%%%%%%%%%%%%%%
%%%%%%%%%%%%%%%%%%%%%%%%%%%%%%%%%%%%%%

\subsection{$\mathcal A^{D_2}_{a}\left(T^{I},\bar{U}^{J}\right)$ and $\mathcal A^{D_2}_{a}\left(T^{I},{U}^{J}\right)$}
The last set of amplitudes in \eqref{eq:2untwmoduli_disk} involves one K\"ahler modulus $T^I$ and one complex structure modulus $U^J$. We anticipate that this kind of amplitudes are zero. For this reason we give less details here in contrast to the previous cases.
\subsubsection*{$\boxed{{\cal A}_{a}\left(T^{I},\bar{U}^{J}\right)}$}
The amplitude which involves the pair ($T,\bar U$), given the explicit form of vertex operator taken from the Section \ref{Toolkit}, is
\begin{equation}
\begin{aligned}
&{\cal A}_{a}\left(T^{I},\bar{U}^{J}\right)=\,g_c^2C_{_{D_2}}\int_{\Hp}\frac{d^2z_1d^2z_2}{V_{CKG}}\langle{\cal W}_{T^I(-1,-1)}(E_1, k_1,z_1, \bar z_1){\cal W}_{\bar U^J(0,0)}(E_2,k_2,z_2, \bar z_2)\rangle\\
%%%%
&=\frac{8g_c^2C_{_{D_2}}}{\al(T^I -\bar T^I)(U^J - \bar U^J)}\int_{\Hp}\frac{d^2z_1d^2z_2}{V_{CKG}}\langle:\tilde\Psi^Ie^{-\phi}e^{ik_1X}(z_1)\bar\Psi^Ie^{-\bar\phi}e^{ik_1\bar X}(\bar z_1):\\&\hspace{5cm}:\Big(i\DP \tilde Z^J + \frac{\al}{2}\big(k_2\psi\big)\tilde\Psi^J\Big)e^{ik_2X}(z_2)\Big(i\bar\DP\bar{\tilde Z}^J + \frac{\al}{2}\big(k_2\bar \psi\big)\bar{\tilde \Psi}^J\Big)e^{ik_2\bar X}(\bar z_2):\rangle
\\
%%%%
&=\frac{8g_c^2C_{_{D_2}}}{\al(T^I {-}\bar T^I)(U^J {-} \bar U^J)}\int_{\Hp}\frac{d^2z_1d^2z_2}{V_{CKG}}\langle:e^{-\phi}(z_1)::e^{-\bar\phi}(\bar z_1):\rangle\left({\cal M}^{(1)}+{\cal M}^{(2)}+{\cal M}^{(3)}+{\cal M}^{(4)} \right)
\end{aligned}\label{eq:disk_TU_-1-100}
\end{equation}
where as one can see in Appendix A.6 of ref.\,\cite{AM:appendice}, all the ${\cal M}^{(i)}$ terms are zero due to the vanishing of the particular two-point functions that enter the ${\cal M}^{(i)}$ definition. So the amplitude is zero when $I=J$, $I\ne J$ and for $a\in\{9,5_I\}$ 
\begin{equation}
\boxed{\mathcal A_a(T^I,\bar U^J) =\,0 }\qquad ( T \leftrightarrow U)\,\,.\label{eq:ATU}
\end{equation}
%%%%%%%%%%%%%%%%%%%%%%%%%%%%%%%%%%%%%%%%%%%%
%%%%%%%%%%%%%%%%%%%%%%%%%%%%%%%%%%%%%%%%%%%%
%\subsection{${\cal A}_{a}\left(T^{I},{U}^{J}\right)$}
%
\subsubsection*{$\boxed{{\cal A}_{a}\left(T^{I},{U}^{J}\right)}$}
The same happens for the second amplitude in this specific set 
\begin{equation}
\begin{aligned}
&{\cal A}_{a}\left(T^{I},{U}^{J}\right)=-\,g_c^2C_{_{D_2}}\int_{\Hp}\frac{d^2z_1d^2z_2}{V_{CKG}}\langle{\cal W}_{T^I(-1,-1)}(E_1, k_1,z_1, \bar z_1){\cal W}_{U^J(0,0)}(E_2,k_2,z_2, \bar z_2)\rangle\\
%%%%
&=-\frac{8g_c^2C_{_{D_2}}}{\al(T^I -\bar T^I)(U^J - \bar U^J)}\int_{\Hp}\frac{d^2z_1d^2z_2}{V_{CKG}}\langle:\tilde\Psi^Ie^{-\phi}e^{ik_1X}(z_1)\bar\Psi^Ie^{-\bar\phi}e^{ik_1\bar X}(\bar z_1):\\&\hspace{5cm}:\Big(i\DP Z^J + \frac{\al}{2}\big(k_2\psi\big)\Psi^J\Big)e^{ik_2X}(z_2)\Big(i\bar\DP\bar{Z}^J + \frac{\al}{2}\big(k_2\bar \psi\big)\bar{\Psi}^J\Big)e^{ik_2\bar X}(\bar z_2):\rangle
\\
%%%%
&=-\frac{8g_c^2C_{_{D_2}}}{\al(T^I {-}\bar T^I)(U^J {-} \bar U^J)}\int_{\Hp}\frac{d^2z_1d^2z_2}{V_{CKG}}\langle:e^{-\phi}(z_1)::e^{-\bar\phi}(\bar z_1):\rangle\left({\cal M}^{(1)}+{\cal M}^{(2)}+{\cal M}^{(3)}+{\cal M}^{(4)} \right)\,.
\end{aligned}\label{eq:disk_TU_-1-100}
\end{equation}
For the same reasons (there is no difference between $I=J$ and $I\ne J$) one finds
\begin{equation}
\boxed{{\cal A}_{a}\left(T^{I},{U}^{J}\right) = 0}\,\,.
\label{eq:TU}
\end{equation}
as one can see from the Appendix A.6 of ref.\,\cite{AM:appendice}.
To be rigorous also in this case one should take the vertex for the NS-NS $T^I_2$ K\"ahler modulus \eqref{eq:VT2}, but the result will be the same. So equations \eqref{eq:TU} and \eqref{eq:ATU} confirm again the structure of the K\"ahler potential \eqref{eq:kahlerP}.
%%%%%%%%%%%%%%%%%%%%%%%%%%%%%%%%%%%%%%%%%%%%
%%%%%%%%%%%%%%%%%%%%%%%%%%%%%%%%%%%%%%%%%%%%
%%%%%%%%%%%%%%%%%%%%%%%%%%%%%%%%%%%%%%%%%%%%
%%%%%%%%%%%%%%%%%%%%%%%%%%%%%%%%%%%%%%%%%%%%
%%%%%%%%%%%%%%%%%%%%%%%%%%%%%%%%%%%%%%%%%%%%
\subsection{$\mathcal A^{RP_2}_{\alpha}\left({T^{I}},\bar{T}^{J} \right)$ and $\mathcal A^{RP_2}_{\alpha}\left(T^{I},{T}^{J}\right)$}
The technology we have applied to the scattering amplitude on $RP_2$ worldsheet for the uncompactified model is useful to understand how one has to deal with the scattering of two moduli on $RP_2$ when the compactified model is considered. The set of the relevant amplitudes involving untwisted moduli calculated are 
\begin{equation}
\small{
\begin{aligned}
{\cal A}_{\alpha}\left(T^{I},\bar{T}^{J}\right) + {\cal A}_{\alpha}\left(T^{I},{T}^{J}\right)&= g_c^2C_{_{RP_2}}\int_{\abs{z}\leq 1}\frac{d^2z_1d^2z_2}{V_{CKG}}\Big(\langle{\cal W}^{\otimes}_{T^I}(z_1, \bar z_1){\cal W}^{\otimes}_{\bar T^J}(z_2, \bar z_2)\rangle +\langle{\cal W}^{\otimes}_{T^I}(z_1, \bar z_1){\cal W}^{\otimes}_{T^J}(z_2, \bar z_2)\rangle\Big)\\
%%%%%
{\cal A}_{\alpha}\left(U^{I},\bar{U}^{J}\right) + {\cal A}_{\alpha}\left(U^{I},{U}^{J}\right)&=g_c^2C_{_{RP_2}}\int_{\abs{z}\leq 1}\frac{d^2z_1d^2z_2}{V_{CKG}}\Big(\langle{\cal W}^{\otimes}_{U^I}(z_1, \bar z_1){\cal W}^{\otimes}_{\bar U^J}(z_2, \bar z_2)\rangle + \langle{\cal W}^{\otimes}_{U^I}(z_1, \bar z_1){\cal W}^{\otimes}_{U^J}(z_2, \bar z_2)\rangle\Big)\\
%%%%%
{\cal A}_{\alpha}\left(T^{I},\bar{U}^{J}\right) + {\cal A}_{\alpha}\left(T^{I},{U}^{J}\right)&=g_c^2C_{_{RP_2}}\int_{\abs{z}\leq 1}\frac{d^2z_1d^2z_2}{V_{CKG}}\Big(\langle{\cal W}^{\otimes}_{T^I}(z_1, \bar z_1){\cal W}^{\otimes}_{\bar U^J}(z_2, \bar z_2)\rangle +\langle{\cal W}^{\otimes}_{T^I}(z_1, \bar z_1){\cal W}^{\otimes}_{U^J}(z_2, \bar z_2)\rangle \Big)\\&\quad+\,(T\leftrightarrow U)\,. \\
\end{aligned}\label{eq:AMPLRP2_New}}
\end{equation}
%But, as when we build the vertex operators for the moduli in the appendix \ref{Toolkit}, we know that since the compactified theory has $\mathcal N =2$ worldsheet supersymmetries we have to take care of the $U(1)$ R-symmetry that fix uniquely the form of the vertex .This is because from the very beginning one can associate the $U(1)$ $Q$-charge to the compactified field in a quasi-arbitrary way \cite{BLT}. As we have said from the 10 dimensional scattering amplitude on RP2 we have understood the mechanism behind the construction of the vertex operators in this case. Now the presence of the $U(1)$ $Q$-charge plays a crucial role, because we have to take under control as the orientifold operator (in particular the worldsheet parity operator $\Omega$) acts on the compactified field in order to preserve the value of the $Q$-charge when the orientifold operator exchange $Q \leftrightarrow \bar Q$. 
Details on the vertex operators construction on $RP_2$ are reported in the Section \ref{Toolkit}.
\subsubsection*{$\boxed{\mathcal A_{\alpha}({T^I},{\bar{T}^J})}$}
The first set of scattering amplitudes, for the pair ($T,\bar T$) massless K\"ahler moduli in the picture (-1,-1; 0,0) start with
%%%%%%%%%%%%%%%%%%%%%%%%%%%
\begin{equation}
\begin{aligned}
\mathcal A_{\alpha}({T^I},{\bar{T}^J}) =&g_c^2C_{_{RP2}}\int_{_{\abs{z}\leq 1}}\frac{d^2z_1d^2z_2}{V_{CKG}}\langle {\cal W}^{\otimes}_{_{T^I({-}1,{-}1)}}(E_1,k_1,z_1,\bar z_1){\cal W}^{\otimes}_{_{\bar T^J(0,0)}}(E_2,k_2,z_2,\bar z_2)\rangle = \sum_{i=1}^4\Lambda_i \\
\end{aligned}\label{eq:RP2_TT_-1-100}
\end{equation}
where the different $\Lambda_i$'s sub-amplitudes are associated to the form of vertex operators, as one can see in Section \ref{Toolkit}. Each $\Lambda $'s sub-amplitude behaves as a single disk scattering amplitude. As an example we have for $\Lambda_1$
\begin{equation}
\begin{aligned}
&\Lambda_1 =-\frac{2g_c^2C_{_{RP_2}}}{\al(T^I - \bar T^I)(T^J - \bar T^J)}\int_{_{\abs{z}\leq 1}}\frac{d^2z_1d^2z_2}{V_{CKG}}\langle:\tilde{\Psi}^Ie^{-\phi}e^{ik_1X}(z_1)\bar\Psi^Ie^{-\bar\phi}e^{ik_1\bar X}(\bar z_1):\\&\hspace{5cm}:\Big(i\DP Z^J + \frac{\al}{2}\big(k_2\psi\big)\Psi^J\Big)e^{ik_2X}(z_2)\Big(i\bar\DP\bar{\tilde Z}^J + \frac{\al}{2}\big(k_2\bar\psi\big)\bar{\tilde\Psi}^J\Big)e^{ik_2\bar X}(\bar z_2):\rangle\\
&=-\frac{2g_c^2C_{_{RP_2}}}{\al(T^I - \bar T^I)(T^J - \bar T^J)}\int_{_{\abs{z}\leq 1}}\frac{d^2z_1d^2z_2}{V_{CKG}}\langle:e^{-\phi}(z_1)e^{-\bar\phi}(\bar z_1):\rangle
\bigg({\cal M}^{(1)}_{_{\Lambda_1}} +{\cal M}^{(2)}_{_{\Lambda_1}} + {\cal M}^{(3)}_{_{\Lambda_1}} + {\cal M}^{(4)}_{_{\Lambda_1}}\bigg)\\
\end{aligned}\label{eq:lambda1_TT_-1-100}
\end{equation}
%%%%%%%%%
where again ${\cal M}^{(j)}_{_{\Lambda_i}}$'s with $(i,j = 1,2,3,4)$ are the different contraction terms that one meets in each $\Lambda_i$'s sub-amplitude, among which for instance we have
\begin{equation}
{\cal M}_{_{\Lambda_1}}^{(1)}=\langle:\tilde{\Psi}^Ie^{ik_1X}(z_1)\bar\Psi^Ie^{ik_1\bar X}(\bar z_1)::i\DP Z^Je^{ik_2X}(z_2)i\bar\DP\bar{\tilde Z}^Je^{ik_2\bar X}(\bar z_2):\rangle\,.
\end{equation}
The ${\cal M}^{(j)}_{_{\Lambda_i}}$'s can be calculated using the two-point functions \eqref{eq:correlator10D_RP2}\, \footnote{Where in  \eqref{eq:correlator10D_RP2} one has to reinsert the reflection matrix $\R$ on the two points correlation function which involves left and right field.} for the uncompactified fields and the two-point functions for the compactified fields \eqref{eq:boscomSfera}, \eqref{eq:fermcomSfera}, \eqref{eq:boscomRp} and \eqref{eq:fermcomRp}.
%\begin{equation} 
%\boxed{
%\begin{aligned}
%&\langle\DP Z^I(z_1)\DP \tilde Z^J(z_2)\rangle= -\frac{\al\delta_{I,J}}{2(z_1 -z_2)^2}\,\,,\quad\langle\bar\DP \bar Z^I(\bar z_1)\bar\DP \bar{\tilde Z}^J(\bar z_2)\rangle =  -\frac{\al\delta_{I,J}}{2(\bar z_1 - \bar z_2)^2}\\\\
%
%&\langle\DP Z^I(z_1)\bar\DP\bar{\tilde Z}^J(\bar z_2)\rangle=-\frac{\al\mathscr R^J_{\alpha}\delta_{I,J}}{2(1 +z_1\bar z_2)^2} \,\,,\quad \langle\DP\tilde Z^I(z_1)\bar\DP\bar Z^J(\bar z_2)\rangle= -\frac{\al\mathscr R^J_{\alpha}\delta_{I,J}}{2(1 + z_1\bar z_2)^2}\\\\
%
%&\langle\Psi^I(z_1)\tilde\Psi^J(z_2)\rangle = \frac{\delta_{I,J}}{(z_1 - z_2)}\,\,,\quad\langle\bar\Psi^I(\bar z_1)\bar{\tilde\Psi}^J(\bar z_2)\rangle= \frac{\delta_{I,J}}{(\bar z_1 - \bar z_2)}\\\\
%
%&\langle\Psi^I(z_1)\bar{\tilde\Psi}^J(\bar z_2)\rangle = \frac{\mathscr R^J_{\alpha}\delta_{I,J}}{(1 +z_1\bar z_2)}\,\,,\quad\langle\tilde\Psi^I(z_1)\bar\Psi^J(\bar z_2)\rangle = \frac{\mathscr R^J_{\alpha}\delta_{I,J}}{(1 +z_1\bar z_2)}
%\end{aligned}}\label{eq:FermBoseComMIX2}
%\end{equation}

The calculation of ${\cal M}^{(j)}_{_{\Lambda_i}}$'s is equal to the disk case, apart from the difference due to the involution $\mathfrak{I}(z)= -1/\bar z$ that characterizes the projective plane. As for the disk cases the list of all $\Lambda$'s sub-amplitudes and the details on their ${\cal M}$'s terms are in  Appendix A.7 of ref.\,\cite{AM:appendice}. Summing the $\Lambda$'s sub-amplitudes having the same Koba-Nielsen factor one gets
\begin{equation}
\begin{aligned}
\Lambda_1 + \Lambda_4 =&\,-\frac{2g_c^2C_{_{RP_2}}}{\al(T^I {-} \bar T^I)(T^J {-} \bar T^J)}\int_{_{\abs{z}\leq 1}}\frac{d^2z_1d^2z_2}{V_{CKG}}\bigg(\frac{(1 + \abs{z_1}^2)(1 +\abs{z_2}^2)}{\abs{1 + z_1\bar z_2}^2}\bigg)^{-\al s}\bigg(\frac{\abs{z_1 - z_2}^2}{\abs{1 +z_1\bar z_2}^2}\bigg)^{-\al \frac{t}{4}}\\
&\bigg(\frac{\al\,\mathscr R^I_{\alpha}\mathscr R^J_{\alpha}}{(1 + \abs{z_1}^2)^2(1 + \abs{z_2}^2)^2} + \frac{\al^2s\,\,\mathscr R^I_{\alpha}\mathscr R^J_{\alpha}}{(1 + \abs{z_1}^2)^2(1 + \abs{z_2}^2)^2} -\frac{\al^2s\,\delta_{I,J}}{(1 + \abs{z_1}^2)(1 + \abs{z_2}^2)\abs{z_2 - z_1}^2}\bigg)
%
%&+\frac{\al\,\mathcal D^I_{\alpha}\mathcal D^J_{\alpha}}{2(1 + \bar z_1z_1)^2(1 + \bar z_2z_2)^2}+ \frac{\al^2s\,\,\mathcal D^I_{\alpha}\mathcal D^J_{\alpha}}{2(1 + \bar z_1z_1)^2(1 + \bar z_2z_2)^2} - \frac{\al^2s\,\delta_{I,J}}{2(1 + \bar z_1z_1)(1 +\bar z_2z_2)(\bar z_1 - \bar z_2)(z_1 - z_2)}\bigg)%\delta^4(k_1 + k_1D + k_2 + k_2D)
\end{aligned}
\label{eq:1lambda23_TT_-1-100}
\end{equation}
\begin{equation}
\begin{aligned}
\Lambda_2 + \Lambda_3 = -&\frac{2g_c^2C_{_{RP_2}}}{\al(T^I {-} \bar T^I)(T^J {-} \bar T^J)}\int_{_{\abs{z}\leq 1}}\frac{d^2z_1d^2z_2}{V_{CKG}}\bigg(\frac{(1 + \abs{z_1}^2)(1 +\abs{z_2}^2)}{\abs{1 + z_1\bar z_2}^2}\bigg)^{-\al s}\bigg(\frac{\abs{z_1 - z_2}^2}{\abs{1 +z_1\bar z_2}^2}\bigg)^{-\al \frac{u}{4}}\\&\bigg(\frac{\al\mathscr R^I_{\alpha}\mathscr R^J_{\alpha}}{( 1+ \abs{z_1}^2)^2(1 + \abs{z_2}^2)^2} +\frac{\al^2s\,\,\mathscr R^I_{\alpha}\mathscr R^J_{\alpha}}{(1 + \abs{z_1}^2)^2(1 + \abs{z_2}^2)^2} -\frac{\al^2s\,\,\delta_{I,J}(\mathscr R^I_a)^2}{(1 + \abs{z_1}^2)(1 + \abs{z_2}^2)\abs{1 + z_1\bar z_2}^2}\bigg)\,.
%+\frac{\al}{2}\frac{1}{(1 + \bar z_1z_1)(1 + \bar z_1z_1)(1 + \bar z_2z_2)^2}+\frac{\al^2}{2}s\frac{1}{(1+ \bar z_1z_1)(1 + \bar z_2z_2)(1 + \bar z_1z_1)(1 + \bar z_2z_2)} -\frac{\al^2}{2}s\frac{\delta^{IJ}}{(1 + \bar z_1z_1)(1 + \bar z_2z_2)(1 + \bar z_1z_2)(1 +\bar z_2z_1)}\bigg]\delta^4(k_1 + k_1D + k_2 + k_2D)
\end{aligned}
\label{eq:lambda23_TT_-1-100}
\end{equation}
The vertices, using the $SU(2)$ symmetry can be fixed at
\begin{equation}
z_1 = 0;\quad \bar z_1= 0;\quad z_2= iy;\quad \bar z_2= -iy
\label{eq:fixpointsRP2}
\end{equation}
and inserting in \eqref{eq:1lambda23_TT_-1-100} and \eqref{eq:lambda23_TT_-1-100} the c-ghost contribution \eqref{eq:c-ghostRP2}, one finds\, \footnote{$(\mathscr R^I_a)^2 = 1$} 
\begin{equation}
\begin{aligned}
\Lambda_1 + \Lambda_4 %&\,-\frac{2g_c^2C_{_{RP2}}}{(T^I {-} \bar T^I)(T^J {-} \bar T^J)}\int_{-1}^{1}dy(1 + y^2)^{-\al s}(y^2)^{-\al \frac{t}{4}}\,4y\bigg\{\frac{(1 + \al s)\mathscr R^I_{\alpha}\mathscr R^J_{\alpha}}{(1 + y^2)^2} -\frac{\al s\,\,\delta_{I,J}}{(1 + y^2)y^2}\bigg\}\\
=& \,-\frac{2g_c^2C_{_{RP_2}}}{(T^I {-} \bar T^I)(T^J {-} \bar T^J)}\int_{0}^{1}dy^2(1 + y^2)^{-\al s}(y^2)^{-\al \frac{t}{4}}\bigg\{\frac{(1 + \al s)\mathscr R^I_{\alpha}\mathscr R^J_{\alpha}}{(1 + y^2)^2} -\frac{\al s\,\,\delta_{I,J}}{(1 + y^2)y^2}\bigg\}\\\\
\Lambda_2 + \Lambda_3 %&\,-\frac{2g_c^2C_{_{RP2}}}{(T^I {-} \bar T^I)(T^J {-} \bar T^J)}\int_{-1}^{1}dy(1 + y^2)^{-\al s}(y^2)^{-\al \frac{u}{4}}\,4y\bigg\{\frac{(1 + \al s)\mathscr R^I_{\alpha}\mathscr R^J_{\alpha}}{(1 + y^2)^2} -\frac{\al s\,\,\delta_{I,J}}{(1 + y^2)}\bigg\}\\
=&\,-\frac{2g_c^2C_{_{RP_2}}}{(T^I {-} \bar T^I)(T^J {-} \bar T^J)}\int_{0}^{1}dy^2(1 + y^2)^{-\al s}(y^2)^{-\al \frac{u}{4}}\bigg\{\frac{(1 + \al s)\mathscr R^I_{\alpha}\mathscr R^J_{\alpha}}{(1 + y^2)^2} -\frac{\al s\,\,\delta_{I,J}}{(1 + y^2)}\bigg\}\,.
\end{aligned}
\label{eq:lambda14-23Fix_TbarT_-1-1-00}
\end{equation}
Based on the definition of the ${}_2F_1$ hypergeometric function \eqref{eq:HY21} in \eqref{eq:lambda14-23Fix_TbarT_-1-1-00}, one obtains
\begin{equation}
\begin{aligned}
&\Lambda_1 + \Lambda_4 = -\frac{2g_c^2C_{_{RP_2}}}{(T^I {-} \bar T^I)(T^J {-} \bar T^J)}\bigg\{\mathscr R^I_{\alpha}\mathscr R^J_{\alpha}\frac{(1 {+} \al s){}_2F_1(\al s {+} 2, -\al t/4 {+} 1; -\al t/4 {+} 2; -1)}{(-\al t/4 {+} 1)}\\
& \hspace{5.5cm}- \delta_{I,J}\frac{\al s{}_2F_1(\al s {+} 1, -\al t/4; -\al t/4 {+} 1; -1)}{(-\al t/4)}\bigg\}\\\\
%%%%%%%%%%%
&\Lambda_2 {+} \Lambda_3 = -\frac{2g_c^2C_{_{RP_2}}}{(T^I {-} \bar T^I)(T^J {-} \bar T^J)}\bigg\{\mathscr R^I_{\alpha}\mathscr R^J_{\alpha}\frac{(1 {+} \al s\big){}_2F_1(\al s {+} 2, {-}\al u/4 {+} 1; {-}\al u/4 {+}2; {-}1)}{(-\al u {+} 1)} \\
&\hspace{5.5cm}- \delta_{I,J}\frac{\al s{}_2F_1(\al s {+} 1, {-}\al u/4 {+} 1; {-}\al u/4 {+} 2; {-}1)}{(-\al u/4 {+} 1)}\bigg\}\,.
\end{aligned}\label{eq:lambda14-23_res_TT_-1-100}
\end{equation}
Using the identity \eqref{HYPid}, one is able to combine the apparently different ${}_2F_1$ hypergeometric functions, as for instance
\begin{equation}
\begin{aligned}
{}_2F_1(\al s {+} 2, -\al u/4 {+} 1; -\al u/4 {+} 2; -1) =& -\frac{(-\al u/4 {+} 1)}{(-\al t/4 {+} 1)}{}_2F_1(\al s {+} 2, -\al t/4 {+} 1; -\al t/4 {+} 2; -1)
\\&
+ \frac{\Gamma(-\al t/4 {+} 1)\Gamma(-\al u/4 {+} 2)}{\Gamma(\al s {+} 2)}
%
%{}_2F_1(\al s {+} 1, -\al u/4 {+} 1; -\al u/4 {+} 2; -1) =& - \frac{(-\al u/4 {+} 1)}{(-\al t/4)}{}_2F_1(\al s {+} 1, -\al t/4; -\al t/4 {+} 1; -1\big)
%\\&
%+ \frac{\Gamma(-\al t/4)\Gamma(-\al u/4 {+} 2)}{\Gamma(\al s {+} 1)}
\end{aligned}\label{eq:HYFident_TT_-1-100}
\end{equation}
and together with the $\Gamma$ function properties allows one to sum all the $\Lambda_i$'s in a final compact result 
%%%%%%%%
\begin{equation}
\begin{aligned}
%&-\frac{2g_c^2C_{_{RP2}}}{(T^I {-} \bar T^I)(T^J {-} \bar T^J)}\bigg\{\mathscr R^I_{\alpha}\mathscr R^J_{\alpha}\frac{(\al s {+}1)\Gamma(-\al t/4 {+} 1)\Gamma(-\al u/4 {+} 2)}{(-\al u/4 {+} 1)\Gamma(\al s {+} 2)}%\\
%
% &\hspace{7.4cm}
% -\delta_{I,J}\frac{(\al s)\Gamma(-\al t/4)\Gamma(-\al u/4 {+} 2)}{(-\al u/4 {+} 1)\Gamma(\al s {+} 1)}\bigg\}\\\\
%
&\boxed{\mathcal A_{\alpha}({T^I},{\bar{T}^J})=-\frac{2g_c^2C_{_{RP_2}}}{(T^I {-} \bar T^I)(T^J {-} \bar T^J)}\bigg({-}\al\frac{u}{4}\bigg)\bigg({-}\al \frac{t}{4}\,\mathscr R^I_{\alpha}\mathscr R^J_{\alpha} - \al s\,\delta_{I,J}\bigg)\frac{\Gamma(-\al t/4)\Gamma(-\al u/4)}{\Gamma(-\al t/4 -\al u/4 {+}1)}}\,\,.
\end{aligned}
\label{eq:TT_result_-1-100RP2}
\end{equation}
%%%%%%%
%%%%%%%%%
We separately consider the diagonal and off-diagonal cases. Performing the expansion of $\Gamma$ as in~\eqref{eq:betaexp}, we get
\begin{itemize}
\item{$I = J$}%\footnote{$(\mathcal D^I_{\alpha})^2=1$}}
\begin{equation}
\begin{aligned}
&\frac{2g_c^2C_{_{RP_2}}}{(T^I {-} \bar T^I)^2}\Big(\al^2\frac{u^2}{16}\Big)\frac{\Gamma(-\al t/4)\Gamma(-\al u/4)}{\Gamma(-\al t/4 -\al u/4 {+}1)} =\\
%
%=\frac{2g_c^2C_{_{RP2}}}{(T^I {-} \bar T^I)^2}\Big(\al^2\frac{u^2}{16}\Big)\bigg\{\frac{1}{\al^2}\frac{16}{tu} - \zeta(2) + O(\al)\bigg\}\\
%
&\boxed{\mathcal A_{\alpha}({T^I},{\bar{T}^I}) =\frac{2g_c^2C_{_{RP_2}}}{(T^I {-} \bar T^I)^2}\bigg\{\frac{u}{t} +\al^2\frac{u^2}{16}\left(- \zeta(2) + O(\al)\right)\Big)\bigg\}}
\end{aligned}\label{eq:IJ_TT_-1-100RP2}
\end{equation}
\end{itemize}
\begin{itemize}
\item{$I\ne J$}
\begin{equation}
\begin{aligned}
&-\frac{2g_c^2C_{_{RP_2}}\mathscr R^I_{\alpha}\mathscr R^J_{\alpha}}{(T^I {-} \bar T^I)(T^J {-} \bar T^J)}\Big(\frac{\al^2\,ut}{16}\Big)\frac{\Gamma(-\al t/4)\Gamma(-\al u/4)}{\Gamma(-\al t/4 -\al u/4 {+}1)}
=\\
%=-\frac{2g_c^2C_{_{RP2}}\mathscr R^I_{\alpha}\mathscr R^J_{\alpha}}{(T^I {-} \bar T^I)(T^J {-} \bar T^J)}\Big(\frac{\al^2\,ut}{16}\Big)\bigg\{\frac{1}{\al^2}\frac{16}{tu} - \zeta(2) + O(\al)\bigg\}\\
%
&\boxed{\mathcal A_{\alpha}({T^I},\bar{T}^J)=-\frac{2g_c^2C_{_{RP_2}}\mathscr R^I_{\alpha}\mathscr R^J_{\alpha}}{(T^I {-} \bar T^I)(T^J {-} \bar T^J)}\bigg\{1 + \al^2\frac{ut}{16}\left(- \zeta(2) + O(\al)\right)\bigg\}}\,\,.
\end{aligned}\label{eq:InoJ_TT_-1-100RP2}
\end{equation}
\end{itemize}
%%%%%%%%%%%%%%%%%%%%%%%%%%%%%%%%%%%%%%%%%%
%%%%%%%%%%%%%%%%%%%%%%%%%%%%%%%%%%%%%%%%%%
%%%%%%%%%%%%%%%%%%%%%%%%%%%%%%%%%%%%%%%%%%
%%%%%%%%%%%%%%%%%%%%%%%%%%%%%%%%%%%%%%%%%%
%\subsection{${\cal A}_{\alpha}\left(T^{I},T^{J}\right)$}
%
\subsubsection*{$\boxed{\mathcal A_{\alpha}({T^I},{{T}^J})}$}
The second amplitude in this set involving the pair ($T^I,T^J$) massless K\"ahler moduli in the picture (-1-1;\,00) reads
\begin{equation}
\begin{aligned}
\mathcal A_{\alpha}({T^I},{{T}^J}) =&g_c^2C_{_{RP_2}}\int_{_{\abs{z}\leq 1}}\frac{d^2z_1d^2z_2}{V_{CKG}}\langle {\cal W}^{\otimes}_{_{T^I({-}1,{-}1)}}(E_1,k_1,z_1,\bar z_1){\cal W}^{\otimes}_{_{T^J(0,0)}}(E_2,k_2,z_2,\bar z_2)\rangle = \sum_{i=1}^4\Lambda_i \\
\end{aligned}\label{eq:RP2_T-T_-1-100}
\end{equation}
where each $\Lambda$'s sub-amplitude admits the same representation as before, i.e behaves like a single disk amplitude. We have for instance
\begin{equation}
\begin{aligned}
&\Lambda_1 =\frac{2g_c^2C_{_{RP_2}}}{\al(T^I - \bar T^I)(T^J - \bar T^J)}\int_{_{\abs{z}\leq 1}}\frac{d^2z_1d^2z_2}{V_{CKG}}\langle:\tilde{\Psi}^Ie^{-\phi}e^{ik_1X}(z_1)\bar\Psi^Ie^{-\bar\phi}e^{ik_1\bar X}(\bar z_1):\\&\hspace{5cm}:\Big(i\DP\tilde{ Z}^J + \frac{\al}{2}\big(k_2\psi\big)\tilde{\Psi}^J\Big)e^{ik_2X}(z_2)\Big(i\bar\DP\bar{ Z}^J + \frac{\al}{2}\big(k_2\bar\psi\big)\bar{\Psi}^J\Big)e^{ik_2\bar X}(\bar z_2):\rangle\\
&=\frac{2g_c^2C_{_{RP_2}}}{\al(T^I - \bar T^I)(T^J - \bar T^J)}\int_{_{\abs{z}\leq 1}}\frac{d^2z_1d^2z_2}{V_{CKG}}\langle:e^{-\phi}(z_1)e^{-\bar\phi}(\bar z_1):\rangle
\bigg({\cal M}^{(1)}_{_{\Lambda_1}} +{\cal M}^{(2)}_{_{\Lambda_1}} + {\cal M}^{(3)}_{_{\Lambda_1}} + {\cal M}^{(4)}_{_{\Lambda_1}}\bigg)\\
\end{aligned}
\label{eq:lambda1_T-T_-1-100}
\end{equation}
with ${\cal M}_{_{\Lambda_1}}^{(1)}$ given by
\begin{equation}
%{\cal M}_{_{\Lambda_1}}^{(1)}=
\langle:\tilde{\Psi}^Ie^{ik_1X}(z_1)\bar\Psi^Ie^{ik_1\bar X}(\bar z_1)::i\DP \tilde{Z}^Je^{ik_2X}(z_2)i\bar\DP\bar{Z}^Je^{ik_2\bar X}(\bar z_2):\rangle\,\,.
\end{equation} 
We collect all the $\Lambda$'s sub-amplitudes and $\mathcal M$'s terms in Appendix A.7 of the ref.\,\cite{AM:appendice}. $SU(2)$ invariance allows one to fix the vertex operators at \eqref{eq:fixpointsRP2} and with the specific c-ghost correlator given in \eqref{eq:c-ghostRP2},  \eqref{eq:RP2_T-T_-1-100} becomes
 \begin{equation}
\begin{aligned}
\Lambda_1 + \Lambda_4 = %&\,\frac{2g_c^2C_{_{RP2}}}{(T^I {-} \bar T^I)(T^J {-} \bar T^J)}\int_{-1}^{1}dy\,(1 + y^2)^{-\al s}(y^2)^{-\al \frac{t}{4}}\,4y\,\bigg\{\frac{(1 + \al s)\mathscr R^I_{\alpha}\mathscr R^J_{\alpha}}{(1 + y^2)^2} -\frac{\al s\,\,\delta_{I,J}(\mathscr R^I_{\alpha})^2}{(1 + y^2)}\bigg\}\\=
& \,\frac{2g_c^2C_{_{RP_2}}}{(T^I {-} \bar T^I)(T^J {-} \bar T^J)}\int_{0}^{1}dy^2\,(1 + y^2)^{-\al s}(y^2)^{-\al \frac{t}{4}}\bigg\{\frac{(1 + \al s)\mathscr R^I_{\alpha}\mathscr R^J_{\alpha}}{(1 + y^2)^2} -\frac{\al s\,\,\delta_{I,J}}{(1 + y^2)}\bigg\}\\
\Lambda_2 + \Lambda_3 =%&\,\frac{2g_c^2C_{_{RP_2}}}{(T^I {-} \bar T^I)(T^J {-} \bar T^J)}\int_{-1}^{1}dy\,(1 + y^2)^{-\al s}(y^2)^{-\al \frac{u}{4}}\,4y\,\bigg\{\frac{(1 + \al s)\mathscr R^I_{\alpha}\mathscr R^J_{\alpha}}{(1 + y^2)^2} -\frac{\al s\,\,\delta_{I,J}(\mathscr R^I_{\alpha})^2}{(1 + y^2)y^2}\bigg\}\\=
&\,\frac{2g_c^2C_{_{RP_2}}}{(T^I {-} \bar T^I)(T^J {-} \bar T^J)}\int_{0}^{1}dy^2\,(1 + y^2)^{-\al s}(y^2)^{-\al \frac{u}{4}}\bigg\{\frac{(1 + \al s)\mathscr R^I_{\alpha}\mathscr R^J_{\alpha}}{(1 + y^2)^2} -\frac{\al s\,\,\delta_{I,J}}{(1 + y^2)y^2}\bigg\}\,\,.
\end{aligned}\label{eq:lambda14-23Fix_TT_-1-1-00}
\end{equation}
In terms of the integral definition of Hypergeometric functions ${}_2F_1$ (see eq.\,\eqref{eq:HY21}) they can be represented by the expressions
\begin{equation}
\begin{aligned}
&\Lambda_1 + \Lambda_4 = \frac{2g_c^2C_{_{RP_2}}}{(T^I {-} \bar T^I)(T^J {-} \bar T^J)}\bigg\{\mathscr R^I_{\alpha}\mathscr R^J_{\alpha}\frac{(1 {+} \al s)\,{}_2F_1(\al s {+} 2, -\al t/4 {+} 1; -\al t/4 {+} 2; -1)}{(-\al t/4 {+} 1)}\\
& \hspace{5.5cm}- \delta_{I,J}\frac{\al s\,\,{}_2F_1(\al s {+} 1, -\al t/4; -\al t/4 {+} 1; -1)}{(-\al t/4 + 1)}\bigg\}\\
%%%%%%%%%%%
&\Lambda_2 {+} \Lambda_3 = \frac{2g_c^2C_{_{RP_2}}}{(T^I {-} \bar T^I)(T^J {-} \bar T^J)}\bigg\{\mathscr R^I_{\alpha}\mathscr R^J_{\alpha}\frac{(1 {+} \al s\big)\,{}_2F_1(\al s {+} 2, {-}\al u/4 {+} 1; {-}\al u/4 {+}2; {-}1)}{(-\al u {+} 1)} \\
&\hspace{5.5cm}- \delta_{I,J}\frac{\al s\,\,{}_2F_1(\al s {+} 1, {-}\al u/4 {+} 1; {-}\al u/4 {+} 2; {-}1)}{(-\al u/4)}\bigg\}
\end{aligned}
\label{eq:lambda14-23_res_TT_-1-100}
\end{equation}
that together with the identity for the Hypergeometric functions such as \eqref{eq:HYFident_TT_-1-100}, the $\Lambda$'s sub-amplitudes can be combined to give
\begin{equation}
\small{\begin{aligned}
&\frac{2g_c^2C_{_{RP_2}}}{(T^I {-} \bar T^I)(T^J {-} \bar T^J)}\bigg\{\mathscr R^I_{\alpha}\mathscr R^J_{\alpha}\frac{(\al s {+}1)\Gamma(-\al t/4 {+} 1)\Gamma(-\al u/4 {+} 2)}{(-\al u/4 {+} 1)\Gamma(\al s {+} 2)}%\\
%
% &\hspace{7.4cm}
 -\delta_{I,J}\frac{(\al s)\Gamma(-\al t/4 + 1)\Gamma(-\al u/4 {+} 1)}{(-\al u/4)\Gamma(\al s {+} 1)}\bigg\}
\end{aligned}}\label{eq:TT_result1_-1-100RP_2}
\end{equation}
yielding the final result
\begin{equation}
\small{\begin{aligned}
&\boxed{{\cal A}_{\alpha}\left(T^{I},T^{J}\right) = \frac{2g_c^2C_{_{RP_2}}}{(T^I - \bar T^I)(T^J - \bar T^J)}\bigg\{\mathscr R^I_{\alpha}\mathscr R^J_{\alpha}\,\al^2\frac{t\,u}{16} + \delta_{I,J}\,\al^2\frac{s\,t}{4}\bigg\}\frac{\Gamma(-\al t/4)\Gamma(-\al u/4)}{\Gamma(-\al t/4 - \al u/4 +1)}}\,\,.
\end{aligned}}\label{eq:TT_result2_-1-100RP_2}
\end{equation}
For the diagonal and off-diagonal case, one finds 
\begin{itemize}
\item{$I=J$}
\begin{equation}
\begin{aligned}
&\frac{2g_c^2C_{_{D_2}}}{(T^I - \bar T^I)^2}\left(-\al^2\frac{t^2}{16}\right)\frac{\Gamma(-\al t/4)\Gamma(-\al u/4)}{\Gamma(-\al t/4 - \al u/4 +1)}\\\\
&\boxed{{\cal A}_{\alpha}\left(T^{I},T^{I}\right) = -\frac{2g_c^2C_{_{RP_2}}}{(T^I - \bar T^I)^2}\bigg\{\frac{t}{u} + \al^2\frac{t^2}{16}\left(- \zeta(2) + O(\al)\right)\bigg\}}
\end{aligned}
\end{equation}
\end{itemize}
\begin{itemize}
\item{$I\ne J$}
\begin{equation}
\begin{aligned}
&\frac{2g_c^2C_{_{D_2}}}{(T^I - \bar T^I)(T^J - \bar T^J)}\left(\al^2\frac{tu}{16}\right)\frac{\Gamma(-\al t/4)\Gamma(-\al u/4)}{\Gamma(-\al t/4 - \al u/4 +1)}\\\\
&\boxed{{\cal A}_{\alpha}\left(T^{I},T^{J}\right) = \frac{2g_c^2C_{_{RP_2}}\mathscr R^I_{\alpha}\mathscr R^J_{\alpha}}{(T^I - \bar T^I)(T^J - \bar T^J)}\bigg\{1 + \al^2\,\frac{t\,u}{16}\left(- \zeta(2) + O(\al)\right)\bigg\}}\,\,.
\end{aligned}
\end{equation}
\end{itemize}
In order to take the string scattering amplitude involving two NS-NS K\"ahler moduli $T^I_2$, as in the disk $D_2$ cases, one needs to consider the definition \eqref{eq:VT22}
and the corresponding string scattering amplitude results for both the cases $I=J$ and $I\ne J$, come from the following linear combination
\begin{equation}
\mathcal A_{\alpha}(T_2^I,T^J_2) = \frac{1}{4}\left({\cal A}_{\alpha}\left(\bar T^{I}, {T}^{J}\right) -{\cal A}_{\alpha}\left(T^{I},{T}^{J}\right) - {\cal A}_{\alpha}\left(\bar T^{I},\bar {T}^{J}\right) + {\cal A}_{\alpha}\left(T^{I},\bar{T}^{J}\right)\right)
\end{equation}
%in particular the $I=J$ and $I\ne$ are equal to 
\begin{itemize}
\item{$I = J$}
\end{itemize}
\begin{equation}
\begin{aligned}
&\boxed{\mathcal A_{\alpha}(T_2^I,T^I_2) = \frac{g_c^2C_{_{RP_2}}}{(T^I - \bar T^I)^2}\bigg\{\frac{u}{t} +\frac{t}{u} + \frac{\al^2}{16}\left(u^2 + t^2\right)\left(- \zeta(2) + O(\al)\right)\bigg\}}\\&
\end{aligned}
\label{eq:TTrp21}
\end{equation}
\begin{itemize}
\item{$I\ne J$} 
\end{itemize}
\begin{equation}
\boxed{\mathcal A_{\alpha}(T_2^I,T^J_2) = -\frac{2g_c^2C_{_{RP_2}}\mathscr R^I_{\alpha}\mathscr R^J_{\alpha}}{(T^I - \bar T^I)(T^J - \bar T^J)}\bigg\{1 + \al^2\,\frac{t\,u}{16}\left(- \zeta(2) + O(\al)\right)\bigg\}}
\label{eq:TTrp2}
\end{equation}
where on $\Omega_9$-planes $\alpha = 9$ and $\mathscr R^I_9 = +1$, while on $\Omega5_I$-planes $\alpha = 5_I$ one has $\mathscr R^I_{5_I} = +1 $ and $\mathscr R^J_{5_I} = -1 $. As expected, only closed string pole channels are allowed. Moreover the results of eq. \eqref{eq:TTrp21} and \eqref{eq:TTrp2} tell us that the contribution of $RP_2$ does not spoil the structure of the K\"ahler potential \eqref{eq:kahlerP} but confirms it. 
%%%%%%%%%%%%%%%%%%%%%%%%%%%%%%%%%%%%%%%%%%
%%%%%%%%%%%%%%%%%%%%%%%%%%%%%%%%%%%%%%%%%%
%%%%%%%%%%%%%%%%%%%%%%%%%%%%%%%%%%%%%%%%%%
\subsection{$\mathcal A^{RP_2}_{\alpha}\left(U^{I},\bar{U}^{J}\right)$ and $\mathcal A^{RP_2}_{\alpha}\left(U^{I},{U}^{J}\right)$}
The second set of amplitudes in \eqref{eq:AMPLRP2_New} involves two massless complex structures $U$ where again the scattering amplitudes are taken in the picture $({-}1,{-}1;0,0)$ and the specific vertex operators can be found in Section \ref{Toolkit}.
\subsubsection*{$\boxed{\mathcal A_{\alpha}(U^I,\bar{U}^J)}$}
The scattering amplitude for the pair ($U^I,\bar U^J$) is 
\begin{equation}
\begin{aligned}
\mathcal A_{\alpha}(U^I,\bar{U}^J)=&g_c^2C_{_{RP_2}}\int_{_{\abs{z}\leq 1}}\frac{d^2z_1d^2z_2}{V_{CKG}}\langle {\cal W}^{\otimes}_{U^I({-}1,{-}1)}(E_1,k_1,z_1,\bar z_1){\cal W}^{\otimes}_{\bar U^J(0,0)}(E_2,k_2,z_2,\bar z_2)\rangle = \sum_{i=1}^4\Lambda_i \\
\end{aligned}\label{eq:RP2_UU_-1-100}
\end{equation}
where one finds, for instance, for the $\Lambda_1$ sub-amplitude 
\begin{equation}
\begin{aligned}
\Lambda_1 =&-\frac{2g_c^2C_{_{RP_2}}}{\al(U^I - \bar U^I)(U^J - \bar U^J)}\int_{_{\abs{z}\leq 1}}\frac{d^2z_1d^2z_2}{V_{CKG}}\langle:\Psi^Ie^{-\phi}e^{ik_1X}(z_1)\bar\Psi^Ie^{-\bar\phi}e^{ik_1X}(\bar z_1):\\&\hspace{4cm}:\Big(i\DP\tilde{Z}^J + \frac{\al}{2}\big(k_2\psi\big)\tilde{\Psi}^J\Big)e^{ik_2X}(z_2)\Big(i\bar\DP\bar{\tilde Z}^J+ \frac{\al}{2}\big(k_2\bar\psi\big)\bar{\tilde\Psi}^J\Big)e^{ik_2 \bar X}(\bar z_2):\rangle\\
=&-\frac{2g_c^2C_{_{RP_2}}}{\al(U^I - \bar U^I)(U^J - \bar U^J)}\int_{_{\abs{z}\leq 1}}\frac{d^2z_1d^2z_2}{V_{CKG}}\langle:e^{-\phi}(z_1)e^{-\bar\phi}(\bar z_1):\rangle\bigg({\cal M}^{(1)}_{_{\Lambda_1}} +{\cal M}^{(2)}_{_{\Lambda_1}} + {\cal M}^{(3)}_{_{\Lambda_1}} + {\cal M}^{(4)}_{_{\Lambda_1}}\bigg)\\
\end{aligned}\label{eq:lambda1_UU_-1-100}
\end{equation}
%%%%%%%%%%%%%%%%%%%%%%%%
and correlators as ${\cal M}^{(1)}_{_{\Lambda_1}}$
\begin{equation}
\begin{aligned}
%{\cal M}^{(1)}_{_{\Lambda_1}}=
&\langle:\Psi^Ie^{ik_1X}(z_1)\bar\Psi^Ie^{ik_1 \bar X}(\bar z_1)::i\DP\tilde{Z}^Je^{ik_2X}(z_2)i\bar\DP\bar{\tilde Z}^{J}e^{ik_2 \bar X}(\bar z_2):\rangle%=\,0\\
\end{aligned}\label{eq:lambda1_UU_1}
\end{equation}
can be calculated using \eqref{eq:correlator10D_RP2}, \eqref{eq:boscomSfera}, \eqref{eq:fermcomSfera}, \eqref{eq:boscomRp} and \eqref{eq:fermcomRp}. The details, together with the other $\Lambda$'s and the corresponding ${\cal M}_{\Lambda}$ correlators are reported in Appendix A.7 of the ref.\,\cite{AM:appendice}. As for the $T\bar T$ case, one takes the sum of  $\Lambda_i$'s sub-amplitudes having the same Koba-Nielsen factor, obtaining
\begin{equation*}
\begin{aligned}
\Lambda_1 + \Lambda_4 = &-\frac{2g_c^2C_{_{RP_2}}}{(U^I {-} \bar U^I)(U^J {-} \bar U^J)}\int_{_{\abs{z}\leq 1}}\frac{d^2z_1d^2z_2}{V_{CKG}}\bigg(\frac{(1 + \abs{z_1}^2)(1 +\abs{z_2}^2)}{\abs{1 + z_1\bar z_2}^2}\bigg)^{-\al s}\bigg(\frac{\abs{z_1 - z_2}^2}{\abs{1 +z_1\bar z_2}^2}\bigg)^{-\al \frac{t}{4}}\\&\hspace{2cm}\bigg(-\frac{\al s\,({\mathscr R}^I_{\alpha})^2\delta_{I,J}}{(1 + \abs{z_1}^2)(1 + \abs{z_2}^2)\abs{1 + z_1\bar z_2}^2} - \frac{\al s\,\delta_{I,J}}{(1 + \abs{z_1}^2)(1 + \abs{z_2}^2)\abs{z_1 -z_2}^2}\bigg)\\
\end{aligned}\label{eq:lambda14_UU_-1-100}
\end{equation*}
\begin{equation}
\begin{aligned}
\Lambda_2 + \Lambda_3 =& -\frac{2g_c^2C_{_{RP_2}}}{(U^I - \bar U^I)(U^J - \bar U^J)}\int_{_{\abs{z}\leq 1}}\frac{d^2z_1d^2z_2}{V_{CKG}}\bigg(\frac{(1 + \abs{z_1}^2)(1 +\abs{z_2}^2)}{\abs{1 + z_1\bar z_2}^2}\bigg)^{-\frac{\al}{2}\frac{s}{2}}\bigg(\frac{\abs{z_1 - z_2}^2}{\abs{1 +z_1\bar z_2}^2}\bigg)^{-\al \frac{u}{4}}\\&\hspace{2.5cm}\bigg(-\frac{\al s\,\delta_{I,J}}{(1 + \abs{z_1}^2)(1 + \abs{z_2}^2)\abs{z_1 - z_2}^2} -\frac{\al s\,({\mathscr R}^I_{\alpha})^2\delta_{I,J}}{(1 + \abs{z_1}^2)(1 +\abs{z_2}^2)\abs{1+z_1\bar z_2}^2}\bigg)\,\,.
\end{aligned}\label{eq:lambda23_UU_-1-100}
\end{equation}
As always the positions of vertices are fixed at \eqref{eq:fixpointsRP2} using the SU(2) symmetry. In this way the $\Lambda_i$'s sub-amplitudes with the right insertion of c-ghost determinant \eqref{eq:c-ghostRP2} read
\begin{equation}
\begin{aligned}
\Lambda_1 + \Lambda_4 =% &-g_s\frac{2\mathcal{D}^I\mathcal{D}^J}{\al(U^I {-} \bar U^I)(U^J {-} \bar U^J)}\int_{-1}^{1}dy\,(1 + y^2)^{-\al s}(y^2)^{-\al \frac{t}{4}}\underbrace{4y}_{_\text{c-ghost}}\bigg\{-\frac{\al^2}{2}s\frac{2\delta^{IJ}}{(1 +y^2)} -\frac{\al^2}{2}s\frac{2\delta^{IJ}}{(1 + y^2)y^2}\bigg\}\\
&-\frac{2g_c^2C_{_{RP_2}}\,\,\delta_{I,J}}{(U^I - \bar U^I)(U^J - \bar U^J)}(\al s)\int_{0}^1dy^2(1 + y^2)^{-\al s}(y^2)^{-\al \frac{t}{4}}\bigg\{-\frac{1}{(1 + y^2)} - \frac{1}{(1 + y^2)y^2}\bigg\}\\
%%%%
\Lambda_2 + \Lambda_3 =%&-g_s\frac{2\mathcal{D}^I\mathcal{D}^J}{\al(U^I {-} \bar U^I)(U^J {-} \bar U^J)}\int_{-1}^{1}dy\,(1 + y^2)^{-\al s}(y^2)^{-\al u/4}\underbrace{4y}_{_\text{c-ghost}}\bigg\{-\frac{\al^2}{2}s\frac{2\delta^{IJ}}{(1 + y^2)y^2} - \frac{\al^2}{2}s\frac{2\delta^{IJ}}{(1 + y^2)}\bigg\}\\
&-\frac{2g_c^2C_{_{RP_2}}\,\,\delta_{I,J}}{(U^I {-} \bar U^I)(U^J {-} \bar U^J)}(\al s)\int_0^1dy^2\,(1 + y^2)^{-\al s}(y^2)^{-\al u/4}\bigg\{-\frac{1}{(1 + y^2)y^2} - \frac{1}{(1 + y^2)}\bigg\}
\end{aligned}
\label{eq:lambda14-23_UU_fix}
\end{equation}
%%%%%%%%%%%
and using \eqref{eq:HY21} for the integral definition of ${}_2F_1$ function one can write 
\begin{equation}
\begin{aligned}
\Lambda_1 + \Lambda_4=&-\frac{2g_c^2C_{_{RP_2}}\,\,\delta_{I,J}}{(U^I - \bar U^I)(U^J - \bar U^J)}(\al s)\bigg\{-\frac{{}_2F_1(\al s {+} 1, -\al t/4 {+}1; -\al t/4 {+}2; -1)}{(-\al t/4 {+}1)}\\&\hspace{5.2cm} - \frac{{}_2F_1(\al s {+}1, -\al t/4; -\al t/4 {+}1; -1)}{(-\al t/4)}\bigg\}\\
%%%%
\Lambda_2 + \Lambda_3=&-\frac{2g_c^2C_{_{RP_2}}\,\,\delta_{I,J}}{(U^I {-} \bar U^I)(U^J {-} \bar U^J)}(\al s)\bigg\{-\frac{{}_2F_1(\al s {+}1, -\al u/4; -\al u/4 {+}1; -1)}{(-\al u/4)}\\&\hspace{4.8cm} -\frac{{}_2F_1(\al s {+}1, -\al u/4 {+}1; -\al u/4 {+} 2; -1)}{(-\al u/4 {+}1)}\bigg\}\,\,.
\end{aligned}\label{eq:lambda_14-23-UU-HYp}
\end{equation}
With these results, summing all the $\Lambda_i$ using the ${}_2F_1$-identity \eqref{HYPid}, the final expression is 
%\begin{equation}
%\small{
%\begin{aligned}
%{}_2F_1(\al s {+}1, -\al u/4; -\al u/4 {+} 1; -1) =& -\frac{(-\al u/4)}{(-\al t/4 {+}1)}{}_2F_1(\al s {+} 1, -\al t/4 {+} 1; -\al t/4 {+} 2;-1)
%\\& 
%+\frac{\Gamma(-\al t/4 {+} 1)\Gamma(-\al u/4 {+}1)}{\Gamma(\al s {+}1)}
%\\\\
%{}_2F_1(\al s {+}1, -\al u/4 {+} 1; -\al u/4 {+}2;-1)=& -\frac{(-\al u/4 {+}1)}{(-\al t/4)}{}_2F_1(\al s {+}1, -\al t/4; -\al t/4 {+} 1; -1)
 %\\&
% + \frac{\Gamma(-\al t/4)\Gamma(-\al u/4 {+} 2)}{\Gamma(\al s {+} 1)}
%\end{aligned}\label{eq:HYI_UU}}
%\end{equation}
%
\begin{equation}
\begin{aligned}
%& \sum_{i=1}^{4}\Lambda_i =-\frac{8g_c^2C_{_{RP2}}\delta_{I,J}}{(U^I {-} \bar U^I)(U^J {-} \bar U^J)}(\al s)\bigg\{-\frac{\Gamma(-\al t/4 {+} 1)\Gamma(-\al u/4 {+}1)}{(-\al u/4)\Gamma(\al s {+}1)} -  \frac{\Gamma(-\al t/4)\Gamma(-\al u/4 {+} 2)}{(-\al u/4 {+}1)\Gamma(\al s {+} 1)}\bigg\}\\\\
%
&\boxed{\mathcal A_{\alpha}(U^I,\bar{U}^J)=\frac{2g_c^2C_{_{RP_2}}\,\,\delta_{I,J}}{(U^I {-} \bar U^I)(U^J {-} \bar U^J)}(\al s)\bigg(-\al \frac{t}{4} -\al \frac{u}{4}\bigg)\frac{\Gamma(-\al t/4)\Gamma(-\al u/4)}{\Gamma(-\al t/4 - \al u/4 {+} 1)}}
\end{aligned}\label{eq:lambdaSum_UU_-1-100}
\end{equation}
with $\alpha\in\{9,5_I\}$. For the diagonal and off-diagonal case one thus finds
\begin{itemize}
\item{$I=J$}
\begin{equation}
\begin{aligned}
%&\frac{2g_c^2C_{_{RP_2}}}{(U^I {-} \bar U^I)^2}(\al^2 s^2)\bigg\{ \frac{1}{\al^2}\frac{16}{tu} - \zeta(2) + O(\al)\bigg\}
%= \frac{8g_c^2C_{_{RP2}}}{(U^I {-} \bar U^I)^2}\bigg(\al^2 \frac{t^2}{16} + \al^2 \frac{u^2}{16} + \al^2\frac{tu}{8}\bigg)\bigg\{ \frac{1}{\al^2}\frac{16}{tu} - \frac{\pi^2}{6} + O((k^2)^3)\bigg\}
%
&\boxed{\mathcal A_{\alpha}(U^I,\bar{U}^I)=\frac{2g_c^2C_{_{RP_2}}}{(U^I {-} \bar U^I)^2}\bigg\{\frac{t}{u} + \frac{u}{t} + 2 +\al^2s^2\left(- \zeta(2) + O(\al)\right)\bigg\}}
\end{aligned}
\label{eq:UU_I=J_res_-1-100}
\end{equation}
\end{itemize}
%%%%%%
\begin{itemize}
\item{$I\ne J$}
\begin{equation}
\boxed{\mathcal A_{\alpha}(U^I,\bar{U}^J) = 0}\,\,.
\label{eq:UU_I/J_res_-1-100}
\end{equation}
\end{itemize}
%%%%%%%%%%%%%%%%%%%%%%%%%%%%%%%%%%%%%%%%%%%%
%%%%%%%%%%%%%%%%%%%%%%%%%%%%%%%%%%%%%%%%%%%%
%%%%%%%%%%%%%%%%%%%%%%%%%%%%%%%%%%%%%%%%%%%%
%\subsection{${\cal A}_{\alpha}\left(U^{I},\bar{U}^{J}\right)$}
%
\subsubsection*{$\boxed{\mathcal A_{\alpha}(U^I,{U}^J)}$}
In this second set of amplitudes that involves the complex structure moduli $U$, the second amplitude, associated to the pair ($U^I,U^J$) is 
\begin{equation}
\mathcal A_{\alpha}(U^I,{U}^J)=g_c^2C_{_{RP_2}}\int_{_{\abs{z}\leq 1}}\frac{d^2z_1d^2z_2}{V_{CKG}}\langle {\cal W}^{\otimes}_{U^I({-}1,{-}1)}(E_1,k_1,z_1,\bar z_1){\cal W}^{\otimes}_{\bar U^J(0,0)}(E_2,k_2,z_2,\bar z_2)\rangle =\sum_{i=1}^4 \Lambda_i \,\,.
\end{equation}
Since the intermediates steps are equal to that of the ($U^I,\bar U^J$) case, here we give only an example of one type of $\Lambda$'s sub-amplitudes and of $\mathcal M$'s contraction terms, i.e.
\begin{equation}
\begin{aligned}
\Lambda_1 =&\frac{2g_c^2C_{_{RP_2}}}{\al(U^I - \bar U^I)(U^J - \bar U^J)}\int_{_{\abs{z}\leq 1}}\frac{d^2z_1d^2z_2}{V_{CKG}}\langle:\Psi^Ie^{-\phi}e^{ik_1X}(z_1)\bar\Psi^Ie^{-\bar\phi}e^{ik_1X}(\bar z_1):\\&\hspace{4cm}:\Big(i\DP{Z}^J + \frac{\al}{2}\big(k_2\psi\big){\Psi}^J\Big)e^{ik_2X}(z_2)\Big(i\bar\DP\bar{Z}^J+ \frac{\al}{2}\big(k_2\bar\psi\big)\bar{\Psi}^J\Big)e^{ik_2 \bar X}(\bar z_2):\rangle\\
=&\frac{2g_c^2C_{_{RP_2}}}{\al(U^I - \bar U^I)(U^J - \bar U^J)}\int_{_{\abs{z}\leq 1}}\frac{d^2z_1d^2z_2}{V_{CKG}}\langle:e^{-\phi}(z_1)e^{-\bar\phi}(\bar z_1):\rangle\bigg({\cal M}^{(1)}_{_{\Lambda_1}} +{\cal M}^{(2)}_{_{\Lambda_1}} + {\cal M}^{(3)}_{_{\Lambda_1}} + {\cal M}^{(4)}_{_{\Lambda_1}}\bigg)
\end{aligned}\label{eq:lambda1_UU_-1-100}
\end{equation}
and
\begin{equation}
{\cal M}^{(1)}_{_{\Lambda_1}}=\langle:\Psi^Ie^{ik_1X}(z_1)\bar\Psi^Ie^{ik_1 \bar X}(\bar z_1)::i\DP{Z}^Je^{ik_2X}(z_2)i\bar\DP\bar{Z}^{J}e^{ik_2 \bar X}(\bar z_2):\rangle
\end{equation}
In contrast to the previous case, all the $\mathcal M$'s terms inside the $\Lambda$'s sub-amplitudes vanish being equal to zero some two-point functions in \eqref{eq:boscomSfera}, \eqref{eq:fermcomSfera}, \eqref{eq:boscomRp} as well as \eqref{eq:fermcomRp}. See Appendix A.7 of ref.\,\cite{AM:appendice} for details. Therefore the result of the amplitudes with $\alpha\in\{9,5_I\}$ in both cases $I=J$ and $I\ne J$ vanishes
\begin{equation}
\boxed{{\cal A}_{\alpha}\left(U^{I},U^{J}\right) =\,0}\,\,.
\label{eq:UUrp2}
\end{equation}
The results \eqref{eq:UU_I=J_res_-1-100},\,\eqref{eq:UU_I/J_res_-1-100} and\,\eqref{eq:UUrp2} confirm that the K\"ahler potential maintains its tree-level form \eqref{eq:kahlerP}.
%%%%%%%%%%%%%%%%%%%%%%%%%%%%%%%%%%%%%%%%%%%%
%%%%%%%%%%%%%%%%%%%%%%%%%%%%%%%%%%%%%%%%%%%%
%%%%%%%%%%%%%%%%%%%%%%%%%%%%%%%%%%%%%%%%%%%%
%%%%%%%%%%%%%%%%%%%%%%%%%%%%%%%%%%%%%%%%%%%%
\subsection{$\mathcal A^{RP_2}_{\alpha}\left(T^I,\bar{U}^J \right)$ and $\mathcal A^{RP_2}_{\alpha}\left(T^I,{U}^J \right)$}
The last set of amplitudes in \eqref{eq:AMPLRP2_New} involves one K\"ahler modulus $T^I$ and one complex structure modulus $U^I$. As before, the specific vertex operators are taken in the picture $(-1-1;\,00)$ and are explicitly written in Section \ref{Toolkit}. We anticipate that this amplitude, as in the disk case, vanishes due to the particular combination of two-point functions for the compactified fields that are involved. Nevertheless for completeness we sketch here just a few of the intermediate steps. For the full details of the calculation the reader can have a look at ref.\,\cite{AM:appendice}. 

\subsubsection*{$\boxed{\mathcal A_{\alpha}(T^I,\bar{U}^J)}$}
The first amplitude that we consider is 
\begin{equation}
\begin{aligned}
\mathcal A_{\alpha}(T^I,\bar{U}^J)=&g_c^2C_{_{RP_2}}\int_{_{\abs{z}\leq 1}}\frac{d^2z_1d^2z_2}{V_{CKG}}\langle {\cal W}^{\otimes}_{T^I({-}1,{-}1)}(E_1,k_1,z_1,\bar z_1){\cal W}^{\otimes}_{\bar U^J(0,0)}(E_2,k_2,z_2,\bar z_2)\rangle =\sum_{i=1}^4\Lambda_i \end{aligned}\label{eq:RP2_TU_-1-100}
\end{equation}
where one finds, for instance, for the sub-amplitude $\Lambda_1$ the expression
\begin{equation}
\begin{aligned}
\Lambda_1 =&\frac{2g_c^2C_{_{RP_2}}}{\al(T^I - \bar T^I)(U^J - \bar U^J)}\int_{_{\abs{z}\leq 1}}\frac{d^2z_1d^2z_2}{V_{CKG}}\langle:\tilde{\Psi}^Ie^{-\phi}e^{ik_1X}(z_1)\bar\Psi^Ie^{-\bar\phi}e^{ik_1\bar X}(\bar z_1):\\&\hspace{4cm}:\Big(i\DP\tilde{Z}^J + \frac{\al}{2}\big(k_2\psi\big)\tilde{\Psi}^J\Big)e^{ik_2X}(z_2)\Big(i\bar\DP\bar{\tilde Z}^J + \frac{\al}{2}\big(k_2\bar\psi\big)\bar{\tilde\Psi}^J\Big)e^{ik_2 \bar X}(\bar z_2):\rangle\\
=&\frac{2g_c^2C_{_{RP_2}}}{\al(T^I - \bar T^I)(U^J - \bar U^J)}\int_{_{\abs{z}\leq 1}}\frac{d^2z_1d^2z_2}{V_{CKG}}\langle:e^{-\phi}(z_1)::e^{-\bar\phi}(\bar z_1):\rangle\bigg({\cal M}^{(1)}_{_{\Lambda_1}} +{\cal M}^{(2)}_{_{\Lambda_1}} + {\cal M}^{(3)}_{_{\Lambda_1}} + {\cal M}^{(4)}_{_{\Lambda_1}}\bigg)
\end{aligned}\label{eq:lambda1_TU_-1-100}
\end{equation}
%%%%%%%%%%%%%%%%%%%%%%%%%%%%%%%LAMBDA1
while for ${\cal M}^{(1)}_{\Lambda_1}$ one has
\begin{equation}
\begin{aligned}
%{\cal M}^{(1)}_{_{\Lambda_1}}=&
\langle:\tilde{\Psi}^Ie^{ik_1X}(z_1)\bar\Psi^Ie^{ik_1\bar X}(\bar z_1)::i\DP\tilde{Z}^Je^{ik_2X}(z_2)i\bar\DP\bar{\tilde Z}^Je^{ik_2 \bar X}(\bar z_2):\rangle\,\,.
\end{aligned}\label{lambda1_TU_1234}
\end{equation}
Considering the correlators  for the uncompactified fields \eqref{eq:correlator10D_RP2} and \eqref{eq:boscomSfera}, \eqref{eq:fermcomSfera}, \eqref{eq:boscomRp}, \eqref{eq:fermcomRp} for the compactified fields, one is able to calculate all the ${\cal M}_{\Lambda}^{(i)}$'s terms inside the $\Lambda$'s and their explicit expression can be found in Appendix A.7 of ref.\,\cite{AM:appendice}. The final result, as anticipated, is zero
%thus, due to the balance of the $(Q,\bar Q)$ charge and the vanishing of particular two point correlators that one mets performing the calculation, the amplitude is  
%
\begin{equation}
\boxed{\mathcal A_{\alpha}(T^I,\bar{U}^J) = 0\,\,\,(T\leftrightarrow U)}
\label{eq:ATUrp2}
\end{equation}
for $I=J$ and $I\ne J$ with $\alpha\in\{9,5_I\}$, as in \eqref{eq:ATU}.

\subsubsection*{$\boxed{\mathcal A_{\alpha}(T^I,{U}^J)}$}
In view of these results one might suspect that also the second amplitude in this set vanishes. Indeed, one can verify that the scattering amplitude for the pair ($T,U$), namely  
\begin{equation}
\mathcal A_{\alpha}(T^I,{U}^J)=g_c^2C_{_{RP_2}}\int_{_{\abs{z}\leq 1}}\frac{d^2z_1d^2z_2}{V_{CKG}}\langle {\cal W}^{\otimes}_{T^I({-}1,{-}1)}(E_1,k_1,z_1,\bar z_1){\cal W}^{\otimes}_{U^J(0,0)}(E_2,k_2,z_2,\bar z_2)\rangle =\sum_{i=1}^4 \Lambda_i
\end{equation}
with $\alpha\in\{9,5_I\}$ gives, for the same reason as said before, a vanishing result for both the case $I=J$ and $I \ne J$ (see Appendix A.7 of ref.\, \cite{AM:appendice}).
\begin{equation}
\boxed{\mathcal A_{\alpha}(T^I,{U}^J) = 0\,\,\,(T\leftrightarrow U)}\,\,.
\label{eq:TUrp2}
\end{equation}
The conclusion of this section is that direct string scattering amplitudes calculations prove that the form of the tree-level K\"ahler potential~\eqref{eq:kahlerP} as well as the associated tree-level K\"ahler metric components needed to write down the kinetic terms for the closed untwisted moduli in the LEEA of the Type IIB orientifold on $\T^6/{\mathbb Z}_2\times{\mathbb Z}_2$ (and \textit{mutatis mutandis} for all the models that have a similar moduli space), is confirmed. 
%%%%%%%%%%%%%%%%%%%%%%%%%%%%%%%%%%%%%%%%%%%%%%%%%%%%%%%%%%%%%%%%%%%%%%%%%%%%%%%%%%%%%%%%%%%%%%%%%%%%%%%%%%%%%%%%%%%%%%%%%%%%%%%%%%%%%%%%%%%%%%%%%%%%%%%%%%%%%%%%%%%%%%%%%%%%%%%%%%%%%%%%%%%%%%%%%%%%%%%%%%%%%%%%%%%%%%%%%%%%%%%%%%%%%%%%%%%%%%%%%%%%%%%%%%%%%%%%%%%%%%%%%%%%%%%%%%%%%%%%%%%%%%%%%%%%%%%%%%%%%%%%%%%%%%%%%%%%%%%%%%%%%%%%%%%%%
\section{Adding $\al^2 \bm R^2$ to LEEA}
\label{eq:Rhigh}\label{sec5}
In this section we want to discuss which kind of terms can be produced if, at the $\al^2$-order in the high derivative expansion of the LEEA, $\bm R^2$ terms like the contraction of two Ricci tensors $R_{\lambda\sigma}R^{\lambda\sigma}$, of Riemann tensors $R_{\lambda\sigma\alpha\beta}R^{\lambda\sigma\alpha\beta}$ and the square of scalar curvature $R^2$ are included. In particular, the question we want to answer is whether this kind of terms are reproduced by the string scattering amplitudes considered in this paper.\footnote{By $\bm R^2$ we generically mean combinations of Riemann, Ricci and scalar curvature that one can construct at this order.}In $g_s$-expansion at sphere level, terms of this type are absent in the action of Type II theories. In Type I theory terms at sphere level like $C\,e^{-2\Phi}{\bm R}^2$ are absent because they are related by the S-duality to terms like $C'\,e^{-\Phi}{\bm R}^2$ in the Heterotic theories, which are clearly absent \cite{Polchinski:1998rr,Polchinski:1995df,Tseytlin:1995bi}.\footnote{In Heterotic case, certain specific dilaton couplings are absent.} %Unoriented theory like Type I contains extended objects as $D$-branes and $\Omega$-planes, and the corresponding actions at tree-level can be derived from the $g_s$-expansion at disk and projective plane level respectively. 
Terms like $\bm R^2$ can be added to the tree-level actions for $D$-branes and $\Omega$-planes, because they are supported by S-duality relation \cite{Tseytlin:1995bi}. More precisely, terms of the form $C\,e^{-\Phi}\bm R^2$ in the $g_s$-expansion correspond to a tree-level string scattering amplitude on the disk or projective plane that, under S-duality, are mapped to a sphere tree-level terms $C'\,e^{-2\Phi}\bm R^2$ allowed in Heterotic $g_s$-expansion \cite{Tseytlin:1995bi,Bachas:1999um,Fotopoulos:2001pt,Fotopoulos:2002wy}.
\\ 
\\
Therefore the Dirac-Born-Infeld (DBI) actions for the $D$-branes and $\Omega$-planes at $\al^2$-order contain the terms \cite{Tseytlin:1995bi,Bachas:1999um,Fotopoulos:2001pt,Fotopoulos:2002wy,Garousi:2006zh}
\begin{equation}
{\cal S}^{_{(D_P,\Omega_P)}}_{DBI}= \al^2\tau_{_{({D_P,\Omega_P})}}\int\,d^{P+1}\xi\,e^{-\Phi}\sqrt{-g}\left\{aR_{\lambda\sigma\alpha\beta}R^{\lambda\sigma\alpha\beta} +bR_{\sigma\beta}R^{\sigma\beta} + cR^2 \cdots \right\}
\label{eq:actionDO}
\end{equation}
where $\xi^{\beta}$ are the (intrinsic) $D$-brane and $\Omega$-plane worldvolume coordinates, $g_{\alpha\beta}$ is the pull-back of the ten-dimensional metric $g_{MN}$ to the worldvolume, $g_{\alpha\beta} = \DP_{\alpha}X^{M}(\xi)\DP_{\beta}X^{N}(\xi)g_{MN}$ with indices $\alpha,\beta$ labelling the directions tangent to the $D$-brane and $\Omega$-plane and finally $\tau$ is a constant which includes the tension of $D$-brane and $\Omega$-plane, respectively, plus other constants. The DBI actions of eq. \eqref{eq:actionDO}, are all in static gauge, in the sense that the worldvolume coordinates  $\xi^{\beta}$ coincide with the string coordinate $X^{M}$ in the $p+1$-directions. Not all the terms in \eqref{eq:actionDO} are the pull-back to the worldvolume of the corresponding bulk terms. Only the (Riemann)$^2$-term  is. As explained in \cite{Bachas:1999um}, at linearised level around flat space, the vanishing (in the vacuum) of bulk Ricci tensor gives on $D$-brane and $\Omega$-plane three independent equations 
\begin{equation}
\begin{aligned}
&R^{L}_{\,\,MLN} = 0\qquad \text{with}\qquad L\in\{\lambda = 0,\dots,p;\,l= p{+}1,\dots,9\}\\
&R^{l}_{\,\,\mu l\nu} = -R^{\lambda}_{\,\,\mu\lambda \nu}\equiv R_{\mu\nu}\,\,;\qquad R^{l}_{\,\,\mu ln} = -R^{\lambda}_{\,\,\mu\lambda n}\equiv R_{\mu n}\,\,;
\qquad R^{l}_{\,\,mln} = -R^{\lambda}_{\,\,m\lambda n}\equiv R_{mn}\,\,.
\end{aligned}
\end{equation}
In order to build (Ricci)$^2$ terms for $D$-brane and $\Omega$-plane action, one can use only the three linearly independent Ricci tensors $R_{\mu\nu}, R_{\mu n}, R_{mn}$. The scalar curvature $R$ can be obtained from $R_{\mu\nu}$ and $R_{mn}$ appropriately contracting indices. The dots in \eqref{eq:actionDO} mean that other terms with tensor components along the orthogonal directions to the worldvolume (for the $D$-brane case only) can enter in general the action \cite{Tseytlin:1995bi,Bachas:1999um,Fotopoulos:2001pt,Fotopoulos:2002wy,Garousi:2006zh}. However, we concentrate our discussion only on the tangent part \eqref{eq:actionDO}. In order to find the right combination in \eqref{eq:actionDO} among $R^{\lambda}_{\,\,\mu\alpha\nu}, R_{\mu\nu}$ and $R$, one can start by first expanding the terms in \eqref{eq:actionDO} using the linearized approximation for which, the spacetime metric is expanded as $g_{\mu\nu} = \eta_{\mu\nu} + \hat{h}_{\mu\nu}$ with $\eta_{\mu\nu}$ the Minkowski metric and $\hat{h}_{\mu\nu}$ the fluctuation around the Minkowski metric, i.e. the graviton field.\,\footnote{$\hat{h}_{\mu\nu} = k_dh_{\mu\nu}$ with $k_d$ the d-dimensional physical gravitational coupling.} After that, one needs to match the terms up to two graviton fields $\hat h\hat h$ with the results of the string scattering amplitudes of eqs. \eqref{eq:alpha2} and \eqref{eq:alpha22} specialized to the case of two gravitons\,\footnote{Specializing $E_{\mu\nu}$ to $E_{(\mu\nu)}$ after the Fourier transformation we identify it to be $h_{\mu\nu}$.} (i.e. $E_{\mu\nu} = h_{\mu\nu}$) emitted and absorbed form $D_P$-brane and $\Omega_P$-plane. Recalling that $g^{\mu\nu} = \eta^{\mu\nu} - \hat{h}^{\mu\nu} + o(h^2)$ \cite{Veltman:1975vx}, expanding up to terms with four derivatives and two gravitons $\hat h$ gives for (Riemann)$^2$
\begin{equation}
\begin{aligned}
aR_{\lambda\sigma\alpha\beta}R^{\lambda\sigma\alpha\beta} &= a\,g_{\lambda\mu}R^{\mu}_{\,\,\sigma\alpha\beta}g^{\sigma\epsilon}g^{\alpha\rho}g^{\beta\gamma}R^{\lambda}_{\,\,\epsilon\rho\gamma}\Big|_{\hat h\hat h} = \frac{a}{4}\Big\{\DP_{\alpha}\DP_{\sigma}{\hat h}_{\beta\lambda} - \DP_{\alpha}\DP_{\lambda}{\hat h}_{\sigma\beta} - \DP_{\beta}\DP_{\sigma}{\hat h}_{\alpha\lambda} + \DP_{\beta}\DP_{\lambda}{\hat h}_{\sigma\alpha}\Big\}\cdot\\
&\Big\{\DP^{\alpha}\DP^{\sigma}{\hat h}^{\beta\lambda} - \DP^{\alpha}\DP^{\lambda}{\hat h}^{\sigma\beta} - \DP^{\beta}\DP^{\sigma}{\hat h}^{\alpha\lambda} + \DP^{\beta}\DP^{\lambda}{\hat h}^{\sigma\alpha}\Big\}
\end{aligned}
\label{eq:Riemann2}
\end{equation}
for (Ricci)$^2$
\begin{equation}
\begin{aligned}
bR_{\sigma\beta}R^{\sigma\beta} &= b\,R^{\lambda}_{\,\,\sigma\lambda\beta}g^{\sigma\rho}g^{\beta\mu}R^{\gamma}_{\,\,\rho\gamma\mu}\Big|_{\hat h\hat h} = \frac{b}{4}\Big\{\DP^{\chi}\DP_{\sigma}{\hat h}_{\beta\chi} + \DP^{\chi}\DP_{\beta}{\hat h}_{\sigma\chi} - \DP^{\chi}\DP_{\chi}{\hat h}_{\sigma\beta}\Big\}\Big\{\DP_{\gamma}\DP^{\sigma}{\hat h}^{\beta\gamma} + \DP_{\gamma}\DP^{\beta}{\hat h}^{\sigma\gamma}\\
&- \DP^{\gamma}\DP_{\gamma}{\hat h}^{\sigma\beta}\Big\}
\end{aligned}
\label{eq:Ricci2}
\end{equation}
and for (R)$^2$
\begin{equation}
cR^2 = c\,(g^{\mu\nu}R_{\mu\nu})(g^{\alpha\beta}R_{\alpha\beta})\Big|_{\hat h\hat h} = c\,\DP^{\mu}\DP^{\nu}{\hat h}_{\mu\nu}\DP^{\alpha}\DP^{\beta}{\hat h}_{\alpha\beta} 
\label{eq:R2}
\end{equation}
where symmetry properties and tracelessness of ${\hat h}$ have been used. Performing in \eqref{eq:actionDO} also the expansion of $\sqrt{-g} = \sqrt{-\eta}\big(1 + o(\Tr(\hat h)\big)$ and integrating by parts the terms \eqref{eq:Riemann2} and \eqref{eq:Ricci2}, one obtains (up to total derivative terms)
\begin{equation}
\begin{aligned}
aR_{\lambda\sigma\alpha\beta}R^{\lambda\sigma\alpha\beta} + bR_{\sigma\beta}R^{\sigma\beta} +cR^2\Big|_{\hat h\hat h} =&  \left(a + \frac{b}{4}\right)\,\DP^{\alpha}\DP_{\alpha}\DP^{\sigma}\DP_{\sigma}{\hat h}_{\beta\lambda}{\hat h}^{\beta\lambda} + \left(2a + \frac{b}{2}\right)\,\DP^{\alpha}\DP_{\alpha}\DP^{\beta}{\hat h}_{\beta\lambda}\DP_{\sigma}{\hat h}^{\sigma\lambda}\\
&+\left(a + \frac{b}{2} + c\right)\,\DP^{\beta}\DP^{\lambda}{\hat h}_{\beta\lambda}\DP_{\sigma}\DP_{\alpha}{\hat h}^{\sigma\alpha}\,.
%
%+\frac{b}{4}\,\DP^{\chi}\DP_{\chi}\DP^{\gamma}\DP_{\gamma}{\hat h}_{\sigma\beta}{\hat h}^{\sigma\beta} +\frac{b}{2}\,\DP^{\sigma}\DP_{\sigma}\DP^{\chi}{\hat h}_{\beta\chi}\DP_{\gamma}{\hat h}^{\beta\gamma} + \frac{b}{2}\,\DP^{\beta}\DP^{\chi}{\hat h}_{\beta\chi}\DP_{\sigma}\DP_{\gamma}{\hat h}^{\sigma\gamma} + c\,\DP^{\mu}\DP^{\nu}{\hat h}_{\mu\nu}\DP^{\alpha}\DP^{\beta}{\hat h}_{\alpha\beta} 
\end{aligned}
\label{eq:expris}
\end{equation}
Symmetrising and transforming to momentum space, one gets
\begin{equation}
\begin{aligned}
aR_{\lambda\sigma\alpha\beta}R^{\lambda\sigma\alpha\beta} + bR_{\sigma\beta}R^{\sigma\beta} +cR^2\Big|_{\hat h\hat h} =& \left(\frac{a}{2} + \frac{b}{8}\right)\,\left(k_1^{\parallel\,2}\right)^2\Tr({\hat h}_1{\hat h}_2)+ \left(a + \frac{b}{4}\right)\,\left(k_1^{\parallel\,2}\right)k_2^{\parallel}{\hat h}_1{\hat h}_2k_1^{\parallel}\\
&+\left(\frac{a}{2} + \frac{b}{4} + \frac{c}{2}\right)\,k_2^{\parallel}{\hat h}_1k_2^{\parallel}k_1^{\parallel}{\hat h}_2k_1^{\parallel} + (1\leftrightarrow 2)\,\,.
\end{aligned}
\label{eq:risexpmom}
\end{equation}
On the other hand the results of scattering amplitudes \eqref{eq:alpha2} and \eqref{eq:alpha22} using the transversality $(k_i\cdot h_i)_\nu = 0$, the condition on the trace $\Tr(h) = 0$ and the reflection matrix equal to $\mathcal R_{\mu\nu} = \eta_{\mu\nu}$ (i.e. only NN directions of the $D$-brane and $\Omega$-plane) are  
\begin{equation}
\mp\al^2\,\tau\zeta(2)\left\{\frac{1}{2}\Tr(h_1h_2)\left(k_1^{\parallel\,2}\right)^2\, + \left(k_1^{\parallel\,2}\right)k^{\parallel}_1h_2h_1k^{\parallel}_2\, + (1\leftrightarrow 2)\right\}
\label{eq:rishh}
\end{equation} 
with $\mp \tau$ equal to $-g_c^2C_{_{D_2}}$ or $g_c^2C_{_{RP_2}}/2$ for disk and projective plane respectively and where $k_1^{\parallel\,2}= -s$ is used (see Appendix \eqref{KIN}).\,\footnote{Eqs.\,\eqref{eq:alpha2} and \eqref{eq:alpha22} involve several terms. Some of them refer to the ortogonal directions of the $D_P$-brane and of the $\Omega_P$-plane. We did not consider those because our analysis is restricted to the tangent directions.} The cases with $D_9$-brane and $\Omega_9$-plane lead to amplitudes that vanish on-shell. Thus in order to match the terms one should use, for instance, the helicity formalism in ten-dimensions \cite{CaronHuot:2010rj}. The cases with $D_P$-brane and $\Omega_P$-plane where $P< 9$ have no problems since in general $k^{\parallel\,2}\ne0$. The first two terms in \eqref{eq:risexpmom} are those that the scattering of two gravitons \eqref{eq:rishh} reproduce, while the third term $k_2^{\parallel}{\hat h}_1k_2^{\parallel}k_1^{\parallel}{\hat h}_2k_1^{\parallel}$ as explained in \cite{Veltman:1975vx}, could produces negative norm states in the theory unless new ghost fields are included. In this case one has to find a solution for the coefficients in \eqref{eq:risexpmom} that cancel this unwanted term. Matching \eqref{eq:risexpmom} with \eqref{eq:rishh} one is able to fix only two coefficients. We chose to solve for $a$ and $c$, finding  
\begin{equation}
a = 1 - \frac{b}{4}\,\,;\qquad c = -1 -\frac{b}{4}\,.
\label{eq:AC}
\end{equation}
Inserting these values in \eqref{eq:actionDO}, the combination between $R^{\lambda}_{\,\,\mu\alpha\nu}, R_{\mu\nu}$ and $R$ is equal to
\begin{equation}
\small{
{\cal S}^{_{({D_P,\Omega_P})}}_{DBI}=\al^2\tau_{_{({D_P,\Omega_P})}}\int\,d^{P+1}X\,e^{-\Phi}\sqrt{-g}\left\{R_{\lambda\sigma\alpha\beta}R^{\lambda\sigma\alpha\beta} - R^2 -\frac{b}{4}\left(R_{\lambda\sigma\alpha\beta}R^{\lambda\sigma\alpha\beta} -4R_{\sigma\beta}R^{\sigma\beta} + R^2\right) \cdots \right\}}
\label{eq:Riemann3}
\end{equation}
the term multiplied by $b/4$ is \textit{Gauss-Bonnet}-type and the coefficient $b$ is unfixed using scattering amplitude of two gravitons, but it could be fixed by the calculation of string scattering amplitude of three gravitons.\footnote{The Gauss-Bonnet term in four-dimensions, as the worldvolume of a $D_3$-brane and $\Omega_3$-plane, is a purely topological term.} Setting for instance $b= -4$ one can find again the result reported in \cite{Bachas:1999um,Fotopoulos:2001pt,Fotopoulos:2002wy,Garousi:2006zh}\footnote{In \cite{Bachas:1999um,Fotopoulos:2001pt,Fotopoulos:2002wy,Garousi:2006zh} they found arguments to fix to zero the coefficient of the Gauss-Bonnet term.}
This analysis is valid in a general $d$-dimensional spacetime, but if one considers the compactification to a $d=4$ spacetime of an higher-dimensional theory, extra terms will appear from the dimensional reduction of the action \eqref{eq:Riemann3}. 

The Type IIB orientifold on $\T^6/{\mathbb Z}_2\times{\mathbb Z}_2$ considered in Section \ref{sec3}, is characterised by the presence of one set of $D_9$-branes on top of $\Omega_9$-plane and three sets of $D_5$-branes on top of $\Omega_5$-planes. In order to compactify to $d=4$ the starting model in ten-dimensions, the set of $D_9$-branes on top of $\Omega_9$-plane have to wrap the full internal $\T^6$-torus, while each set of $D_{5_I}$-branes on top of $\Omega_{5_I}$-plane have to wrap one $\T^2_I$-torus respectively, with $I = 1,2,3$. 
In Type IIB orientifold on $\T^6/{\mathbb Z}_2\times{\mathbb Z}_2$, the action of each element of ${\mathbb Z}_2\times{\mathbb Z}_2$ works as if it was a single $\mathbb Z_2$ that leaves invariant one of $\T^2_I$ tori while it flips the other $\T^2_{J\ne I}$ tori. Since we are focused on the moduli of the untwisted sector, and in particular on the geometric moduli which parametrise the $\T^2_I$ torus which the $\theta_I$ orbifold element leaves invariant respectively, one can approximate locally the internal manifold by
\begin{equation}
\theta_I \T^6 \rightarrow \T^2_I\times \frac{(\T^2_J\times \T^2_K)}{\mathbb{Z}_2} \sim \T^2_I\times K_3\,.
\end{equation}
The resulting LEEA at the $\al$-order can be found in \cite{Berg:2005ja},\footnote{Adding in \cite{Berg:2005ja} the DBI-action for the $\Omega_P$-planes also.} setting to zero the open string moduli (Wilson lines)\footnote{Up to now the modifications of axion dilaton and k\"ahler moduli definitions due to the presence of open string moduli are not considered \cite{Antoniadis:1996vw,Berg:2005ja}.}. In this way the LEEA of the Type IIB orientifold on $\T^6/{\mathbb Z}_2\times{\mathbb Z}_2$ can be interpreted as a superposition of three copies of the Type I on $\T^2\times K_3$ model\footnote{The Type I on $\T^2\times K_3$ model has one set of $D_9$-branes on top of $\Omega_9$-plane and one set of $D_5$-branes on top of $\Omega_5$-plane, and has $N=2$ supersymmetry in four- dimensions} \cite{Antoniadis:1996vw}. For simplicity we consider the set of $D_9$-branes, $\Omega_9$-plane and one set of $D_5$-branes, $\Omega_5$-plane. We add to the $\al$-order action in \cite{Berg:2005ja,Antoniadis:1996vw} the $\al^2$-order terms that arise from the compactification of \eqref{eq:Riemann3} on $\T^2\times K_3$ and match these with the results coming from the scattering amplitudes involving untwisted moduli in Section \ref{sec4} at the $\al^2$-order. The compactification process can be dealt with in two steps, first we go from ten-dimensions to six-dimensions using the $K_3$, then we go from six-dimensions to four-dimensions using the $T^2$. Only the set of $D_9$-branes, $\Omega_9$-plane is subject to the $K_3$ compactification\footnote{Up to a redefinition of the six-dimensional dilaton for the $D_5$-branes, $\Omega_5$-plane.}, while the $\T^2$ compactification involves all the sets of $D$-branes and $\Omega$-planes. In the first step the bulk metric factorises as $g_{\mu\nu}^{(6)}\times G_{_{K_3}}^{(4)}$, with $g^{(6)}_{\mu\nu}$ the six-dimensional worldvolume metric and $G_{_{K_3}}^{(4)}$ the $K_3$-metric. Using the static gauge for the DBI-actions, one has the same splitting for the worldvolume metric of the $D_9$-branes and $\Omega_9$-plane. The equation \eqref{eq:Riemann3} specialised to the $D_9$-branes, $\Omega_9$-plane case on $g_{\mu\nu}^{(6)}\times G_{_{K_3}}^{(4)}$ reads
\begin{equation}
\small{
\begin{aligned}
{\cal S}^{_{({D_9,\Omega_9})}}_{DBI}=&\,\al^2\tau_{(D_9,\Omega_9)}\int\,d^6X\,e^{-\Phi_6}\sqrt{-g^{(6)}}\nu^2\bigg\{\left[R_{\lambda\sigma\alpha\beta}R^{\lambda\sigma\alpha\beta} - R^2 -\frac{b}{4}\left(R_{\lambda\sigma\alpha\beta}R^{\lambda\sigma\alpha\beta} -4R_{\sigma\beta}R^{\sigma\beta} + R^2\right)\right]\left(g^{(6)}\right)\\
&+ \bm R^2 \left(g^{(6)},G_{_{K_3}}^{(4)}\right) + \bm R^2\left(G_{_{K_3}}^{(4)}\right)\cdots \bigg\}
\end{aligned}}
\label{eq:Riemann4}
\end{equation}
where in the round brackets we have specified the metric dependences of the $\bm R^2$ terms, $\nu$ is the $K_3$ volume modulus and $\Phi_6$ the dilaton in six-dimensions which is related to the ten-dimensional dilaton as $\exp(-2\Phi_6)= \exp(-2\Phi_{10})\nu^4$ \cite{Antoniadis:1996vw}. From eq. \eqref{eq:Riemann4} we consider only the $\bm R^2 \left(g^{(6)}\right)$ since we are interested neither in the scattering of $K_3$ moduli (which can be thought of as the blowing-up of the twisted moduli of $\T^6/{\mathbb Z}_2\times{\mathbb Z}_2$) nor on the scattering of mixed $K_3$ moduli with the $T^2$ moduli. Then, the compactification on $\T^2$ of the $\bm R^2 \left(g^{(6)}\right)$ in \eqref{eq:Riemann4} and of \eqref{eq:Riemann3} specialised to $D_5$-branes and $\Omega_5$-plane, with the splitting of the worldvolume metric $g^{(6)} = g_{\mu\nu}^{(4)} \times G^{(2)}_{mn}$  where $g_{\mu\nu}^{(4)}$ is the four dimensional worldvolume metric and $G_{mn}^{(2)}$ is the $\T^2$ torus metric \eqref{eq:torusmetric}, are 
\begin{equation}
\small{
\begin{aligned}
{\cal S}^{_{({D_9,\Omega_9})}}_{DBI} &+ {\cal S}^{_{({D_5,\Omega_5})}}_{DBI}=\,\al^2\bigg\{\tau_{(D_9,\Omega_9)}\int\,d^4X\,e^{-\Phi_4}\nu^2 + \tau_{(D_5,\Omega_5)}\int\,d^4X\,e^{-\Phi_4}\nu^{-2}\bigg\}\left(\sqrt{G^{(2)}}\right)^{\frac{1}{2}}\sqrt{-g^{(4)}}
\bigg\{\bm R^2{ \left(g^{(4)}\right)} \\
 & + \bm R^2\left(G^{(2)}\right)+\bigg[R_{LQAB}R^{LQAB} - R^2 -\frac{b}{4}\bigg(R_{LQAB}R^{LQAB} -4R_{QB}R^{QB} + R^2\bigg)\bigg]\left(g^{(4)},G^{(2)}\right){+ {\cdots} }\bigg\}\,.
\end{aligned}}
\label{eq:Riemann5}
\end{equation}

In eq. \eqref{eq:Riemann5} $\Phi_4$ is the four-dimensional dilaton related to the six-dimensional dilaton by $\exp(-2\Phi_4)= \exp(-2\Phi_6)\sqrt{G^{(2)}}$. Moreover, the $\bm R^2 \left(g^{(4)}\right){+} \bm R^2\left(G^{(2)}\right)$ contains only contractions between worldvolume indices ($\{\mu,\nu\}$) or $\T^2$-torus indices ($\{m,n\}$), respectively, while the terms in the square bracket contain all the possible contractions with mixed indices, $L=\{\lambda,l\}$. The terms that we want to compare with the string scattering amplitudes of Section \ref{sec4} can be extracted from the square bracket terms of \eqref{eq:Riemann5}. The mixed contractions of indices are in turn, for (Riemann)$^2$ 
\begin{equation}
\begin{aligned}
R_{LQAB}R^{LQAB} &= G_{ln}R^{n}_{\,\,\sigma\alpha\beta}g^{\sigma\epsilon}g^{\alpha\rho}g^{\beta\gamma}R^{l}_{\,\,\epsilon\rho\gamma} + g_{\lambda\mu}R^{\mu}_{\,\,\sigma\alpha b}g^{\sigma\epsilon}g^{\alpha\rho}G^{bh}R^{\lambda}_{\,\,\epsilon\rho h} + G_{ln}R^{n}_{\,\,\sigma\alpha b}g^{\sigma\epsilon}g^{\alpha\rho}G^{bh}R^{l}_{\,\,\epsilon\rho h}\\
&+g_{\lambda\mu}R^{\mu}_{\,\,\sigma a\beta}g^{\sigma\epsilon}G^{aq}g^{\beta\gamma}R^{\lambda}_{\,\,\epsilon q\gamma} + G_{ln}R^{n}_{\,\,\sigma a\beta}g^{\sigma\epsilon}G^{aq}g^{\beta\gamma}R^{l}_{\,\,\epsilon q\gamma} + g_{\lambda\mu}R^{\mu}_{\,\,\sigma ab}g^{\sigma\epsilon}G^{aq}G^{bh}R^{\lambda}_{\,\,\epsilon qh}\\
&+G_{ln}R^{n}_{\,\,\sigma ab}g^{\sigma\epsilon}G^{aq}G^{bh}R^{l}_{\,\,\epsilon qh} + g_{\lambda\mu}R^{\mu}_{\,\,r\alpha\beta}G^{re}g^{\alpha\rho}g^{\beta\gamma}R^{\lambda}_{\,\,e\rho\gamma} + G_{ln}R^{n}_{\,\,r\alpha\beta}G^{re}g^{\alpha\rho}g^{\beta\gamma}R^{l}_{\,\,e\rho\gamma}\\
&+g_{\lambda\mu}R^{\mu}_{\,\,r\alpha b}G^{re}g^{\alpha\rho}G^{bh}R^{\lambda}_{\,\,e\rho h} + G_{ln}R^{n}_{\,\,r\alpha b}G^{re}g^{\alpha\rho}G^{bh}R^{l}_{\,\,e\rho h} + g_{\lambda\mu}R^{\mu}_{\,\,ra\beta}G^{re}G^{aq}g^{\beta\gamma}R^{\lambda}_{\,\,eq\gamma}\\
&+G_{ln}R^{n}_{\,\,ra\beta}G^{re}G^{aq}g^{\beta\gamma}R^{l}_{\,\,eq\gamma} + g_{\lambda\mu}R^{\lambda}_{\,\,rab}G^{re}G^{aq}G^{bh}R^{\mu}_{\,\,eqh}
\end{aligned}
\label{eq:riemann-open}
\end{equation}
for (Ricci)$^2$ 
\begin{equation}
\begin{aligned}
R_{QB}R^{QB} &= R^{\lambda}_{\,\,\sigma\lambda\beta}g^{\sigma\rho}g^{\beta\gamma}R^{s}_{\,\,\rho s\gamma} + R^{l}_{\,\,\sigma l\beta}g^{\sigma\rho}g^{\beta\gamma}R^{\chi}_{\,\,\rho\chi\gamma} + R^{l}_{\,\,\sigma l\beta}g^{\sigma\rho}g^{\beta\gamma}R^{s}_{\,\,\rho s\gamma} + R^{\lambda}_{\,\,\sigma\lambda b}g^{\sigma\rho}G^{bh}R^{\chi}_{\,\,\rho\chi h}\\
&+R^{\lambda}_{\,\,\sigma\lambda b}g^{\sigma\rho}G^{bh}R^{s}_{\,\,\rho sh} + R^{l}_{\,\,\sigma lb}g^{\sigma\rho}G^{bh}R^{\chi}_{\,\,\rho\chi h} + R^{l}_{\,\,\sigma lb}g^{\sigma\rho}G^{bh}R^{s}_{\,\,\rho sh} + R^{\lambda}_{\,\,q\lambda\beta}G^{qp}g^{\beta\gamma}R^{\chi}_{\,\,p\chi\gamma}\\
&+R^{\lambda}_{\,\,q\lambda b}G^{qp}g^{\beta\gamma}R^{s}_{\,\,ps\gamma} + R^{l}_{\,\,ql\beta}G^{qp}g^{\beta\gamma}R^{\chi}_{\,\,p\chi\gamma} + R^{l}_{\,\,ql\beta}G^{qp}g^{\beta\gamma}R^{s}_{\,\,ps\gamma} + R^{\lambda}_{\,\,q\lambda b}G^{qp}G^{bh}R^{\chi}_{\,\,p\chi h}\\
&+R^{\lambda}_{\,\,q\lambda b}G^{qp}G^{bh}R^{s}_{\,\,p sh} + R^{l}_{\,\,qlb}G^{qp}G^{bh}R^{\chi}_{\,\,p \chi h}
\end{aligned} \label{eq:ricci-open}
\end{equation}
and for (R)$^2$
\begin{equation}
\begin{aligned}
R^2 &= G^{qb}R^{\lambda}_{\,\,q\lambda b}g^{\rho\gamma}R^{\chi}_{\,\,\rho\chi\gamma} + g^{\sigma\beta}R^{\lambda}_{\,\,\sigma\lambda\beta}g^{\rho\gamma}R^{s}_{\,\,\rho s\gamma} + G^{qb}R^{l}_{\,\,ql b}g^{\rho\gamma}R^{\chi}_{\,\,\rho\chi\gamma} + g^{\sigma\beta}R^{l}_{\,\,\sigma l \beta}g^{\rho\gamma}R^{\chi}_{\,\,\rho\chi\gamma}\\
&+G^{qb}R^{\lambda}_{\,\,q\lambda b}g^{\rho\gamma}R^{s}_{\,\,\rho s\gamma} +g^{\sigma\beta}R^{l}_{\,\,\sigma l\beta}g^{\rho\gamma}R^{s}_{\,\,\rho s\gamma} + G^{qb}R^{l}_{\,\,qlb}g^{\rho\gamma}R^{s}_{\,\,\rho s\gamma} + g^{\sigma\beta}R^{\lambda}_{\,\,\sigma\lambda\beta}G^{ph}R^{\chi}_{\,\,p\chi h}\\
&+G^{qb}R^{\lambda}_{\,\,q\lambda b}G^{ph}R^{\chi}_{\,\,p\chi h} + g^{\sigma\beta}R^{\lambda}_{\,\,\sigma\lambda\beta}G^{ph}R^{s}_{\,\,psh} + G^{qb}R^{\lambda}_{\,\,q\lambda b}G^{ph}R^{s}_{\,\,psh}
+ g^{\sigma\beta}R^{l}_{\,\,\sigma l \beta}G^{ph}R^{\chi}_{\,\,p\chi h}\\
&+ G^{qb}R^{l}_{\,\,q l b}G^{ph}R^{\chi}_{\,\,p\chi h} + + g^{\sigma\beta}R^{l}_{\,\,\sigma l \beta}G^{ph}R^{s}_{\,\,p s h}\,\,.
\end{aligned} \label{eq:r2-open}
\end{equation}
Using two-point string scattering amplitudes for the untwisted moduli computed in Section \ref{sec4}, we can extract from \eqref{eq:riemann-open}, \eqref{eq:ricci-open} and \eqref{eq:r2-open} terms which involve four derivatives and two untwisted moduli using the linearised approximation for the spacetime (worldvolume) metric $g_{\mu\nu} = \eta_{\mu\nu} + O(\hat{h}_{\mu\nu})$, recalling that no mixed components of the metric $\tilde{g}_{\mu n}$ are present.\footnote{These $\tilde{g}_{\mu n}$ components are related to the \textit{graviphoton} that we set to zero.} With these approximations, several terms in \eqref{eq:riemann-open}, \eqref{eq:ricci-open} and \eqref{eq:r2-open} are zero due to the vanishing of   some of the Christoffel symbols. For instance, one finds
\begin{equation}
\Gamma^{l}_{\epsilon\rho} = \frac{1}{2}G^{ld}\left(\DP_{\epsilon}\tilde{g}_{\rho d} + \DP_{\rho}\tilde{g}_{\epsilon d} - \DP_{d} g_{\epsilon\rho}\right) = 0\,.
\end{equation}
The non vanishing terms in \eqref{eq:riemann-open}, \eqref{eq:ricci-open} and \eqref{eq:r2-open} that can contain terms starting with four derivatives and two untwisted moduli  have the expression (after integration by parts)
\begin{equation}
\begin{aligned}
&R_{LQAB}R^{LQAB}\mapsto \left(\DP_{\sigma}\DP_{\alpha}G_{bl}\right) G^{bh}G^{le}\left(\DP^{\sigma}\DP^{\alpha}G_{he}\right)\\
&R_{QB}R^{QB}\mapsto \frac{1}{4}\left(\DP_{\sigma}\DP_{\alpha}G_{bl}\right) G^{bh}G^{le}\left(\DP^{\sigma}\DP^{\alpha}G_{he}\right) + \frac{1}{4}G^{qb}\left(\DP_{\sigma}\DP_{\alpha}G_{qb}\right)G^{ph}\left(\DP^{\sigma}\DP^{\alpha}G_{ph}\right)\\
&R^2\mapsto G^{qb}\left(\DP_{\sigma}\DP_{\alpha}G_{qb}\right)G^{ph}\left(\DP^{\sigma}\DP^{\alpha}G_{ph}\right)\,.
\end{aligned}\label{eq:GGexp}
\end{equation}
From the above equations one can verify that the terms in the round brackets in the second line of \eqref{eq:Riemann5} cancel each other at this order, leaving the $b$ coefficient still undetermined. The $\al^2$-order terms that can be read off from the scattering amplitudes of untwisted moduli (geometric moduli) in Section $\ref{sec4}$ are as follows. For two K\"ahler moduli $T$ (or better, for the imaginary part of $T$, i.e. $T_2$) one gets, using eqs. \eqref{eq:TT}, \eqref{eq:TbarT}, \eqref{eq:TTrp21} and \eqref{eq:TTrp2}
\begin{equation}
\mathcal A_{a/\alpha}(T_2^I,T^I_2)
= -\al^2\hat{C}_{_{D_2/RP_2}}\zeta(2)\,\frac{\left(u^2 + t^2\right)}{(T^I - \bar T^I)^2}\,\,;\quad A_{a/\alpha}(T_2^{I},T^{J}_2)
= \al^2\hat{C}_{_{D_2/RP_2}}\zeta(2)\,\frac{2\,tu\,\mathscr R^I_{a/\alpha}\mathscr R^J_{a/\alpha}}{(T^I - \bar T^I)(T^J - \bar T^J)}
\label{eq:TTalpha2}
\end{equation}
where $\hat{C}_{_{D_2}}=g_c^2C_{_{D_2}}/4$ and $\hat{C}_{_{RP_2}}= g_c^2C_{_{RP_2}}/16$.
For two complex structure $U$ using eqs. \eqref{eq:Disk-UbarU-res1}, \eqref{eq:Disk-UbarU-res2}, \eqref{eq:UU}, \eqref{eq:UU_I=J_res_-1-100}, \eqref{eq:UU_I/J_res_-1-100} and \eqref{eq:UUrp2}, one finds
\begin{equation}
\mathcal A_{a/\alpha}(U^I,\bar U^I) = -\al^2\hat{C}_{_{D_2/RP_2}}\zeta(2)\,\frac{32\,s^2}{(U^I {-} \bar U^I)^2}\,\,;\,\,\mathcal A_{a/\alpha}(U^I,\bar U^J) = 0\,\,;\,\, A_{a/\alpha}(U^I, U^J) = A_{a/\alpha}(U^I, U^I) = 0
\label{eq:UUalpha2}
\end{equation}
and finally, for one K\"ahler modulus $T$ and one complex structure $U$ recalling eqs. \eqref{eq:ATU}, \eqref{eq:TU}, \eqref{eq:ATUrp2}, \eqref{eq:TUrp2}, one has
\begin{equation}
\mathcal A_{a/\alpha}(T^I,\bar U^J) = \mathcal A_{a/\alpha}(T^I,\bar U^I)= 0\,\,;\quad \mathcal A_{a/\alpha}(T^I, U^J) = \mathcal A_{a/\alpha}(T^I, U^I) = 0
\label{eq:UTalpha2}
\end{equation}
with $a$ and $\alpha$ labelling the $D$-brane and $\Omega$-plane respectively.
At this point, to make more clear the match between the string scattering amplitudes in Section 4 and the non vanishing combination in the LEEA \eqref{eq:Riemann5}
\begin{equation}
\left[R_{LQAB}R^{LQAB} - R^2\right]\left(g^{(4)},G^{(2)}\right)
\label{eq:alpha2R}
\end{equation} 
we proceed as follows. In order to capture the scattering 
for the imaginary ($T_2$) part of the K\"ahler modulus $T$, we recall that one needs to construct the vertex operator \eqref{eq:VT2}. Using this vertex operator definition, the scattering amplitude for the utwisted moduli $T^I_2$ is given by the eq. \eqref{eq:amplitudedef} which gives as results eq.\,\eqref{eq:TTalpha2}. On the same line, we can build the vertex operators for the real ($U_1$) and the imaginary ($U_2$) part of the complex structure $U$, that are
\begin{equation}
\begin{aligned}
&
\mathcal W_{{U_1^I}{(q,\bar q)}}(E,z,\bar z, k) = \frac{1}{2}\left(\mathcal W_{{U^I}{(q,\bar q)}}(E,z,\bar z, k) + \mathcal W_{{\bar U^I}{(q,\bar q)}}(E,z,\bar z, k)\right)\\
&\mathcal W_{{U_2^I}{(q,\bar q)}}(E,z,\bar z, k) = -\frac{i}{2}\left(\mathcal W_{{U^I}{(q,\bar q)}}(E,z,\bar z, k) - \mathcal W_{{\bar U^I}{(q,\bar q)}}(E,z,\bar z, k)\right)\,.
\end{aligned}\label{eq:VU2}
\end{equation}
The resulting expressions for the scattering amplitudes of the real $U_1$ and the imaginary $U_2$ part of the complex structure moduli $U$, using eq. \eqref{eq:VU2} respectively, are given by
\begin{equation}
\begin{aligned}
&\mathcal A_{a/\alpha}(U_1^I,U^J_1) = \frac{1}{4}\left({\cal A}_{a/\alpha}\left(\bar U^{I}, {U}^{J}\right) +{\cal A}_{a/\alpha}\left(U^{I},{U}^{J}\right) + {\cal A}_{a/\alpha}\left(\bar U^{I},\bar {U}^{J}\right) + {\cal A}_{a/\alpha}\left(U^{I},\bar{U}^{J}\right)\right)\\
&\mathcal A_{a/\alpha}(U_2^I,U^J_2) = \frac{1}{4}\left({\cal A}_{a/\alpha}\left(\bar U^{I}, {U}^{J}\right) -{\cal A}_{a/\alpha}\left(U^{I},{U}^{J}\right) - {\cal A}_{a/\alpha}\left(\bar U^{I},\bar {U}^{J}\right) + {\cal A}_{a/\alpha}\left(U^{I},\bar{U}^{J}\right)\right)\,.
\end{aligned}
\label{eq:AmplitudeU1U2}
\end{equation}
With the help of equations \eqref{eq:UUalpha2} one can verify that in eq. \eqref{eq:AmplitudeU1U2} the only non vanishing contributions are those with $I=J$
\begin{equation}
\mathcal A_{a/\alpha}(U_1^I,U^I_1) = -\al^2\hat{C}_{_{D_2/RP_2}}\zeta(2)\,\frac{32\,s^2}{2(U^I {-} \bar U^I)^2}\,;\quad\mathcal A_{a/\alpha}(U_2^I,U^I_2)=-\al^2\hat{C}_{_{D_2/RP_2}}\zeta(2)\,\frac{32\,s^2}{2(U^I {-} \bar U^I)^2}\,.
\label{eq:AU1}
\end{equation}
The above equation means that there is no mixing between real (imaginary) $U_1^I$ ($U_2^I$) moduli coming from different $T^2_I$ tori. Moreover, using eq.\eqref{eq:UUalpha2} one can verify that there is no mixing between the real $U_1$ and the imaginary $U_2$ part of the complex structure $U$ because the associated scattering amplitude vanishes both for $I=J$ and $I\ne J$
\begin{equation}
\mathcal A_{a/\alpha}(U_1^I,U^J_2) = 0.
\label{eq:AU2}
\end{equation}
As well as using eqs.\eqref{eq:VT2},\eqref{eq:VU2} and \eqref{eq:UTalpha2} tell us that even at $\al^2$-order there is no mixing between the K\"ahler modulus $T$ (i.e. $T_2$) and the complex structure $U$ owing to the vanishing of the amplitudes still both for $I=J$ and $I\ne J$
\begin{equation}
\mathcal A_{a/\alpha}(T_2^I,U^J_1) = 0\,;\quad\mathcal A_{a/\alpha}(T_2^I,U^J_2)=0\,.
\label{eq:AU3}
\end{equation}
The string scattering amplitude results \eqref{eq:TTalpha2}, \eqref{eq:AU1}, \eqref{eq:AU2} and \eqref{eq:AU3} are in agreement with the non vanishing $\al^2$ term \eqref{eq:alpha2R} in eq.\eqref{eq:Riemann5}. The expression of eq.\eqref{eq:alpha2R} exploiting \eqref{eq:GGexp}, is
\begin{equation}
\begin{aligned}
&R_{LQAB}R^{LQAB} - R^2 \mapsto \left(\DP_{\sigma}\DP_{\alpha}G_{bl}\right) G^{bh}G^{le}\left(\DP^{\sigma}\DP^{\alpha}G_{he}\right) - G^{qb}\left(\DP_{\sigma}\DP_{\alpha}G_{qb}\right)G^{ph}\left(\DP^{\sigma}\DP^{\alpha}G_{ph}\right)\\
&{=} \left(\DP_{\sigma}\DP_{\alpha}\left(G^I\right)_{bl}\right)\!{\left(G^I\right)}^{bh}\!{\left(G^I\right)}^{le}\!\left(\DP^{\sigma}\DP^{\alpha}\left(G^I\right)_{he}\right)
{-} \frac{1}{2}{\left(G^I\right)}^{qb}\!\left(\DP_{\sigma}\DP_{\alpha}\!\left(G^I\right)_{qb}\right)\!{\left(G^J\right)}^{ph}\!\left(\DP^{\sigma}\DP^{\alpha}\left(G^J\right)_{ph}\right) {+} (I{\leftrightarrow} J)
\end{aligned}\label{eq:riemann6}
\end{equation}
with $G^I$ the metric of the $\T^2_I$-torus \eqref{eq:torusmetric}. We point out that terms in \eqref{eq:TTalpha2} with $I\ne J$ can be reproduced only by the scalar curvature squared $R^2$ term. The same terms with $I\ne J$ would be reproduced also from (Riemann)$^2$ term which involves metric tensors mutually contracted. But since we deal with a metric of the $\T^6$-torus that factorises ($\T^6 = \bigotimes_{I=1}^{3}\T^2_I$), there is no mixing between different $T^2_I$ tori. So there are no contributions from (Riemann)$^2$ term when $I\ne J$.
At this point, we consider only the terms with two derivatives acting on the same modulus $\varphi\in\{T_2,U_1,U_2\}$. The expression for the derivatives of the metric components $G^I$ are
\begin{equation}
\begin{aligned}
&\DP_{\sigma}\DP_{\alpha}\left(G_{[2I+1][2I+1]}\right)\bigg|_{\DP\DP\varphi} = \left(\frac{\DP_{\sigma}\DP_{\alpha}T_2^I}{U_2^I} -\frac{T_2^I\DP_{\sigma}\DP_{\alpha}U_2^I}{(U_2^I)^2}\right)\\
&\DP_{\sigma}\DP_{\alpha}\left(G_{[2I+1][2I+2]}\right)\bigg|_{\DP\DP\varphi} = \left(\frac{U_1^I\DP_{\sigma}\DP_{\alpha}T_2^I}{U_2^I} + \frac{T_2^I\DP_{\sigma}\DP_{\alpha}U_1^I}{U_2^I} -\frac{T_2^IU_1^I\DP_{\sigma}\DP_{\alpha}U_2^I}{(U_2^I)^2}\right)\\
&\DP_{\sigma}\DP_{\alpha}\left(G_{[2I+2][2I+2]}\right)\bigg|_{\DP\DP\varphi} = \left(\frac{\abs{U^I}^2\DP_{\sigma}\DP_{\alpha}T_2^I}{U_2^I}+ \frac{2T_2^IU_1^I\DP_{\sigma}\DP_{\alpha}U_1^I}{U_2^I}+ \frac{T_2^I\left((U^I_2)^2 -(U_1^I)^2\right)\DP_{\sigma}\DP_{\alpha}U_2^I}{(U_2^I)^2}\right)\,.
\end{aligned}
\label{eq:2derivaG}
\end{equation}
Using the above expressions \eqref{eq:2derivaG} in \eqref{eq:riemann6}, one can verify that the non vanishing terms are
\begin{equation}
\begin{aligned}
&\left(T_2^I,T_2^I\right) : -\frac{(\DP_{\sigma}\DP_{\alpha}T^I_2)(\DP^{\sigma}\DP^{\alpha}T^I_2)}{(T^I - \bar T^I)^2}\,\,;\quad \left(T_2^{I},T_2^{J}\right) : \frac{(\DP_{\sigma}\DP_{\alpha}T^I_2)(\DP^{\sigma}\DP^{\alpha}T^J_2)}{(T^I - \bar T^I)(T^J - \bar T^J)}\\
&\left(U_1^I,U_1^I\right) : -\frac{(\DP_{\sigma}\DP_{\alpha}U^I_1)(\DP^{\sigma}\DP^{\alpha}U^I_1)}{(U^I - \bar U^I)^2}\,\,;\quad \left(U_2^I,U_2^I\right) : -\frac{(\DP_{\sigma}\DP_{\alpha}U^I_2)(\DP^{\sigma}\DP^{\alpha}U^I_2)}{(U^I - \bar U^I)^2}
\end{aligned}
\label{eq:finalcomp}
\end{equation}
while all the other combinations are zero. The terms in eq. \eqref{eq:finalcomp} exhibit the same moduli dependencies as the scattering amplitudes results \eqref{eq:TTalpha2}, \eqref{eq:AU1}, \eqref{eq:AU2} and \eqref{eq:AU3}, aside from constant factors.
%%%%%%%%%%%%%%
%%%%%%%%%%%%
%%%%%%%%%%%%
\section{Conclusion}\label{sec6}
In this work we have perturbatively analyzed the LEEA for Type IIB orientifold models using information from string scattering calculation. We have focused our attention on tree-level string scattering amplitudes involving only closed strings as external states on the disk $D_2$ and projective-plane $RP_2$ worldsheet. Indeed, in the study of unoriented models, the simultaneous presence of extended objects like $D_P$-branes and $\Omega_P$-planes, under specific conditions, make the given theory well define. Two-point closed sting scattering amplitudes are the object of interest because, after the one-point functions, they are the first non vanishing contributions on both worldsheets. An exceptional case is the zero-point function on the projective-plane that is non zero owing to the properties of its CKG $SU(2)$. In this paper we have summarized in a pedagogical way two-point disk \cite{Garousi:1996ad, Hashimoto:1996kf, Hashimoto:1996bf, Garousi:2006zh} and projective-plane \cite{Garousi:2006zh} calculations. In these calculations only NS-NS external states are considered because we are interested in higher derivative \textit{curvature} corrections to LEEA of $D_P$-branes and $\Omega_P$-planes, i.e. on $\al^2$-order (curvature)$^2$ terms. This was done by ``matching''  string scattering amplitudes involving two-gravitons with the most general linear combination of (curvature)$^2$ terms that the Dirac-Born-Infeld action of $D_P$-branes and $\Omega_P$-planes can contain at this order
%%%%%%%%%%%%%%%%%%
\begin{equation}
{\cal S}^{_{({D_P,\Omega_P})}}_{DBI}=\al^2\tau_{_{({D_P,\Omega_P})}}\int\,d^{P+1}\chi\,e^{-\Phi}\sqrt{-g}\left\{aR_{\lambda\sigma\alpha\beta}R^{\lambda\sigma\alpha\beta} +bR_{\sigma\beta}R^{\sigma\beta} + cR^2 \dots \right\}\,.
\label{eq:gravi2}
\end{equation}
%%%%%%%%%%%
We have proved that ``matching'' implies for the free parameters in the previous equation the values displayed in eq.\eqref{eq:AC} which yield 
%%%%%%%%%%%%%%
\begin{equation}
\small{
{\cal S}^{_{({D_P,\Omega_P})}}_{DBI}=\al^2\tau_{_{({D_P,\Omega_P})}}\int\,d^{P+1}X\,e^{-\Phi}\sqrt{-g}\left\{R_{\lambda\sigma\alpha\beta}R^{\lambda\sigma\alpha\beta} - R^2 -\frac{b}{4}\left(R_{\lambda\sigma\alpha\beta}R^{\lambda\sigma\alpha\beta} -4R_{\sigma\beta}R^{\sigma\beta} + R^2\right) \dots \right\}}\,.
\end{equation}
%%%%%%%%%%%%%%
At this stage the $b$ parameter remains unfixed. Maybe it can be fixed by computing the two-point string scattering amplitudes with two-gravitinos as external states and matching with the supersymmetric extension of the \eqref{eq:gravi2}. Alternatively one can resort to the much more complicated three-point string scattering amplitude with three-gravitons and matching it with \eqref{eq:gravi2} expanded up to three-linear terms in the graviton fields. We don't have specific arguments to prove these procedures  but we claim that should be interest follows these lines.  

We have then focused our analysis on a specific Type IIB orientifold model, i.e. Type IIB orientifold on $\T^6/{\mathbb Z}_2\times{\mathbb Z}_2$ since we wanted to extend the two-point disk calculation with untwisted moduli made in \cite{Lust:2004cx} to the case where also the projective-plane contribution is considered. In order to prepare the reader to the new developments concerning the case of the projective plane, we have for completeness reviewed the disk case~\cite{Lust:2004cx}, giving in Section~\ref{Toolkit} some space to the techniques aimed at the construction of vertex operators for the untwisted moduli and to the two-point correlators for the $D_2$ disk as well as $RP_2$ real projective plane worldsheets. String scattering amplitudes with two-untwisted moduli are derived in Section \ref{scattering_sec}. The studying of $\al$-expansion of these scattering amplitudes is also performed, and it was found that there are no $\al$-order corrections at LEEA by projective-plane calculations\,\footnote{In general there are arguments like in \cite{Berg:2014ama} where do not excluded the presence of these corrections at tree-level.} confirming the tree-level structure of the K\"ahler potential for the untwisted moduli which then reads
%%%%%%%%%
\begin{equation}
k_4^2\,\mathcal K =  - \ln\prod_{I=1}^3(T^I - \bar T^I) - \ln\prod_{I=1}^3(U^I -\bar U^I).
%\label{eq:kahlerP}
\end{equation}
%%%%%%%%%
Finally in Section \ref{eq:Rhigh} we compared the $\al^2$-order terms arising from the string scattering amplitudes derived in Section \ref{sec4} with the terms emerging from the compactification on $\T^6/{\mathbb Z}_2\times{\mathbb Z}_2$ of \eqref{eq:gravi2}. We have checked in this way that the only non-vanishing untwisted moduli contributions are
%%%%%%%%%
\begin{equation}
\begin{aligned}
&\left(T_2^I,T_2^I\right) : -\frac{(\DP_{\sigma}\DP_{\alpha}T^I_2)(\DP^{\sigma}\DP^{\alpha}T^I_2)}{(T^I - \bar T^I)^2}\,\,;\quad \left(T_2^{I},T_2^{J}\right) : \frac{(\DP_{\sigma}\DP_{\alpha}T^I_2)(\DP^{\sigma}\DP^{\alpha}T^J_2)}{(T^I - \bar T^I)(T^J - \bar T^J)}\\
&\left(U_1^I,U_1^I\right) : -\frac{(\DP_{\sigma}\DP_{\alpha}U^I_1)(\DP^{\sigma}\DP^{\alpha}U^I_1)}{(U^I - \bar U^I)^2}\,\,;\quad \left(U_2^I,U_2^I\right) : -\frac{(\DP_{\sigma}\DP_{\alpha}U^I_2)(\DP^{\sigma}\DP^{\alpha}U^I_2)}{(U^I - \bar U^I)^2}
\end{aligned}
\end{equation}
%%%%%%%%%
in agreement with string scattering amplitude results \eqref{eq:TTalpha2}, \eqref{eq:AU1}, \eqref{eq:AU2} and \eqref{eq:AU3}. A similar analysis will be made for two-point string scattering amplitudes for twisted moduli at the orbifold point were the CFT is well define. In this framework a careful evaluation of vertex operators and two-point correlators is required in order to understand whether some new features could appear. Finally, we recall that all the details on the calculations can be found in \cite{AM:appendice}.

\section*{\centering\normalsize{Acknowledgements}}
We would like to thank G. Pradisi for initial discussions about the topics of this paper, without which this work could not even have started. We are very grateful to M. Bianchi and G.C. Rossi for several suggestions, discussions and a careful reading of the manuscript. We were partially supported by the MIUR-PRIN contract 2015MP2CX4002 “Non-perturbative aspects of gauge theories and strings”
%%%%%%%%%%%%%%
%%%%%%%%%%%%
%%%%%%%%%%%%
\begin{appendices}
\chapter{}

\section{}
%%%%%%%%%%%%%%
%%%%%%%%%%%%
%%%%%%%%%%%%
%
%
\subsection{Kinematics: Closed strings as open strings}\label{KIN}
We would like show in this Appendix that a closed string can be described using open strings \cite{Garousi:1996ad,Hashimoto:1996kf,Hashimoto:1996bf,Garousi:2006zh,Lust:2004cx,Stieberger:2009hq}. This is possible owing to the presence of extended objects like $D$-branes (oriented models which contain open string sector) and/or $\Omega$-planes (unoriented models), that project the total momentum $k$ of the closed string in its parallel $k^{\parallel}$ and ortogonal $k^{\perp}$ components to the extended objects, respectively. The conservation of the momentum $k$ works only in the parallel direction \cite{BLT,Lust:2004cx,Garousi:1996ad}, because only in the worldvolume of the extended object the Poincar\'e group is unbroken. Thus one has
\begin{equation}
\begin{aligned}
k^{_M} &= (k^{\parallel})^{\mu} + (k^{\perp})^{m} = \bigg(\frac{k^{_M}}{2} + \frac{(\R{\cdot}k)^{_M}}{2}\bigg)^{\parallel} +  \bigg(\frac{k^{_M}}{2} - \frac{(\R{\cdot}k)^{_M}}{2}\bigg)^{\perp}\quad\left[\begin{split}&M\in\{0,{\dots},d\}; \mu\in\{0,{\dots},p\}\\&m\in\{{p+1},{\dots},{d-p}\}\end{split}\right]\\\\ \R &:= \R^{MN} = \begin{pmatrix} [\eta^{\mu\nu}]^{\parallel}&0\\0&[-\delta^{mn}]^{\perp}\end{pmatrix}\qquad\big[\textbf{Neumann direct.}:[\eta^{\mu\nu}]^{\parallel} ; \textbf{Dirichlet direct.}: [-\delta^{mn}]^{\perp}\big]\,\,.
\end{aligned}\label{eq:closedAsopen}
\end{equation}
\\
From this point of view it seems that the two projection of $k$ can be described by two open-like-string independent momenta $\{k/2 ; \R k/2\}$. Thus the question that arise is: in which way the  closed string mass-squared is split between these two open-like-string? We can answer this question by looking at the mass-shell condition for $k$ which reads
\begin{equation}
\begin{aligned}
- m^2 = k^2 &= (k^{\parallel} + k^{\perp})^2 = \bigg(\frac{k}{2} + \frac{\R{\cdot}k}{2}\bigg)^2_{\parallel} +  \bigg(\frac{k}{2} - \frac{\R{\cdot}k}{2}\bigg)^2_{\perp} + 2\bigg(\frac{k}{2} + \frac{\R{\cdot}k}{2}\bigg)_{\parallel}{\cdot} \bigg(\frac{k}{2} - \frac{\R{\cdot}k}{2}\bigg)_{\perp}\\&= 2\bigg(\frac{k}{2}\bigg)^2 + 2\bigg(\frac{\R{\cdot}k}{2}\bigg)^2\,.
\end{aligned}\label{eq:massCloseAsOp}
\end{equation}
Thinking the mass as a two-component vector, one can split the closed string mass-squared as follows
\begin{equation}
\begin{aligned} 
-m^2= - (\overrightarrow{m_a} + \overrightarrow{m_b})^2= 2\bigg(\frac{k}{2}\bigg)^2 + 2\bigg(\frac{\R{\cdot}k}{2}\bigg)^2;\,\,\,\overrightarrow{m_a} =\begin{pmatrix}\frac{m}{\sqrt 2}\\0\end{pmatrix};\,\,\, \overrightarrow{m_b}= \begin{pmatrix}0\\\frac{m}{\sqrt 2}\end{pmatrix}. 
\end{aligned}
\end{equation}

This decomposition allows to define the following components $-m_a^2 \equiv -\frac{m^2}{2} = 2\big(\frac{k}{2}\big)^2$ and $ -m_b^2 \equiv -\frac{m^2}{2} = 2\big(\frac{\R{\cdot}k}{2}\big)^2$.
We thus see that each open string-like momentum $\{k/2 ; {\R}k/2\}$ brings one half of the original closed string mass $m^2$.
In the case of two massless closed string states, described by $\{k_1,k_2\}$ momenta, one can write
\begin{equation}
\begin{aligned}
k_1 &= k_1^{\parallel} + k_1^{\perp} = \bigg(\frac{k_1}{2} + \frac{\R{\cdot}k_1}{2}\bigg)_{\parallel} + \bigg(\frac{k_1}{2} - \frac{\R{\cdot}k_1}{2}\bigg)_{\perp}\\k_2 &= k_2^{\parallel} + k_2^{\perp} = \bigg(\frac{k_2}{2} + \frac{\R{\cdot}k_2}{2}\bigg)_{\parallel}+\bigg(\frac{k_2}{2} - \frac{\R{\cdot}k_2}{2}\bigg)_{\perp}
\end{aligned}
\end{equation}
but  only the parallel component of each $k_i$ is constrained by momentum conservation, thus only this part enters directly in the construction of the kinematical invariants, giving
\begin{equation}
\small{\begin{aligned}
\sum_{i=1}^{2} k_i^{\parallel} &= k_1 + \R k_1 + k_2 + \R k_2 = 0\,; \quad
\#_{s_{ij}}= \begin{pmatrix} 4\\2\end{pmatrix} = 6 \qquad[i,j\in\{1,\bar 1,2,\bar 2\}; i\ne j ; i<j]\\
%%%%%%%%%%%%%%%%%%
s_{1\bar1}&= - \bigg(\frac{k_1}{2} + \frac{\R k_1}{2}\bigg)^2 \hspace{3mm}= -\bigg(\frac{k_1}{2}\bigg)^2 - \bigg(\frac{\R k_1}{2}\bigg)^2 - \frac{k_1{\cdot}\R k_1}{2} \hspace{2mm}= \frac{m_1^2}{4} + \frac{m_1^2}{4} - \frac{k_1{\cdot}\R k_1}{2}\hspace{2mm}=  - \frac{k_1{\cdot}\R k_1}{2} \\
s_{12}&= - (k_1 + k_2)^2 \hspace{1cm}= -k_1^2 - k_2^2 -2k_1{\cdot}k_2 \hspace{2.2cm}=  m_1^2 + m_2^2 -2k_1{\cdot}k_2 \hspace{6mm}= -2 k_1{\cdot}k_2\\
s_{1\bar2}&= - (k_1 + \R k_2)^2 \hspace{7mm}= - k_1^2 - (\R k_2)^2 -2k_1{\cdot}\R k_2 \hspace{1.2cm}= m_1^2 + m_2^2 - 2k_1{\cdot}\R k_2 \hspace{3mm}=  -2k_1{\cdot}\R k_2\\
s_{\bar12}& = - (\R k_1 + k_2)^2 \hspace{7mm}= -(\R k_1)^2 - k_2^2 -2\R k_1{\cdot}k_2 \hspace{1.2cm}= m_1^2 + m_2^2 - 2\R k_1{\cdot}k_2\hspace{3mm}=  - 2\R k_1{\cdot}k_2\\
s_{\bar1\bar2}& = -(\R k_1 + \R k_2)^2 \,\,\,\,\,\,= -(\R k_1)^2 -(\R k_2)^2 -2
\R k_1{\cdot}\R k_2 \,\,\,\,= 
m_1^2 + m_2^2 - 2\R k_1{\cdot}\R k_2 =  - 2\R k_1{\cdot}\R k_2\\
s_{2\bar2}& = -\bigg(\frac{k_2}{2} + \frac{\R k_2}{2}\bigg)^2 \hspace{3mm}= -\bigg(\frac{k_2}{2}\bigg)^2 - \bigg(\frac{\R k_2}{2}\bigg)^2 -\frac{k_2{\cdot}\R k_2}{2} \hspace{2mm}= \frac{m_2^2}{4} + \frac{m_2^2}{4} - \frac{k_2{\cdot}\R k_2}{2}\,\,\,\,= - \frac{k_2{\cdot}\R k_2}{2}\\&
\end{aligned}}\label{eq:kinInv_2tC}
\end{equation}

where all the $\R k_i/2$ momentum are now label with $\bar i$. Using the momentum  conservation one can eliminate $\bar 2$, obtaining
\begin{equation}
\begin{aligned}
\#_{s_{i\bar 2}} =& 3\quad\textbf{linear dependent k.i.}\,\, [i\in\{1,\bar 1,2\}]\\ 
\#_{s_{lh}}=& \begin{pmatrix}3\\2\end{pmatrix}= 3\quad \textbf{linear independent k.i.}\,\,[l,h\in\{1,\bar 1,2\}; l\ne h; 1<\bar1<2]
\end{aligned}
\end{equation}

\begin{equation}
\begin{aligned}
s_{1\bar 2} &= -(k_1 {+} \R k_2)^2 = {-}(\R k_1 {+} k_2)^2 \equiv s_{\bar 12}:=u\,;\quad
s_{\bar1\bar2} = {-} (\R k_1 {+} \R k_2)^2 = {-}(k_1 {+} k_2)^2 \equiv s_{12}:=t\\
s_{2\bar2} &= -\bigg(\frac{k_2}{2} + \frac{\R k_2}{2}\bigg)^2= -\bigg(\frac{k_1}{2} + \frac{\R k_1}{2}\bigg)^2 \equiv s_{1\bar1}:= s\,\,.
\end{aligned}\label{eq:KinMomCons2tC}
\end{equation}
\\
The on-shellness condition for $\bar2$ reads
\begin{equation}
\begin{aligned}
\boxed{4\,s_{1\bar 1} + s_{12} + s_{\bar 12} \equiv s + \frac{u}{4} + \frac{t}{4} = 0}
\end{aligned}\label{eq:onshellness_2tC}
\end{equation}
where $s$ is the open string channel, while both $t$ and $u$ are closed string channels. 

%%%%%%%%%%%%%%%%%%%%%%%%%%%%%%%%%%%
%%%%%%%%%%%%%%%%%%%%%%%%%%%%%%%%%%%
%%%%%%%%%%%%%%%%%%%%%%%%%%%%%%%%%%%
%%%%%%%%%%%%%%%%%%%%%%%%%%%%%%%%%%%
%%%%%%%%%%%%%%%%%%%%%%%%%%%%%%%%%%%
%%%%%%%%%%%%%%%%%%%%%%%%%%%%%%%%%%

\end{appendices}

\end{document}